\documentclass{aa}  
\usepackage{hyperref}
\hypersetup{
colorlinks=true,
citecolor=blue,
linkcolor=blue,
urlcolor=blue
}
\usepackage[nameinlink,capitalise]{cleveref}
\usepackage{amsfonts,amsmath,amssymb}
\usepackage{url}
\usepackage{float}
\usepackage{subcaption}
\usepackage{natbib}
\usepackage{ulem}
\usepackage{graphicx}
\usepackage{stfloats}

\usepackage{txfonts}
\usepackage{bm}

\newcommand{\hi}{\textsc{Hi}}

\let\Gamma\varGamma
\let\Delta\varDelta
\let\Theta\varTheta
\let\Lambda\varLambda
\let\Xi\varXi
\let\Pi\varPi
\let\Sigma\varSigma
\let\Upsilon\varUpsilon
\let\Phi\varPhi
\let\Psi\varPsi
\let\Omega\varOmega

\begin{document}

   \title{Beyond power spectrum to unveil systematics on \hi\ intensity maps}

\author{
    Pauline Gorbatchev\inst{1, 2} \and
    Jean-Luc\ Starck\inst{2, 3} \and
    Stefano Camera\inst{4, 5, 6, 7} \and
    Marta Spinelli\inst{4, 8}
}

\institute{
    Department of Physics, University of Crete, Greece
    \and
    Institutes of Computer Science and Astrophysics, Foundation for Research and Technology - Hellas (FORTH), Greece \\
    \email{pauline.gorbatchev@ia.forth.gr}
    \and
    Université Paris-Saclay, Université Paris Cité, CEA, CNRS, AIM, 91191, Gif-sur-Yvette, France
    \and
    Dipartimento di Fisica, Universit\`a degli Studi di Torino, via P.\ Giuria 1, 10125 Torino, Italy
    \and
    INFN -- Istituto Nazionale di Fisica Nucleare, via P.\ Giuria 1, 10125 Torino, Italy
    \and
    INAF -- Istituto Nazionale di Astrofisica, Osservatorio Astrofisico di Torino, strada Osservatorio 20, 10025 Pino Torinese, Italy
    \and
    Department of Physics \& Astronomy,
    University of the Western Cape, Cape
    Town 7535 , South Africa
    \and
    Observatoire de la Cote d’Azur, Laboratoire Lagrange, Bd de l’Observatoire, CS 34229, 06304 Nice cedex 4, France
}

   \date{Received Month XX, XXXX; accepted Month XX, XXXX}

  \abstract
{\hi\ intensity mapping is a promising technique to probe large-scale structure, traditionally analyzed via two-point statistics such as the angular power spectrum. This latter technique has proven very powerful but may miss key non-Gaussian information present in the signal.}
{We extend the starlet $\ell_1$-norm, a multi-scale higher-order statistic previously applied to weak lensing maps, to the brightness temperature fluctuations of the \hi\ density field. The \hi\ signal is highly non-Gaussian at late times ($z < 1$) due to nonlinear structure growth, motivating the use of advanced summary statistics.}
{We simulated full-sky \hi\ lognormal brightness temperature maps using CAMB and GLASS, generating 10,000 realizations with associated cosmological parameters. We extracted both the starlet $\ell_1$-norm and angular power spectrum from these maps. Using the \texttt{JaxILI} framework, we performed neural density estimation for implicit likelihood inference. The analysis considered simulated maps incorporating realistic noise and telescope beam, capturing the impact of observational effects on parameter inference. In this work, we focus on the redshift range $0.4 < z < 0.45$, chosen to match the interval already targeted by existing MeerKLASS observations. We also assess the sensitivity of these statistics to observational systematics, highlighting their potential for identifying and mitigating contaminants in \hi\ intensity maps. }
{The starlet $\ell_1$-norm significantly outperforms the angular power spectrum in constraining cosmological parameters, achieving almost a 3x improvement in the figure of merit relative to the angular power spectrum by capturing non-Gaussian features missed by two-point statistics. Moreover, our results suggest that the starlet $\ell_1$-norm is robust to several of the systematic effects included in our simulations.}
{Our findings highlight the potential of multi-scale higher-order statistics such as the starlet $\ell_1$-norm to enhance cosmological inference from future \hi\ intensity mapping surveys.}

\maketitle

\section{Introduction}

    The upcoming SKA Observatory\footnote{\url{https://www.skao.int/en}.} (SKAO) will play a central role in advancing cosmology by enabling large-volume surveys, wide–sky-coverage surveys of the large-scale structure through neutral hydrogen (HI) intensity mapping \citep{Battye2013, Bull15}. This technique measures the integrated 21 cm emission from unresolved galaxies, providing an efficient means to trace the underlying matter distribution across cosmic time \citep{Pritchard2012, Ansari2012, Santos2015}. In particular, intensity mapping at intermediate redshifts, such as \( z < 1 \), probes the Universe during the onset of accelerated expansion and captures the nonlinear growth of structure \citep{Peebles1980, Bernardeau2002}. The SKAO is expected to probe neutral hydrogen emission up to $z \approx 3$, whereas MeerKLASS \citep{Wang21, Cunnington22, MeerKLASS} will reach $z \approx 1.45$ with UHF band. Since the MeerKLASS survey in the L band ($z \simeq 0.40{-}0.46$) has already yielded initial observational results, we adopt observational settings that emulate the characteristics of this survey in the present analysis.

Traditional two-point statistics have been the standard tool for extracting cosmological information from \hi\ intensity mapping surveys, as they effectively capture the Gaussian part of the signal \citep{Pourtsidou2017, VillaescusaNavarro2018}. For full-sky or sufficiently wide surveys, for which statistical isotropy is a valid approximation, the angular power spectrum, $C_\ell$, provides the natural and convenient statistical description of the field. However, this approach assumes that the signal is close to Gaussian, which is no longer valid at low redshifts. Higher-order statistics and wavelet-based summaries have been shown to improve parameter constraints during the epoch of reionization \citep{2020_EoR_3PCF, 2025_EoR_BN_BS} and cosmic dawn \citep{2025_WST_CD_EoR}, yet they remain largely unexplored for post-reionization intensity mapping, where the power spectrum remains the principal tool for analysis. At \( z \sim 0.425 \) where MeerKLASS data is available, the \hi\ distribution is shaped by nonlinear gravitational evolution and complex astrophysical processes, resulting in a strongly non-Gaussian 21 cm brightness temperature field. In this regime, the power spectrum alone cannot capture the full statistical content of the data, and much of the cosmological information encoded in higher-order correlations is effectively lost.

To recover this lost information, we explore an alternative summary statistic: the starlet \( \ell_1 \)-norm, a multi-scale, higher-order measure based on the isotropic undecimated wavelet transform (\citealt{Starck_2006}). The starlet transform decomposes the signal into spatially localized features across multiple angular scales, and the \( \ell_1 \)-norm captures the strength of these features in a way that is particularly sensitive to non-Gaussian structures such as filaments, clusters, and voids, key signatures of nonlinear structure formation. This approach has already proven effective in weak lensing studies, where it has been shown to outperform traditional two-point statistics and peaks statistics in constraining cosmological parameters, particularly in non-Gaussian regimes \citep{Ajani2021, Ajani2023}. These successes motivate the application of this approach to \hi\ intensity mapping, where similar nonlinear features are present. In this context, and given the improving quality of \hi\ intensity-mapping data, we focus on simulations, anticipating that forthcoming observations will contain non-Gaussian information.

In this work, we perform a detailed comparison of cosmological parameter constraints derived from the angular power spectrum \( C_\ell \) and the starlet \( \ell_1 \)-norm using full-sky \hi\ intensity map simulations at redshift \( z = 0.425 \) (using \texttt{GLASS}\footnote{\label{GLASS}\url{https://GLASS.readthedocs.io/stable/}} and \texttt{CAMB}\footnote{\label{camb}\url{https://camb.info}}), which are designed to mimic the MeerKLASS survey and can be easily extended to the type of data expected from upcoming \hi\ intensity mapping surveys, including the SKA-Mid (\citealt{redbook, Battye2013, Santos2015}). We aim to assess the ability of these two summary statistics to constrain three key cosmological parameters: the cold dark matter density, \( \Omega_{\rm c} \), the dimensionless Hubble parameter, \( h \), and the amplitude of primordial scalar perturbations, \( A_{\rm s} \). To obtain these constraints, we employ a simulation-based inference (SBI) approach (\citealt{Cranmer2020, Alsing2019}), which not only avoids assuming a predefined likelihood but also reduces computational costs compared to traditional likelihood-based methods. This is particularly advantageous when dealing with the complex, non-Gaussian features of the 21 cm signal, as it allows the summary statistics to be directly matched to the data without needing a full likelihood model. The starlet \( \ell_1 \)-norm, in particular, captures these non-Gaussian features more effectively than the power spectrum (\citealt{Ajani2023}), making it a promising tool for cosmological parameter estimation in the context of intensity mapping.

The paper is organized as follows. Section \ref{sec:sims} details the simulations and \hi\ modeling used in this work, along with the setup for the full-sky intensity maps. Section \ref{sec:summary} describes the summary statistics. Section \ref{sec:sbi} provides a detailed description of the SBI methodology used for cosmological parameter estimation. In Sect. \ref{sec:results} we present the results of our parameter estimation, comparing the constraints obtained from the angular power spectrum and the starlet \( \ell_1 \)-norm. The validation of the inference pipeline is presented in Appendix \ref{sec:validation}. Finally, we discuss the implications of these results for future \hi\ intensity mapping surveys, such as those planned for the SKAO, and conclude in Sect. \ref{sec:conclusion}.

\section{Simulations for \hi\ maps}\label{sec:sims}
To simulate \hi\ intensity maps and model their angular power spectra, we developed a forward model based on the generation of correlated lognormal matter fields using the \texttt{GLASS} framework. The linear matter power spectrum was computed using \texttt{CAMB}, configured with cosmological parameters: a reduced Hubble constant, $h$, cold dark matter density parameter, $\Omega_{\rm c}$, baryonic matter density parameter, $\Omega_{\rm b}$, and primordial amplitude and spectral index, \( A_{\rm s} \) 
and \(n_{\rm s}\), respectively, including nonlinear corrections via setting to \texttt{True} the \texttt{NonLinear\_both} flag.
We computed the angular power spectra, $C_\ell^{\delta\delta}$, for a set of radial shells with fixed comoving thickness, which we set to $50\,\mathrm{Mpc}$.
Each shell was described by a redshift-dependent window function, $W_i(z)$, and the angular spectra were computed using the \texttt{GLASS} interface to \texttt{CAMB}. These spectra were then used to generate correlated Gaussian random fields on the sphere, which were subsequently transformed into lognormal fields to account for the non-Gaussianity of the matter distribution on small scales.

The \hi\ brightness temperature field was constructed by weighting the lognormal density fields with the mean \hi\ brightness temperature, $\bar{T}_{\rm b}(z)$, and the \hi\ bias, $b_{{\hi}}(z)$, both modeled as redshift-dependent functions following \citep[see, e.g.,][]{Battye2013,Bull15,Casas2022};
namely,
\begin{equation}
   \bar{T}_{\rm b}(z) = 189\,\Omega_{{\hi}}(z)\, h\,\frac{(1 + z)^2\,H_0}{H(z)}\,\mathrm{mK}\;, 
\end{equation} 
with \( \Omega_{{\hi}}(z) = 4 \times 10^{-4}\,(1 + z)^{0.6} \). 
The \hi\ bias was parameterized as 
\begin{equation}
    b_{{\hi}}(z) = 0.6 + 0.3\,(1 + z)\;.
\end{equation} 
The redshift range is represented by a single tomographic bin, covering the full interval. Redshift uncertainties were modeled as Gaussian errors with a standard deviation of $\sigma_{z,0} = 0.0001$ scaled as $\sigma_z(z) = \sigma_{z,0}\,(1 + z)$.
For this bin, we constructed the \hi\ brightness temperature fluctuation map via
\begin{equation}
    \delta T_{{\hi}}(\hat{\bm n}) = \bar{T}_{\rm b}(z) \, b_{{\hi}}(z) \, \delta(\hat{\bm n}, z)\;,
\end{equation}
where \(\hat{\bm n}\) is an angular position on the sky, and $\delta T_{\rm HI}(\hat{\bm n})$ is the simulated map. The final output consists of a full-sky HEALPix map $\delta T_{{\hi}}(\hat{\bm n})$, representing the spatially varying \hi\ brightness–temperature fluctuations across the sky. This procedure provides a fast and flexible means of generating tomographic \hi\ intensity maps that capture both large-scale correlations and redshift evolution.
Figure \ref{fig:mollweide} displays full-sky lognormal maps obtained for different cosmologies at a fixed realization, whereas Fig. \ref{fig:map_realisations} shows maps corresponding to different realizations at the same cosmology.

\subsection{Telescope beam}
To simulate a more realistic observational setup, we modeled the telescope's finite angular resolution by convolving the sky map with a Gaussian beam. We studied two beam cases, $1.34^\circ$ in a MeerKLASS L-band single-dish-like configuration  \citep{Sia,Carucci25} and $0.5^\circ$ for future interferometric configurations (e.g., SKA-Mid with 15m dish diameter\footnote{\url{https://www.skao.int/en/science-users/118/ska-telescope-specifications}}). While these values are representative of the MeerKLASS beam, they span a broader range to assess the effect of beam smoothing. This beam smoothing was implemented by applying a Gaussian filter in harmonic space. The corresponding beam-transfer function is
\begin{equation}
b_\ell = \exp\left[-\frac{1}{2}\,\ell\,(\ell+1)\,\sigma_{\mathrm{beam}}^2\right]\;,
\end{equation}
where $\sigma_{\mathrm{beam}} = \theta_{\mathrm{FWHM}}/\sqrt{8 \ln 2}$ is the standard deviation of the Gaussian in radians and $\theta_{\mathrm{FWHM}} = 0.5^\circ$. This filtering suppresses power on angular scales smaller than the beam size, effectively smoothing out features finer than the beam resolution. We show in Fig. \ref{fig:mollweide2}b an example of a beamed map.

\subsection{Instrumental noise}

\begin{figure*}
    \centering
    \includegraphics[width=\textwidth]{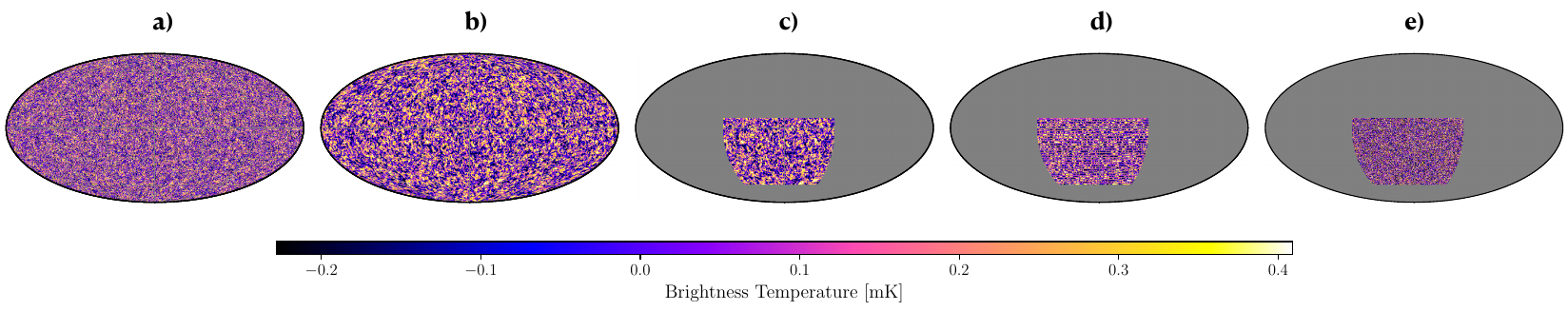}
    \caption{Mollweide projections of the simulated \hi\ intensity mapping signal at $z = 0.425$ for fiducial cosmology with systematics.
\textit{Panel a}: Clean simulated map.
\textit{Panel b}: Clean map after convolution with the telescope beam.
\textit{Panel c}: Same map masked with $f_{\rm sky} = 0.19$.
\textit{Panel d}: Same map with multiplicative stripe patterns.
\textit{Panel e}: Same map after addition of instrumental noise. }
    \label{fig:mollweide2}
\end{figure*}

We modeled the thermal noise per pixel as uncorrelated Gaussian noise with zero mean and variance given by (root mean square temperature fluctuation)
\begin{equation}
\sigma_{\rm N}^2 = \frac{T_{\mathrm{sys}}^2}{2 \, t_{\mathrm{pix}} \, \Delta \nu}\;,
\end{equation}
where  \( \Delta \nu \) is the frequency bandwidth in hertz, given by
\begin{equation}
    \Delta \nu = \nu_{\rm high} - \nu_{\rm low} = \frac{1420\,\mathrm{MHz}}{1+z_{\min}} - \frac{1420\,\mathrm{MHz}}{1+z_{\max}}\;.
\end{equation}
We chose \(z_{\min}=0.4\) and \(z_{\max}=0.45\) to mimic MeerKLASS L-band measurements. Then, \( t_{\mathrm{pix}} = t_{\mathrm{obs}} \, N_{\mathrm{dish}} \, \Omega_{\mathrm{pix}}/\Omega_{\mathrm{survey}} \) is the effective integration time per pixel,
\( N_{\mathrm{dish}} \) is the number of dishes (64 for MeerKlass-like survey), and \( t_{\mathrm{obs}} \) the total observing time, which we set to $2\,000$ hours (corresponding to $7.2\times10^6$ seconds). Finally, \( T_{\mathrm{sys}} \) is the system temperature defined as \citep[see, e.g.,][]{Sheean2021}
  \begin{equation}
    T_{\rm sys} \approx T_{\rm rec} + T_{\rm CMB}+ 10\,\left[\frac{1420\,\mathrm{MHz}}{(1+z)\,400\,\mathrm{MHz} }\right]^{-2.75}\,\mathrm{K}\;.
\end{equation}
The first term corresponds to the temperature of the receiver, which we set to $20\,\mathrm{K}$, the second term is the CMB contribution, corresponding to $2.7\,\mathrm{K}$, and the last term is the sky temperature contribution due to Galactic synchrotron emission. We show in Fig. \ref{fig:mollweide2}e a map with instrumental noise applied.

\begin{table}[]
    \centering
    \caption{Instrumental specifications for MeerKLASS-like survey.}
    \begin{tabular}{ccccc}
        \hline
        $T_{\text{sys}}^{\text{rec}}$ [K] & $f$ & $t_{\text{obs}}$ [s] & $N_{\text{dish}}$ & $\Omega_{\text{pix}}$ [deg$^2$] \\
        \hline
        20 & 1 & 7\,200\,000 & 64 & 0.00328 \\
        \hline
    \end{tabular}
    \label{tab:your_label}
\end{table}

\subsection{Survey area}
\begin{figure}
    \centering
    \includegraphics[width=0.35\textwidth]{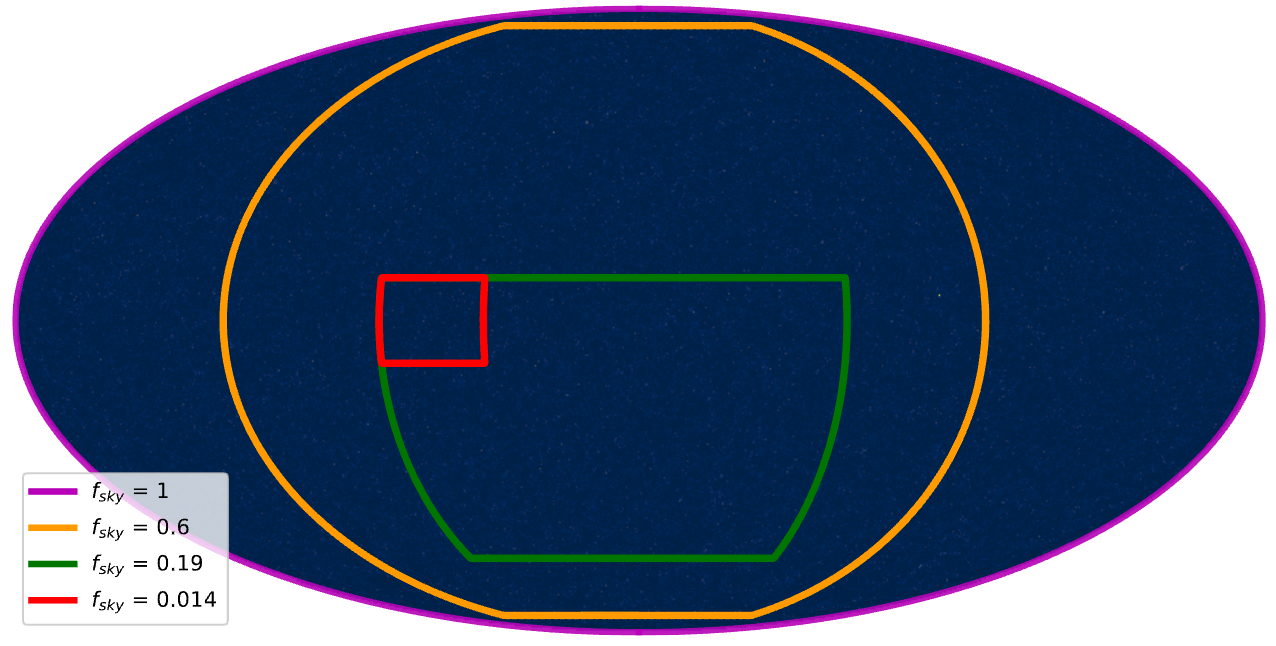}
    \caption{Sky map with masks. The contours show the mask edges: 
    magenta for $f_\mathrm{sky}=1$, 
    orange for $f_\mathrm{sky}=0.6$, 
    green for $f_\mathrm{sky}=0.19$, 
    and red for $f_\mathrm{sky}=0.014$.}
    \label{fig:masks}
\end{figure}

In this analysis, we considered different possible values for the sky coverage ($f_{\rm sky}$) of an intensity mapping survey. 
To mimic these different coverages, we applied a binary mask to the full-sky map. This operation is expressed as
\begin{equation}
\delta T'_{\rm HI}(\hat{\bm n}) = \delta T_{\rm HI}(\hat{\bm n}) \, \mathcal{M} (\hat{\bm{n}})\;,
\end{equation}
where $\delta T_{\rm HI}(\hat{\bm n})$ is the simulated \hi\ intensity map, $\mathcal{M} (\hat{\bm{n}})$ is the binary mask taking values $1$ for retained pixels and $0$ for excluded pixels, and $\delta T'_{\rm HI}(\hat{\bm n})$ is the masked map. We show in Fig. \ref{fig:masks} the different masks we use in Sect. \ref{sec:results}. We employed three masks with different sky fractions: the first used $f_{\rm sky} = 0.60$, representing a case with optimistic sky coverage. The second adopted $f_{\rm sky} = 0.19$,  a value that current intensity mapping surveys are already attaining. The third, with $f_{\rm sky} = 0.014$, represents the sky coverage of the already published results of the MeerKLASS collaboration \citep{Wang21, MeerKLASS}.

\subsection{Systematic stripe pattern}
With the aim of assessing the added value of starlet \(\ell_1\)-norm statistics in detecting and mitigating systematic effects, we considered a controlled toy model of large-scale scanning-related artifacts, implemented as sinusoidal modulations of the \hi\ brightness temperature across the sky, and studied their impact on the analysis pipeline. 
These sinusoidal effect can be motivated by the possible presence of azimuth-dependent systematics coming from the particular scanning strategy of MeerKLASS where the 64 MeerKAT antennas are moved back and forward in azimuth at constant elevation \citep{Wang21}. Such systematics could translate into a declination-dependent pattern on the sky.
We modeled an artificial large-scale systematic effect in the form of a sinusoidal modulation along the colatitude, $\theta$, of the HEALPix pixels and we considered two possible cases. This prescription produces alternating bright and dark bands across the sphere, which we refer to as a ``stripe'' pattern.

\subsubsection{Multiplicative stripes}
We first considered a multiplicative systematic. This results in an observed \hi\ temperature on the sky of the form
\begin{equation}
\delta T_{\mathrm{sys}}(\hat{\bm n}) = \delta T_{\rm HI}(\hat{\bm n}) \,\left[ 1  + A \, \sin(\omega \, \theta )\right]\;,
\end{equation}
where $A$ is the modulation amplitude and $\omega$ controls the stripe frequency. This pattern was applied over the entire map.  We show in Fig. \ref{fig:mollweide2}d a masked map with multiplicative stripe patterns, and in Fig. \ref{fig:maps_sys}b the same map zoomed in.

\subsubsection{Additive stripes}
We also introduced an additive sinusoidal modulation along the colatitude, $\theta$, of the HEALPix pixels. In this case, the observed map is described as
\begin{equation}
\delta T_{\mathrm{sys}}(\hat{\bm{n}}) = \delta T_{{\hi}}(\hat{\bm{n}}) + A \, \sin(\omega \, \theta)\;,
\end{equation}
where now the angular frequency, $\omega = 2\,\pi / \Delta \theta_{\mathrm{stripe}}$, is set by the chosen stripe width, $\Delta \theta_{\mathrm{stripe}}$.  The amplitude, $A$, is scaled such that the effect corresponds to a small perturbation relative to the signal, allowing us to assess the sensitivity of our analysis pipeline to this type of coherent, large-scale systematic.

\subsection{Contamination from foreground residuals}
The separation of the much stronger foreground emission from the pristine HI signal is one of the key challenges of intensity mapping as at $\sim 1\,\mathrm{GHz}$ it is $5$--$6$ orders of magnitude brighter than the cosmological \hi\ \citep[see, e.g.,][]{Ansari2012,Alonso2014,Wolz2014,Carucci2024,Mertens,Sia}.
In this work, we do not simulate the full procedure of foreground cleaning but we explore the impact of foreground residuals in the HI maps in a simplistic scenario: we assume that the residual foreground emission in the cleaned maps strongly correlated with the original foreground contaminants. We again explore two cases: the residual presence of diffuse Galactic synchrotron and extragalactic radio point sources. 

\subsubsection{Diffuse Galactic synchrotron residuals}
We included a Galactic synchrotron residual component in the mock \hi\ intensity mapping maps. We show an example of a map in Fig. \ref{fig:maps_sys}c.  The residual was constructed by scaling a template of diffuse Galactic synchrotron emission -- obtained with \texttt{PYSM 3}\footnote{\url{https://pysm3.readthedocs.io/en/latest/}} \citep{2021JOSS....6.3783Z} following the method described in \cite{Sia} -- by factors of $0.0005\%$, $0.001\%$, and $0.002\%$ of the original map. When recast in terms of the underlying \hi\ signal, these correspond, respectively, to $0.09$, $0.18$, and $0.36$ times the RMS of the \hi\ temperature field. 
A suppression by this factor therefore yields residual fluctuations at the $\sim 0.1\,\mathrm{mK}$ level (for the $0.0005\%$ case), comparable to the expected amplitude of the \hi\ signal, and provides a controlled toy model for the level of contamination that could remain after foreground removal techniques. This approach allows us to explore the potential bias induced by low-level Galactic synchrotron residuals.

\subsubsection{Extragalactic point source residuals}
\begin{figure*}
    \centering
    \includegraphics[width=0.8\textwidth]{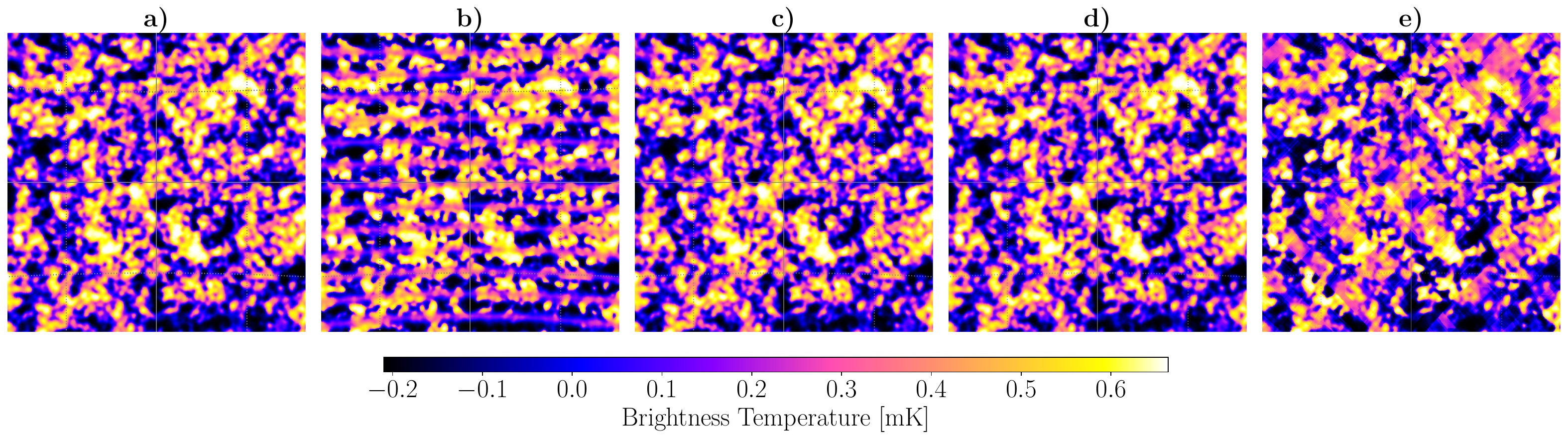}
    \caption{Zoomed maps of different systematic effects + beam (0.5 deg).
\textit{Panel a}: \hi\ beamed map with no systematic effect. \textit{Panel b}: \hi\ beamed map with multiplicative stripe stripe patterns. \textit{Panel c}: \hi\ map with residual Galactic synchrotron contamination beamed. \textit{Panel d}: \hi\ map with residual extragalactic point source contamination beamed. \textit{Panel e}: \hi\ beamed map with multiplicative X-shaped scanning patterns.}
        
    \label{fig:maps_sys}
\end{figure*}

\begin{figure*}
    \centering
    \includegraphics[width=0.7\textwidth]{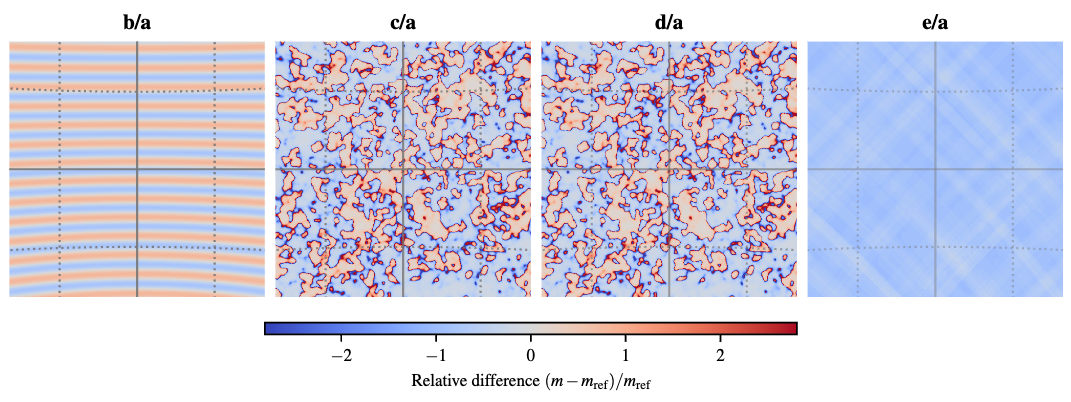}
    \caption{Relative difference of maps from Fig. \ref{fig:maps_sys} with respect to a.}
        
    \label{fig:maps_sys_diff}
\end{figure*}

To evaluate the impact of unresolved extragalactic point source contamination on the \hi\ intensity maps, we included a residual extragalactic point source component in the mock data. We show an example of a map in Fig. \ref{fig:maps_sys}d. The residual map was generated using a template of extragalactic radio point sources -- obtained with \texttt{PYSM 3} \citep{2021JOSS....6.3783Z} as in \cite{Battye2013}, \cite{Alonso2014}, and \cite{Sia} -- and its amplitude was scaled by factors of $0.001\%$, $0.0025\%$, and $0.005\%$ of the original template. When recast in terms of the underlying \hi\ signal, these levels correspond, respectively, to $0.005$, $0.013$, and $0.026$ times the RMS of the \hi\ temperature field. This corresponds to residual fluctuations well below the characteristic \hi\ brightness fluctuations, representing a realistic level of contamination expected after standard point source subtraction or masking. This setup provides again a controlled framework to assess the potential bias induced by residual extragalactic point source in the foreground-cleaned maps. Figures \ref{fig:mollweide2} and \ref{fig:maps_sys} display the effects discussed previously. Figure \ref{fig:maps_sys_diff} shows the relative difference of each case with systematic effects in Fig. \ref{fig:maps_sys} with respect to the map with no systematic effects. Additionally, Fig. \ref{fig:PS_S} shows the angular power spectra of Galactic synchrotron emission and extragalactic point sources, for intermediate contamination levels of 0.001\% and 0.0025\%, respectively, together with the \hi\ clean map prior to the inclusion of any instrumental or observational effects.

\subsection{Scanning patterns}\label{subsec:Xshape}
As was mentioned previously, the scanning strategy of the MeerKLASS survey consists of a synchronised azimuthal sweep of all antennas while the elevation is kept fixed. In addition, the same sky patch is typically observed both during its rise and its set. This combination produces a characteristic X-shaped pattern in the noise \citep{Wang21}.

We generated synthetic coverage-weighted \hi\ maps by applying X-shaped scanning patterns on our masked lognormal maps. Each X had a width of $30^\circ$ and thickness of $1^\circ$ and was placed iteratively on randomly selected pixels, with a 10$\%$ probability of producing a “heavy X spot” consisting of multiple overlapping Xs. The placement was repeated until the coverage map fully sampled the input lognormal map. The resulting map was normalized by its maximum value to produced a weighted mask (see Fig. \ref{fig:scan_map_masks}) with values up to 1, where higher values correspond to regions with more overlapping X. This weighted mask was then applied to our lognormal maps, producing a coverage-dependent weighted map that reflects scanning pattern and redundancy (see Fig. \ref{fig:scan_map} for multiplicative mask and Fig. \ref{fig:scan_map_additive} for an additive mask normalized by 0.05 to achieve patterns on the order of magnitude of the \hi\ signal, and the extracted summary statistics are shown in Figs. \ref{fig:comp_scan} and \ref{fig:comp_scan_add}). 

\section{Summary statistics}\label{sec:summary}
\subsection{Angular power spectrum}

The angular power spectrum, \( C_\ell \), provides a second-order statistical characterization of the temperature fluctuations in \hi\ intensity maps. It quantifies the variance of the map as a function of angular scale and is particularly well suited for probing Gaussian features in the cosmological signal. The spherical harmonic coefficients, \( a_{\ell m} \), were estimated from the map using the \texttt{anafast} function from the \texttt{healpy} package. Under the assumption of statistical isotropy, the angular power spectrum was defined as
\begin{equation}
C_\ell = \frac{1}{2\,\ell + 1} \sum_{m = -\ell}^{\ell} |a_{\ell m}|^2\;,
\end{equation}
which represents the average variance of the harmonic coefficients at a given multipole, \( \ell \). This estimator is unbiased under full-sky, noise-free conditions and can be efficiently computed up to a maximum multipole, \( \ell_{\max} \approx 2 N_{\text{side}} - 1 \), where \( N_{\text{side}} \) is the HEALPix resolution parameter. The \( C_\ell \) were directly computed from our simulated map described in Sect. \ref{sec:sims}.

\begin{figure}
    \centering 
    \includegraphics[width=0.85\columnwidth]{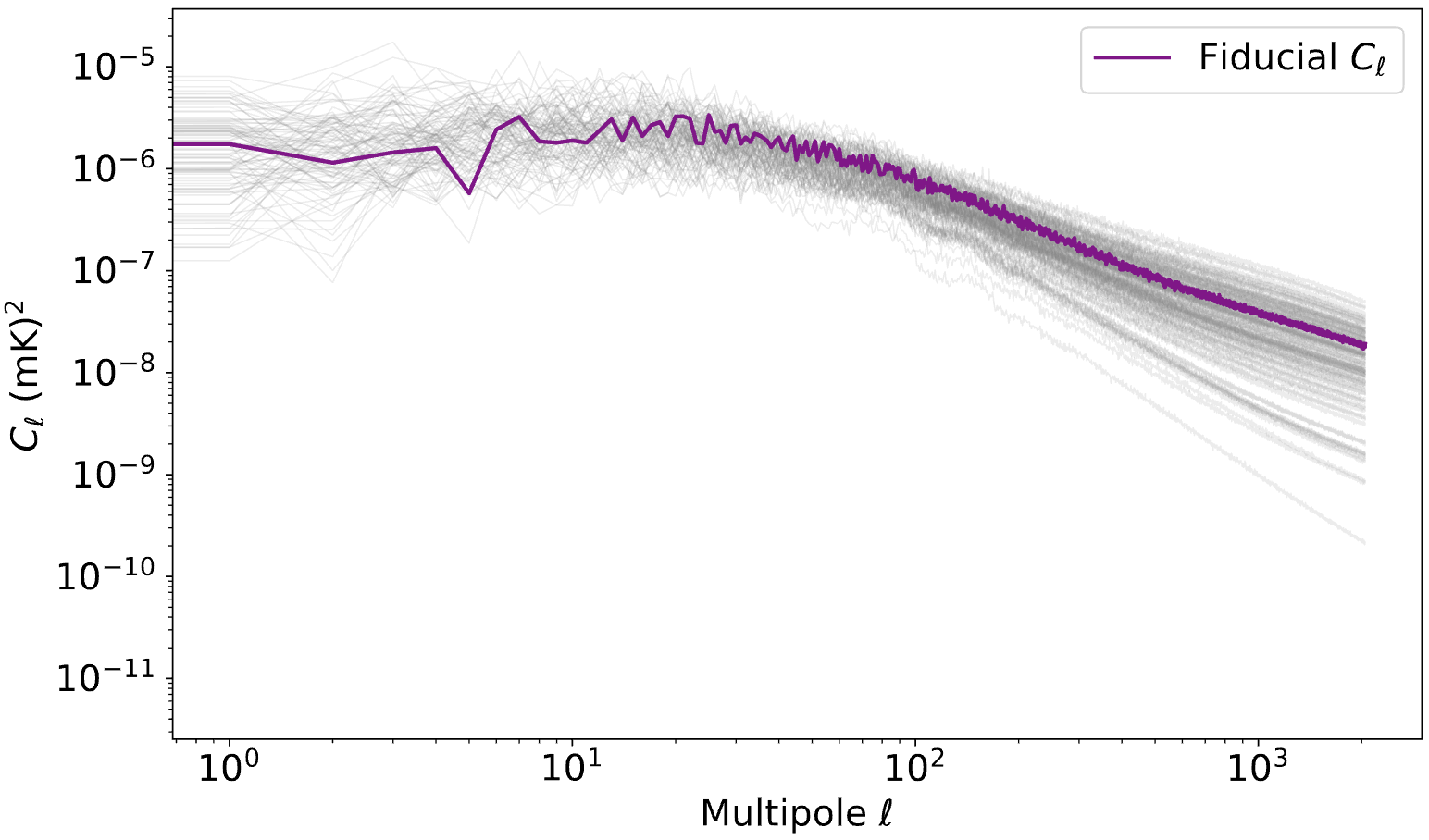}
    \caption{Variability of angular power spectrum over 100 randomly selected realizations with different cosmologies (gray) varying as in Table \ref{tab:example_table} with fiducial \( C_\ell \) (purple).}
    \label{fig:Cl}
\end{figure}

Figure \ref{fig:Cl} shows the variability of the angular power spectrum across realizations and cosmologies. For MeerKLASS-like simulations, we set the lower multipole bound to \( \ell_{\min} = 10 \). This is because the largest angular scales (\( \ell \lesssim 10 \)) are particularly susceptible to contamination from foreground residuals and instrumental systematics such as \(1/f\) noise \citep{2021MNRAS,foreground} and calibration drifts, which are challenging to model and remove accurately. Moreover, the combination of beam effects and partial sky coverage introduces significant mode coupling and leakage at low \( \ell \), which can bias the power spectrum estimation. By excluding these large-scale modes, we focused on the more reliable intermediate-to-small angular scales, where the cosmological signal is better preserved and the systematics are more controllable.
 These spectra were subsequently used as direct summary statistics in our SBI pipeline.

\subsection{Starlet $\ell_1$-norm}

To access non-Gaussian information encoded in the \hi\ maps, we computed the $\ell_1$-norm of the isotropic undecimated wavelet (a.k.a.\ starlet) coefficients on the sphere. This statistic quantifies the sparsity of the signal across multiple angular scales and is sensitive to structures that are not captured by the angular power spectrum alone.
\begin{figure}
    \centering
    \includegraphics[width=0.6\columnwidth]{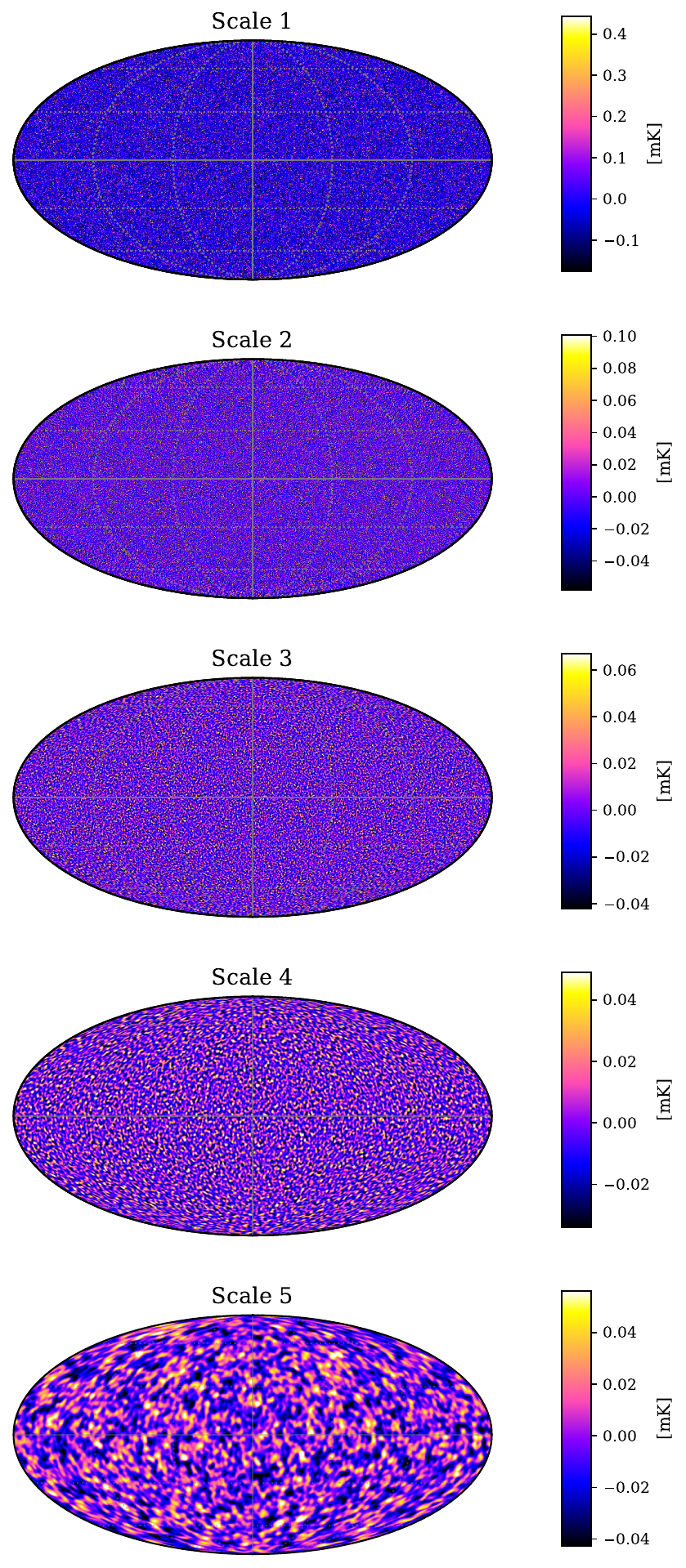}
    \caption{Starlet decomposition on five scales (including coarse scale) of a simulated lognormal map.}
    \label{fig:starlet_decomposition}
\end{figure}

Let \( M(\hat{\bm n}) \) denote the \hi\ temperature fluctuation map on the sphere, where \( \hat{\bm n} \) is the sky direction. The spherical starlet transform \citep{Starck_2006} decomposes this map into a set of wavelet coefficients, \( w_j(\hat{\bm n}) \), at each scale, \( j = 1, \dots, J \), and a coarse residual map, \( c_J(\hat{\bm n}) \), as follows \citep{Ajani2021}:
\begin{equation}
M(\hat{\bm n}) = c_J(\hat{\bm n}) + \sum_{j=1}^J w_j(\hat{\bm n})\;.
\end{equation}
Each wavelet scale, \( w_j(\hat{\bm n}) \), captures angular features localized in a specific band of multipoles. The coefficients were normalized by a scale-dependent factor, \( \mathcal{N}_j \), to account for varying amplitudes across scales \footnote{The scale-dependent normalization factor, \( \mathcal{N}_j \), was computed by applying a first-generation starlet transform to a delta-function image by the \texttt{CosmoStat} package.}.  We show an example of decomposition of a clean \hi\ map on five starlet scales including the coarse scale in \cref{fig:starlet_decomposition}. The normalized coefficients were then binned in the signal-to-noise ratio (S/N) space.

\begin{figure}
    \centering
    \includegraphics[width=0.85\linewidth]{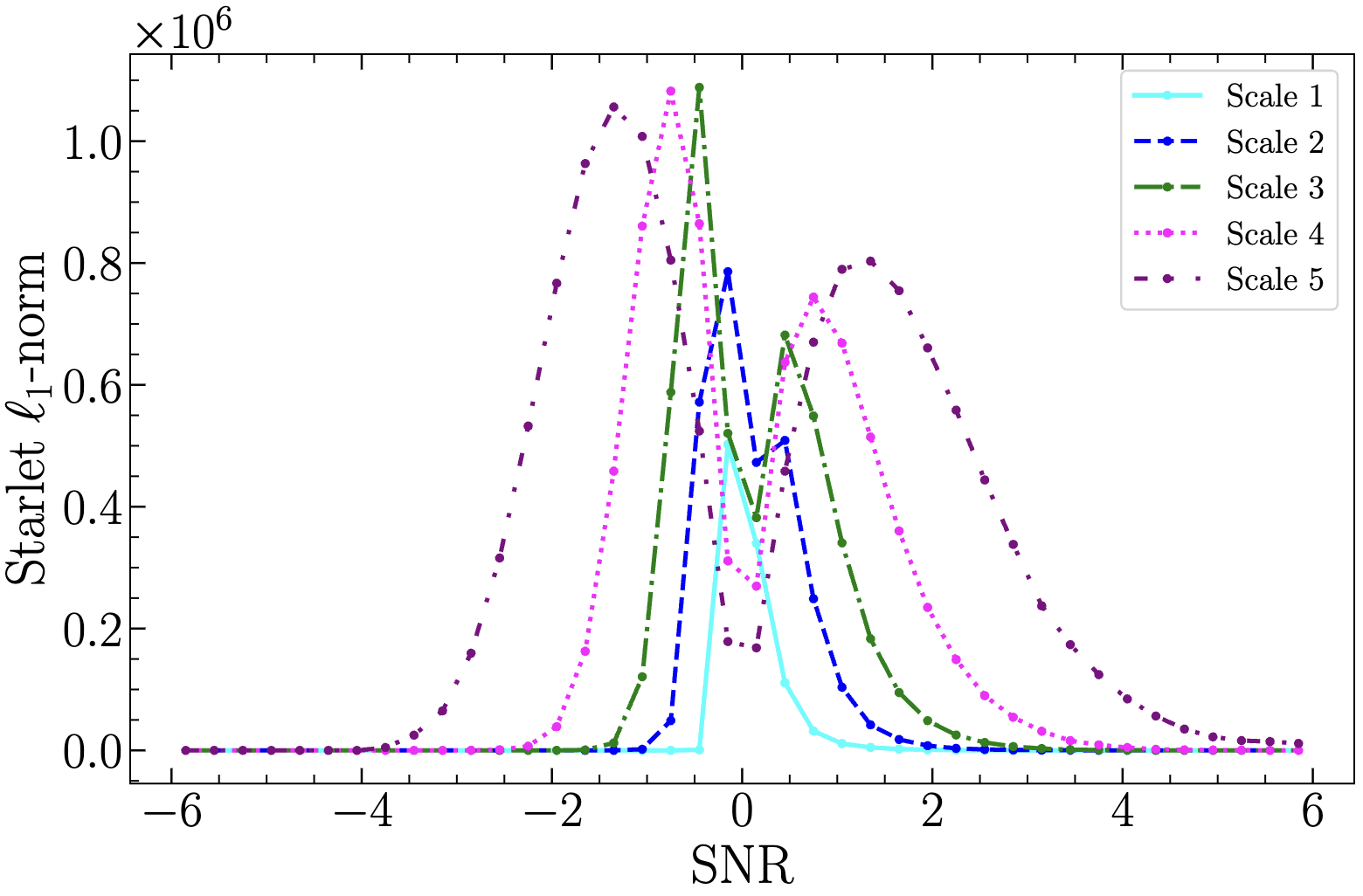}
    \caption{
        $\ell_1$-norms of the starlet coefficients as a function of the bin thresholds for different wavelet scales. 
        Each curve corresponds to a wavelet scale, showing how the distribution of coefficients changes across scales.
        The bins correspond to S/N thresholds or coefficient magnitudes, depending on the normalization.
    }
    \label{fig:l1norm_bins_scales}
\end{figure}

We defined the $\ell_1$-norm of the normalized coefficients at scale \( j \) within S/N bin \( k \) as
\begin{equation}
\ell_1^{(j,k)} = \sum_{\hat{\bm n} \in \mathcal{B}_k} \left| \frac{w_j(\hat{\bm n})}{\mathcal{N}_j} \right|\;,
\end{equation}
where \( \mathcal{B}_k \) is the set of pixels whose normalized coefficient falls in the \( k \)-th bin, namely
\begin{equation}
\mathcal{B}_k = \left\{ \hat{\bm n} \,\middle|\, T_k \leq \frac{w_j(\hat{\bm n})}{\mathcal{N}_j} < T_{k+1} \right\}\;,
\end{equation}
and \( T_k \) are the S/N bin edges. The full summary statistic is the matrix \( \ell_1^{(j,k)} \in \mathbb{R}^{J \times K} \), where \( K \) is the number of bins and \( J \) is the number of wavelet scales. We show the plot of the full summary statistic in \cref{fig:l1norm_bins_scales}.

\begin{figure}
    \centering
    \includegraphics[width=0.85\columnwidth]{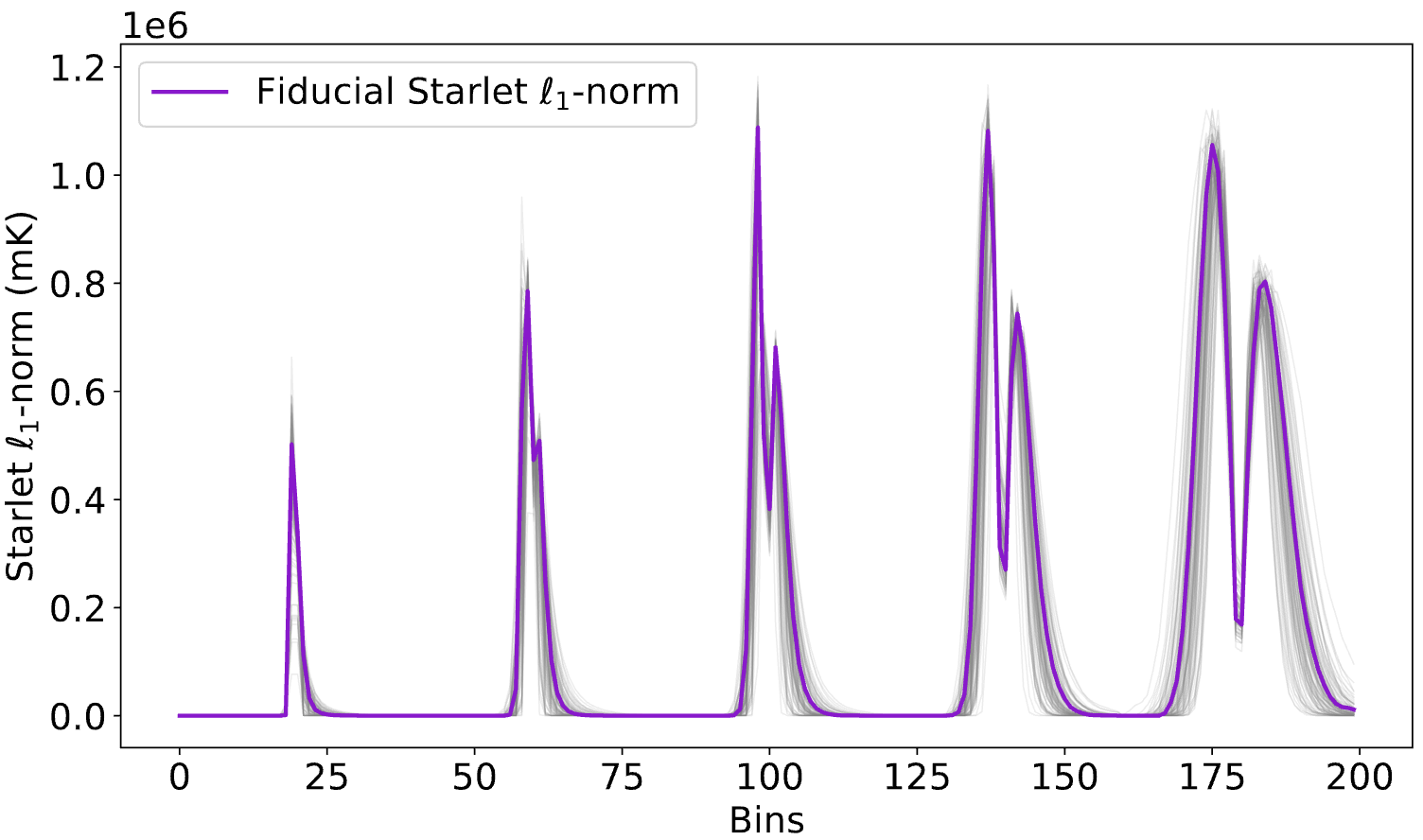}
    \caption{Variability of starlet $\ell_1$-norm over 100 randomly selected realizations with different cosmologies (gray) varying as in Table \ref{tab:example_table} with fiducial starlet $\ell_1$-norm (purple).}
    \label{fig:l1_norm}
\end{figure}

By capturing non-Gaussian structure in the field, the starlet $\ell_1$-norm helps to lift degeneracies that may be unresolved by second-order statistics such as the angular power spectrum. We used the \texttt{MRstarlet} class from the \texttt{CosmoStat}\footnote{\url{https://github.com/CosmoStat/cosmostat/}} package to perform the starlet decomposition and the computation of the $\ell_1$-norm. Figure \ref{fig:l1_norm} shows the variability of the starlet $\ell_1$-norm across realizations and cosmologies. Note that the plots shown in this section correspond to the five-scale Starlet $\ell_1$-norm, while both the five-scale and eight-scale Starlet $\ell_1$-norms are used in the full analysis.
Simulation-based inference pipelines are particularly well suited for the starlet $\ell_1$-norm, as they do not rely on an explicit likelihood function, unlike traditional MCMC methods. 

\section{Simulation-based inference with \texttt{JaxILI}}\label{sec:sbi}

We performed Bayesian inference over three cosmological parameters -- a reduced Hubble constant, $h$, a cold dark matter density parameter, $\Omega_{\rm c}$, and a primordial amplitude and spectral index, \( A_{\rm s} \) -- using SBI. Given the analytical intractability of the likelihood function for the starlet $\ell_1$-norm, we adopted a likelihood-free approach via the \texttt{JaxILI} library,\footnote{\url{https://jaxili.readthedocs.io/en/latest/}} a lightweight framework for implicit likelihood inference (ILI) built on top of \texttt{JAX}, \texttt{Flax}, and \texttt{Optax} that allows scalable and efficient posterior estimation.

Our summary statistics consist of two components: (i) the angular power spectrum, $C_\ell$, and (ii) the starlet $\ell_1$-norm (as defined in Sect. \ref{sec:summary}) computed from the simulated \hi\ full sky maps. The latter is particularly well suited for capturing both Gaussian and non-Gaussian information from the maps. We treated the forward model $\bm{x} = \mathcal{S}(\bm{\theta})$ as a black box, where $\bm{\theta} = ( \Omega_{\rm c}, h, A_{\rm s})$.

\begin{table}
    \centering
    \caption{Prior ranges and mock data values for the cosmological parameters.}
    \label{tab:example_table}
    \begin{tabular}{c c c}
        \hline
        Parameter & Prior Range & Mock Data \\
        \hline
        $\Omega_{\rm c}$ & $[0.15, 0.35]$ & $0.25$ \\
        $h$ & $[0.55, 0.85]$ & $0.70$ \\
        $10^9\,A_{\rm s}$ & $[1,3]$ & $2$ \\
        \hline
    \end{tabular}
\end{table}

\texttt{JaxILI} enabled us to train a conditional neural density estimator -- specifically a normalizing flow -- to approximate the posterior $p(\bm{\theta} \mid \bm{x})$. We generated a set of simulations across the prior range, computed the associated summary statistics, and used these to train the flow using the implicit likelihood inference objective. The implementation leverages \texttt{JAX}'s automatic differentiation and just-in-time compilation capabilities,\footnote{\url{https://docs.jax.dev/en/latest/quickstart.html}} ensuring both speed and scalability. The density estimator was implemented as a masked auto regressive flow (\citealt{Papamakarios2017}), trained on \(10\,000\) simulated pairs $(\bm{\theta}, \bm{x})$ sampled using a Latin hypercube. The dataset was divided into $7000$ training samples, $2000$ validation samples, and $1000$ test samples. Training continued over $2000$ epochs, with early stopping based on validation loss. After convergence, posterior samples conditioned on $\bm{x}_{\mathrm{obs}}$ were drawn to construct confidence contours using \texttt{GetDist}.

Before forming the training pairs, we preprocessed the summary statistics by removing all columns that exhibit zero variance across cosmologies, as these do not provide additional information and may introduce numerical instability. For the starlet $\ell_1$-norm, this preprocessing results, for example, in 18 and 35 points for the five- and eight-scale decompositions, respectively, compared to 200 and 320 points in the original scale representation. This depends on the setting of the instrumental effects used in the map generation for the dataset. Figure \ref{tab:table_bins} lists the number of bins used for the SBI in the different beam, mask, and noise scenarios. By contrast, the $C_\ell$ estimator comprises of 2037 points and contains no columns with zero variance. The smaller dimensionality of the starlet $\ell_1$-norm enables substantially faster evaluation and training relative to $C_\ell$. This preprocessing ensures that SBI reflects only meaningful variability, improving both robustness and interpretability. The validation of the inference pipeline for both summary statistics is presented in Appendix \ref{sec:validation} for the clean map-full sky case as a representative example.

In our study, we note that SBI allows parallelization in both the generation of simulation maps and the extraction of summary statistics, thereby reducing computation time. In contrast, traditional MCMC methods are inherently sequential, requiring simulations, summary statistic estimation, and likelihood evaluation at each step. Additionally, in SBI the same set of training pairs and the trained model can be reused across multiple inference tasks. From an environmental perspective, the CO$_2$ footprint per dataset is lower in SBI, whereas MCMC often generates higher CO$_2$ emissions per posterior sample. This makes SBI particularly advantageous for studies requiring repeated inference over many datasets, offering both computational efficiency and environmental sustainability.

\section{Results}
\label{sec:results}

We investigated how the inclusion of realistic observational effects influence constraints on cosmological parameters obtained from both the angular power spectrum, $C_\ell$, and the starlet $\ell_1$-norm.
We first examined how constraints vary with the observed sky fraction in clean maps. Secondly we investigated the effect of the telescope beam in masked maps. Then, we explored improvements achievable by increasing the number of starlet scales, and finally we assessed the sensitivity and robustness of the starlet $\ell_1$-norm to systematic effects.
This section demonstrates that the starlet $\ell_1$-norm consistently outperforms $C_\ell$ by capturing non-Gaussian information and that its performance can be further enhanced through multi-scale analysis.

\subsection{Sky fraction variation with in clean maps}
\begin{figure*}
\sidecaption
\includegraphics[width=0.71\columnwidth]{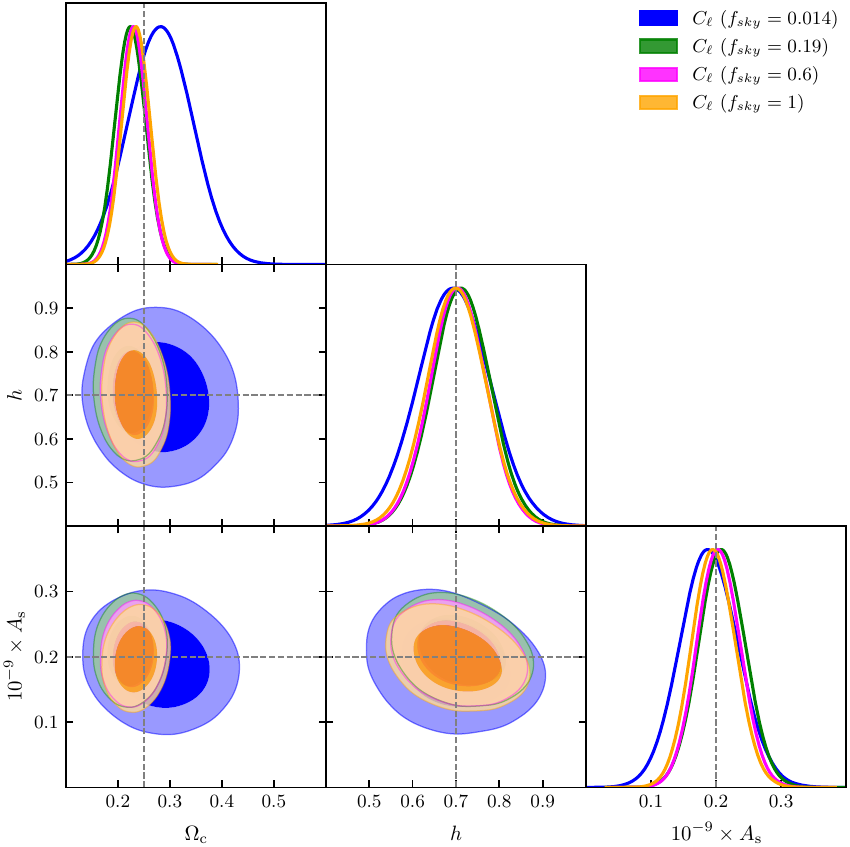}
\includegraphics[width=0.71\columnwidth]{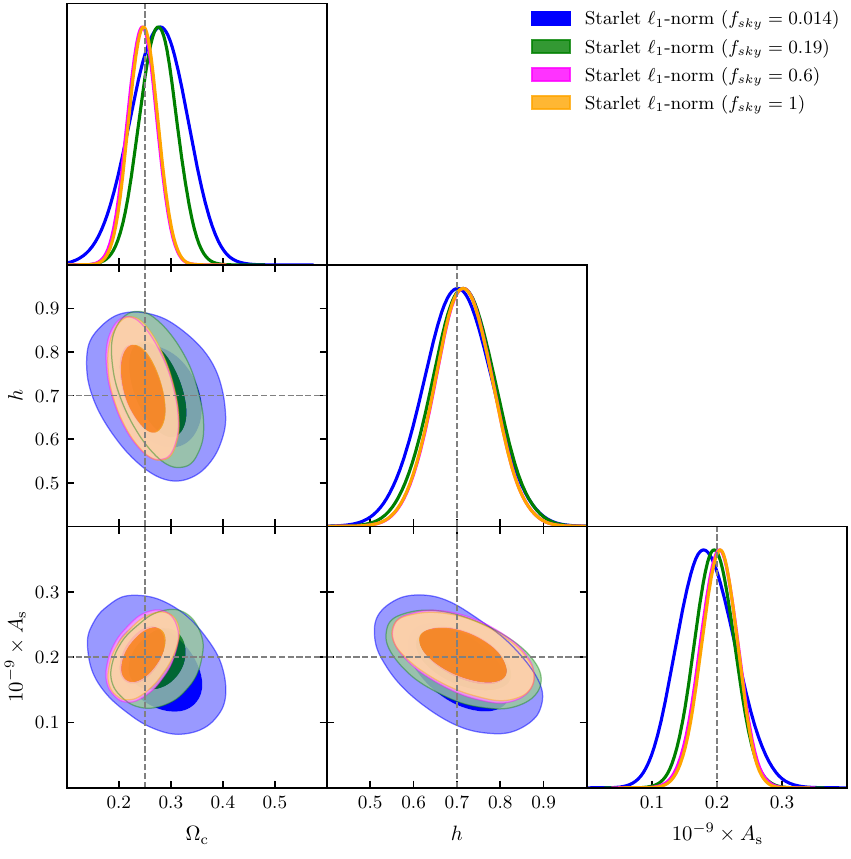}
\caption{
    Posterior distributions for $\Omega_{\rm c}$, $h$, and $A_{\rm s}$ obtained from the angular power spectrum, $C_\ell$ (\textit{left}), and starlet $\ell_1$-norm (\textit{right}) for different sky fractions, $f_{\rm sky}$ (blue: $0.014$, green: $0.19$, magenta: $0.6$, orange: $1.0$). 
        Contours correspond to $68\%$ and $95\%$ confidence levels.
    }
    \label{fig:clean}
\end{figure*}

In this subsection, we assess how varying the observed sky fraction, $f_{\rm sky}$, affects parameter constraints from clean maps. Three masks with sky fractions of $f_{\rm sky} = 0.60$, $0.19$, and $0.014$ are used, corresponding to optimistic coverage, typical current surveys, and the published MeerKLASS results \citep{Wang21, MeerKLASS}, respectively. Figure \ref{fig:clean} shows posterior distributions for $\Omega_{\rm c}$, $h$, and $A_{\rm s}$ derived from $C_\ell$ and the starlet $\ell_1$-norm for the different $f_{\rm sky}$ cases shown in Fig. \ref{fig:masks}.
As expected, the constraining power decreases with smaller sky fractions due to the reduced number of independent modes, although the degradation is not strictly linear in $f_{\rm sky}$. This reflects the fact that the information content depends not only on the sky area but also on mask-induced mode coupling and the scale dependence of each statistic.
Across all $f_{\rm sky}$ values, the starlet $\ell_1$-norm consistently outperforms the angular power spectrum, yielding tighter constraints and smaller degeneracy regions (see Table \ref{tab:fom_sky} for normalized figure of merit (FoM) values calculated as shown in Appendix \ref{sec:FoM}). The impact of masking on $C_\ell$ becomes significant only in the really conservative case ($f_{\rm sky}=0.014$) representing already published results of the MeerKLASS collaboration \citep{MeerKLASS}, where strong mode coupling erases much of the harmonic information.

Interestingly, for $f_{\rm sky}=0.19$, both statistics provide comparable constraints. Here, moderate masking redistributes small-scale information across multipoles, partly compensating for the lack of higher-order correlations usually missed by $C_\ell$. This accidental balance vanishes for larger or smaller sky fractions.

Because the starlet $\ell_1$-norm is derived from a multi-scale, jointly localized decomposition in real and harmonic space, it is inherently less sensitive to partial-sky mode coupling and shows a stronger correlation with $f_{\rm sky}$. This property allows it to retain most of the cosmological information even under strong masking, unlike $C_\ell$, whose sensitivity is directly tied to sky coverage. The results suggests that the non-Gaussian information captured by the starlet transform adds substantial constraining power beyond two-point statistics.

We also note that the parameter $h$ exhibits increased scatter in both $C_\ell$ and starlet $\ell_1$-norm analyses (Figs. \ref{fig:recovery_Cl} and \ref{fig:recovery_l1}). This behavior is consistent with expectations, as $h$ has a weaker imprint on the summary statistics compared to $\Omega_{\rm c}$ and $A_{\rm s}$, making it intrinsically less informative in this analysis.
In the following, we adopt $f_{\rm sky}=0.19$, representative of a realistic survey footprint.

\begin{table}
    \centering
    \caption{FoM normalized to the $C_\ell$ baseline. 
    }
    \label{tab:fom_sky}
    \begin{tabular}{l c c c}
        \hline
        Statistic & $f_\mathrm{sky}$ & Normalized FoM & Improvement vs $C_\ell$ \\
        \hline
        $C_\ell$ & 1.0 & 1.00 & — \\
        S$\ell_1$ & 1.0 & 1.59 & 1.59$\times$ \\
        \hline
        $C_\ell$ & 0.6 & 1.00 & — \\
        S$\ell_1$ & 0.6 & 1.32 & 1.32$\times$ \\
        \hline
        $C_\ell$ & 0.19 & 1.00 & — \\
        S$\ell_1$ & 0.19 & 1.07 & 1.07$\times$ \\
        \hline
        $C_\ell$ & 0.014 & 1.00 & — \\
        S$\ell_1$ & 0.014 & 1.82 & 1.82$\times$ \\
        \hline
    \end{tabular}
    \tablefoot{Values show overall FoM and improvement factors for different $f_\mathrm{sky}$.}
\end{table}

\subsection{Impact of telescope beam}
\begin{figure*}
    \sidecaption
    \includegraphics[width=0.71\columnwidth]{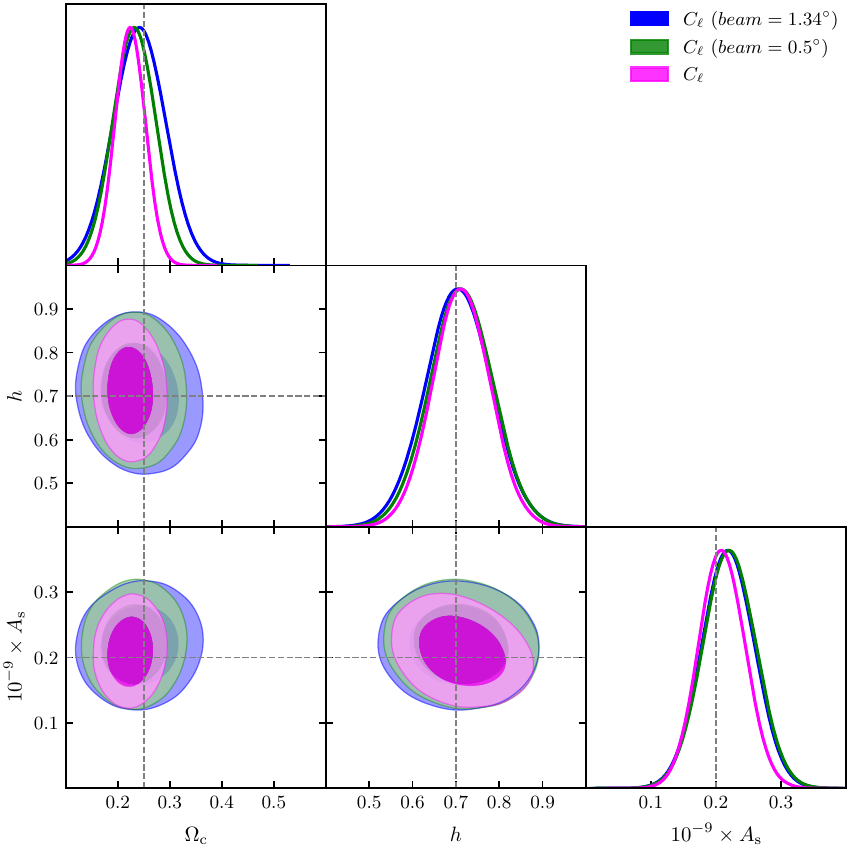}
    \includegraphics[width=0.71\columnwidth]{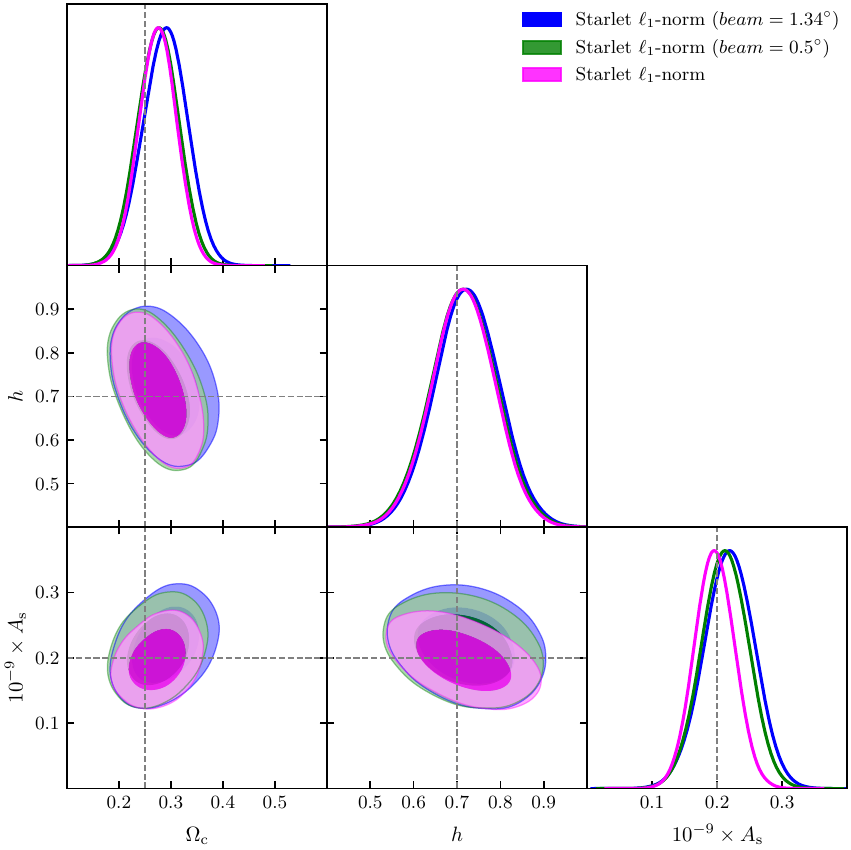}
    \caption{
    Posterior comparison between the angular power spectrum (\textit{left}) and the starlet $\ell_1$-norm (\textit{right}) estimators for different beam sizes, all with masked maps of $f_{\mathrm{sky}} = 0.19$. 
    The solid lines show the 1D marginalized posterior distributions, while the shaded regions correspond to the $68\%$ and $95\%$ confidence contours for the parameter pairs $(\Omega_{\rm c}, h, 10^{-9} \times A_{\rm s})$. 
    Blue for a Gaussian beam of FWHM $= 1.34^\circ$. 
    Green show the same estimators for a beam of FWHM $= 0.5^\circ$ and magenta correspond to the clean masked case without telescope beam.
    }
    \label{fig:Cl_vs_l1_beam}
    
\end{figure*}

\begin{figure*}
    \sidecaption
    \includegraphics[width=0.71\columnwidth]{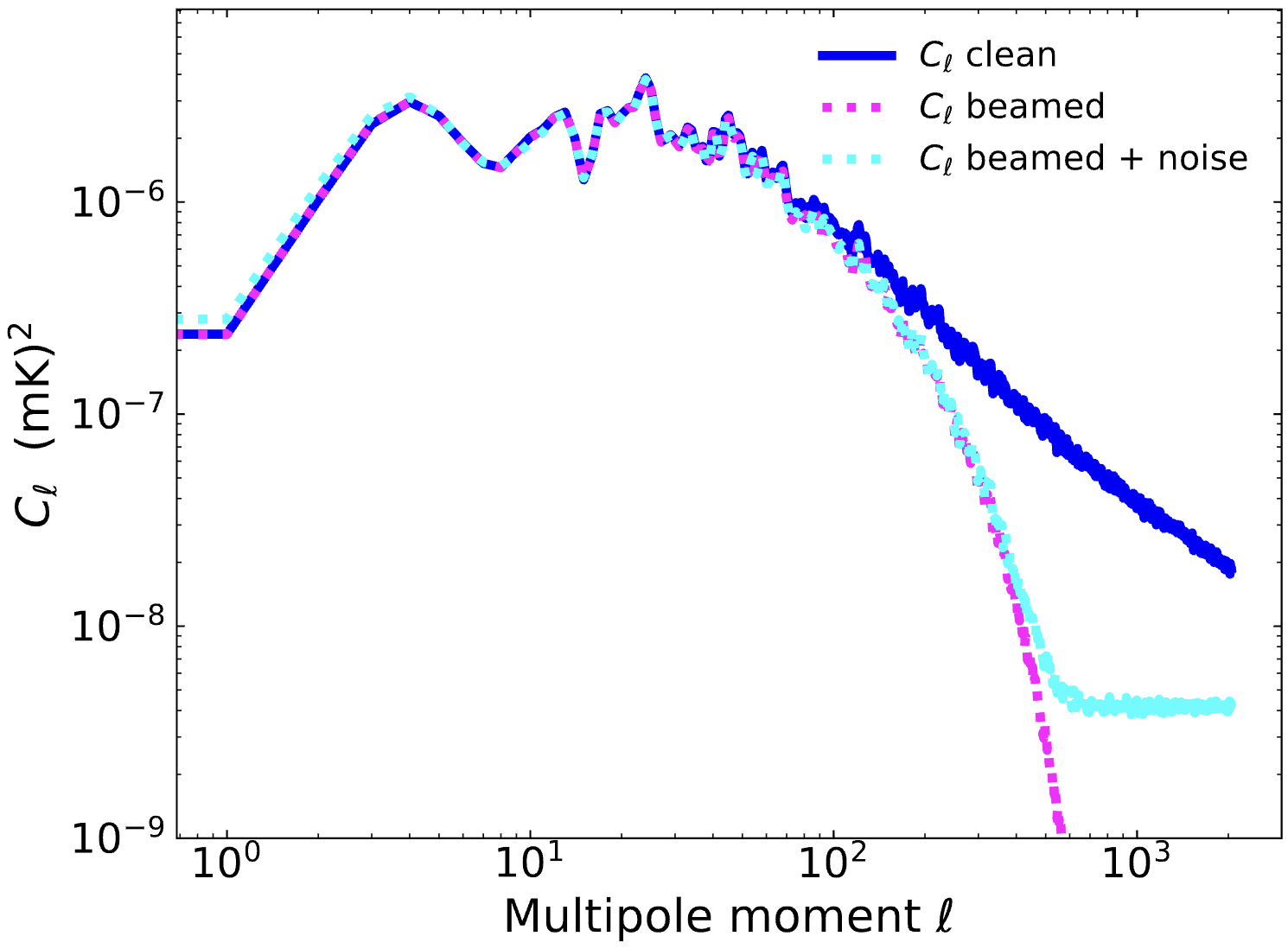}
    \includegraphics[width=0.71\columnwidth]{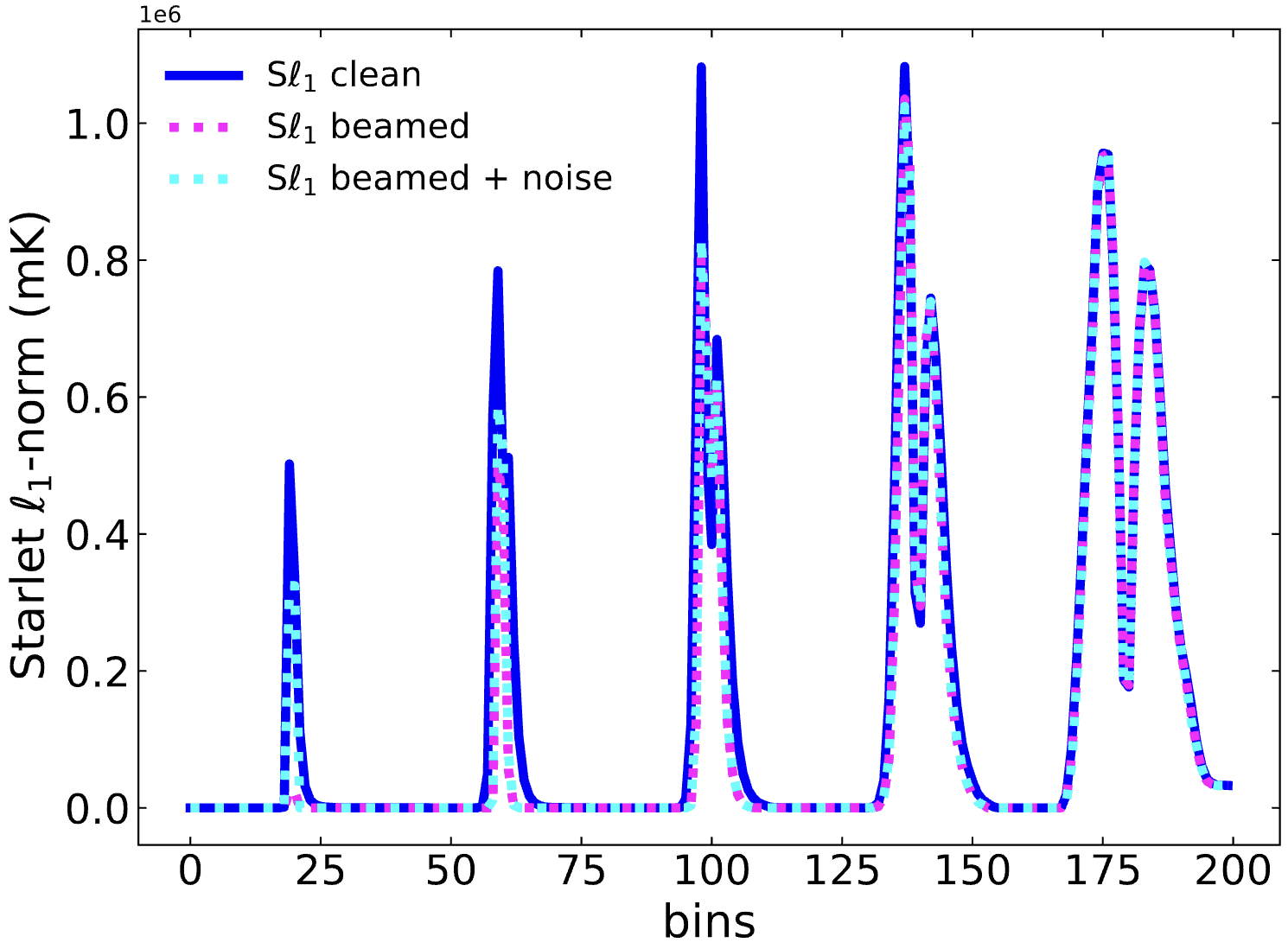}
    \caption{
     C$_\ell$ (\textit{left}), starlet $\ell_1$-norm (\textit{right}) for clean map (blue), beamed map $0.5^\circ$ (magenta), and beamed + map with instrumental noise (cyan) all maps masked at $f_{\rm sky}$=0.19. 
    }
    \label{fig:SS_1}
\end{figure*}

We investigated the influence of smoothing due to the telescope beam on the recovered cosmological parameters. Figure \ref{fig:Cl_vs_l1_beam} illustrates the effect of beam smoothing on parameter constraints. In the absence of a beam, we have shown that the starlet $\ell_1$-norm yields similar constraints than the standard $C_\ell$ estimator. 
Increasing the beam size systematically broadens the posterior contours, with a stronger degradation observed for the power spectrum. Beam convolution suppresses small-scale fluctuations, erasing part of the information contained in the high-$\ell$ modes to which $C_{\ell}$ is most sensitive, because these modes carry most of the constraining power.
The starlet $\ell_{1}$-norm remains more robust, as it quantifies the amplitude of localized features across multiple scales, and therefore retains sensitivity to structural contrast even after smoothing. As a result, while both estimators are affected by beam convolution, the $\ell_{1}$-norm consistently delivers tighter and more stable constraints (see Table \ref{tab:fom_beam}), demonstrating its enhanced ability to recover cosmological information under realistic observational conditions.
In the following, we adopt an intermediate beam size of $0.5^\circ$ full width at half maximum (FWHM), which preserves a significant fraction of the small-scale information while remaining representative of realistic observational conditions. Figure \ref{fig:SS_1} shows the impact of beam and instrumental noise on summary statistics. In the left panel ($C_\ell$), the telescope beam (magenta) smooths small-scale fluctuations, while instrumental noise (cyan) produces a plateau at high $\ell$. In the right panel (starlet $\ell_1$-norm), the beam reduces the power of the first three small scales, whereas noise (cyan) adds power at small scales, yielding values different from the original signal (blue).

\begin{table}
    \centering
    \caption{FoM normalized to the $C_\ell$ baseline. 
    }
    \label{tab:fom_beam}
    \begin{tabular}{l c c c}
        \hline
        Statistic & beam & Normalized FoM & Improvement vs $C_\ell$ \\
        \hline
        $C_\ell$ & no & 1.00 & — \\
        S$\ell_1$ & no & 1.07 & 1.07$\times$ \\
        \hline
        $C_\ell$ & 0.5 deg & 1.00 & — \\
        S$\ell_1$ & 0.5 deg & 1.34 & 1.34$\times$ \\
        \hline
        $C_\ell$ & 1.34 deg & 1.00 & — \\
        S$\ell_1$ & 1.34 deg & 1.40 & 1.40$\times$ \\
        \hline
    \end{tabular}
    \tablefoot{Values show overall FoM and improvement factors for different beam values.}
\end{table}

\subsection{Improving starlet $\ell_1$-norm with additional scales}

\begin{figure*}
    \sidecaption
    \includegraphics[width=0.71\columnwidth]{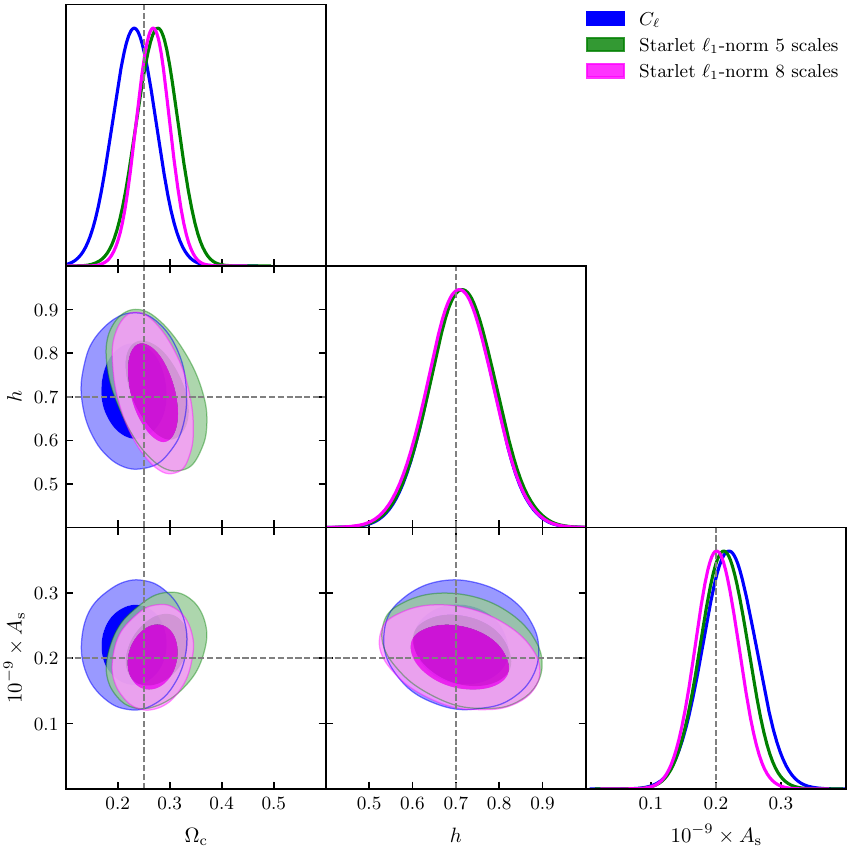}
    \includegraphics[width=0.71\columnwidth]{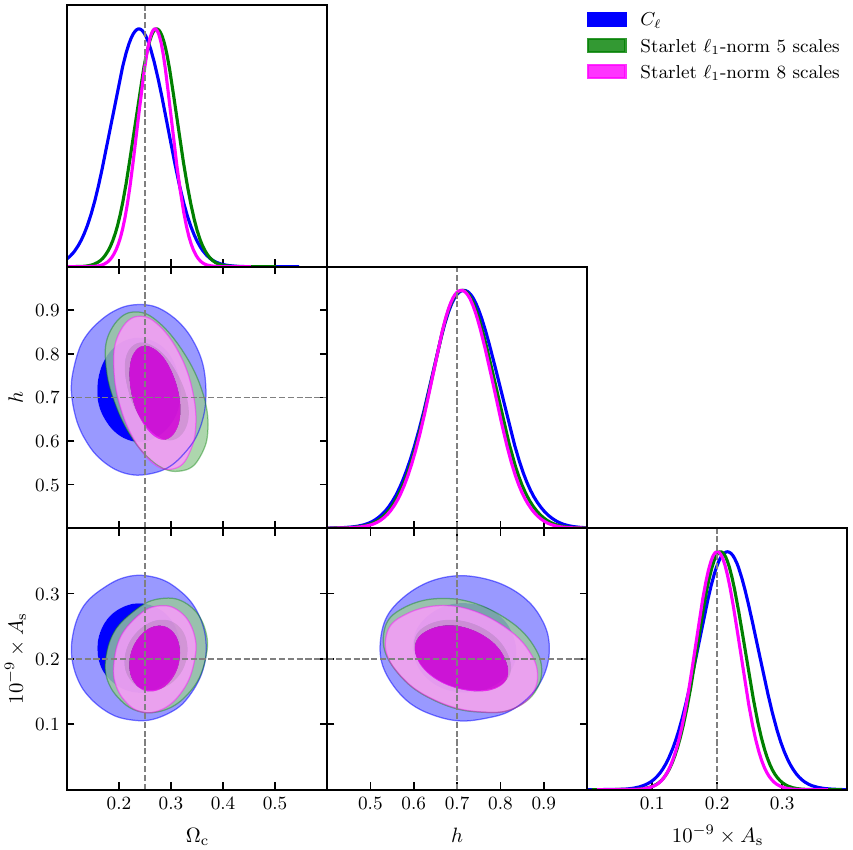}
    \caption{
    Posterior distributions for $\Omega_{\rm c}$, $h$, and $A_{\rm s}$ comparing the standard power spectrum, $C_\ell$ (blue), to the starlet $\ell_1$-norm statistic with five scales (magenta) and with eight scales (green), all computed from beamed $0.5^\circ$ maps with $f_{\mathrm{sky}} = 0.19$ and noiseless case (\textit{left}), or the case with instrumental noise (\textit{right}). 
    }
    \label{fig:l1_scales}
\end{figure*}

We explored the constraining power of the starlet $\ell_1$-norm by increasing the number of decomposition scales. We have shown that the $\ell_1$-norm already provides tighter constraints than the standard power spectrum, and adding more scales further sharpens these constraints. This improvement arises because additional scales capture a wider range of structural information, allowing the $\ell_1$-norm to exploit both fine and coarse features simultaneously. As shown in Fig. \ref{fig:l1_scales} (left panel), increasing the number of starlet scales from five to eight scales \footnote{Angular sizes of the starlet $\ell_1$-norm scales: for the five-scale decomposition, [3.7, 5.6, 11.2, 22.3, 60] arcmin; for the eight-scale decomposition, [5.6, 6.3, 11, 17.5, 25, 35, 45, 60] arcmin. The 3.7 arcmin scale is not included in the eight-scale decomposition because it is largely dominated by instrumental noise. We verified that including or excluding this scale does not affect the cosmological parameter constraints inferred with the SBI.} produces visibly tighter parameter posteriors, demonstrating the potential of multi-scale wavelet statistics for enhanced cosmological inference. Figure \ref{fig:l1_scales} (right panel) shows that, when instrumental noise is included, the constraints from the starlet $\ell_1$-norms remain comparable to the noiseless case, whereas the angular power spectrum exhibits a clear degradation in constraining power. This confirms the robustness of the starlet $\ell_1$-norms against instrumental noise.
The corresponding FoM values, normalized to the $C_\ell$ baseline, are listed in Table \ref{tab:fom_scales}.

\begin{table}
    \centering
    \caption{FoM normalized to the $C_\ell$ baseline. 
    }
    \label{tab:fom_scales}
    \begin{tabular}{l c c}
        \hline
        \multicolumn{3}{c}{Noiseless case} \\
        \hline
        Statistic & Normalized FoM & Improvement vs $C_\ell$ \\
        \hline
        $C_\ell$ & 1.00 & — \\
        S$\ell_1$ (5 scales) & 1.34 & 1.34$\times$ \\
        S$\ell_1$ (8 scales) & 1.79 & 1.79$\times$ \\
        \hline
        \multicolumn{3}{c}{Case with instrumental noise} \\
        \hline
        Statistic & Normalized FoM & Improvement vs $C_\ell$ \\
        \hline
        $C_\ell$ & 1.00 & — \\
        S$\ell_1$ (5 scales) & 2.09 & 2.09$\times$ \\
        S$\ell_1$ (8 scales) & 2.81 & 2.81$\times$ \\
        \hline
    \end{tabular}
    \tablefoot{Values show overall FoM and improvement factors for noiseless and maps with instrumental noise.}
\end{table}

\subsection{Beamed, masked maps affected by systematic stripe patterns}
\begin{figure*}
    \centering
    \begin{minipage}{0.25\textwidth}
        \includegraphics[width=\textwidth]{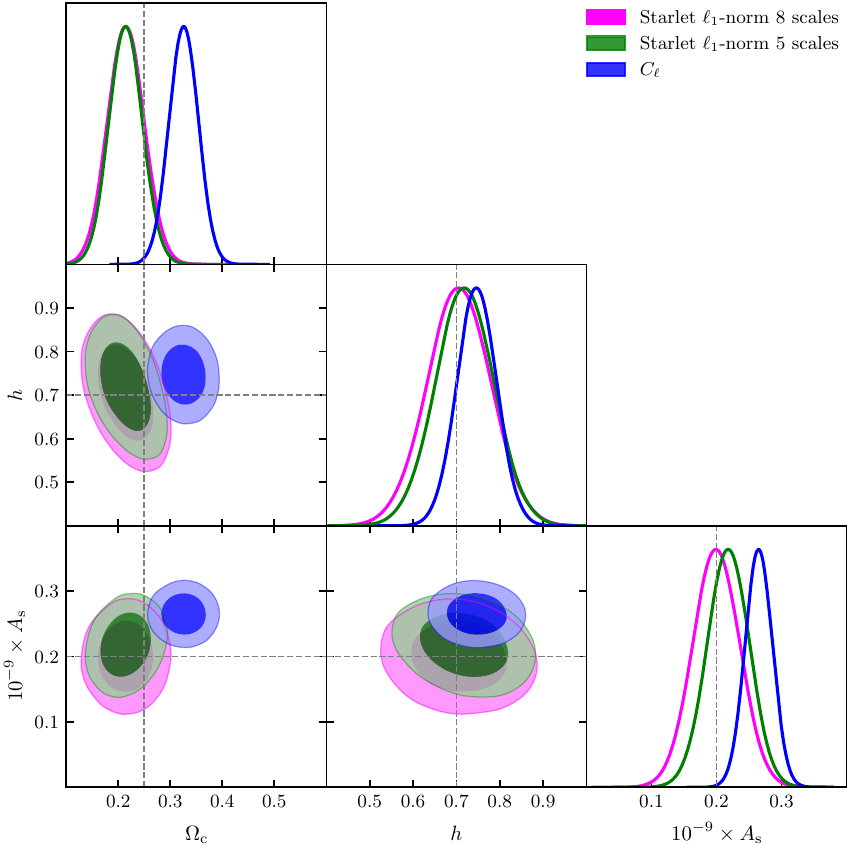}
    \end{minipage}%
    \begin{minipage}{0.25\textwidth}
        \includegraphics[width=\textwidth]{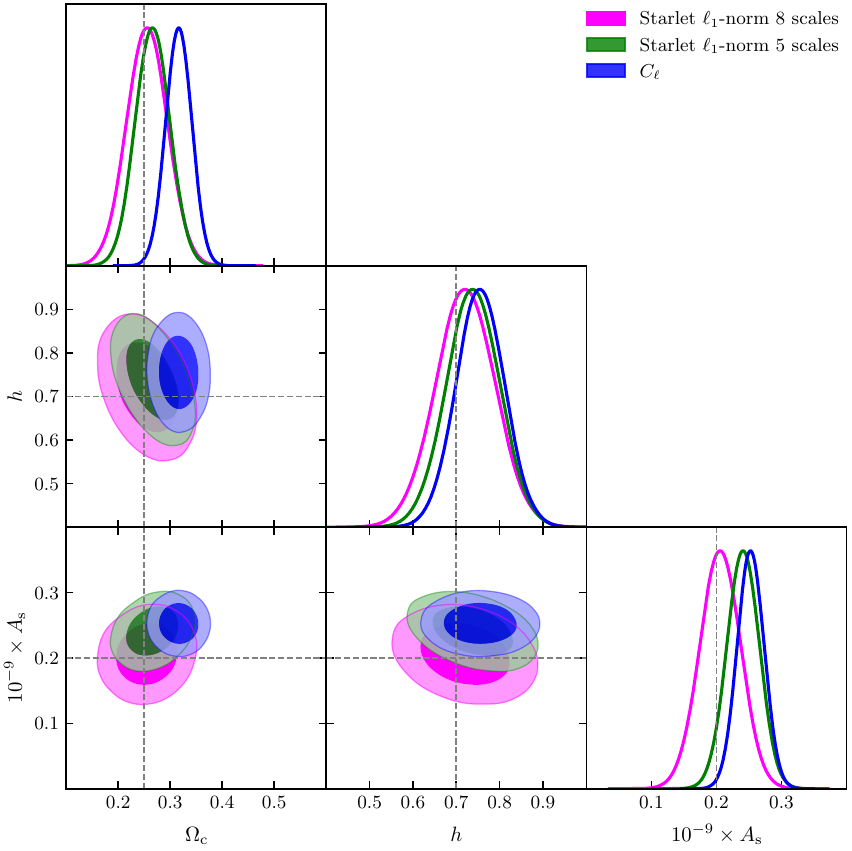}
    \end{minipage}%
    \begin{minipage}{0.25\textwidth}
        \includegraphics[width=\textwidth]{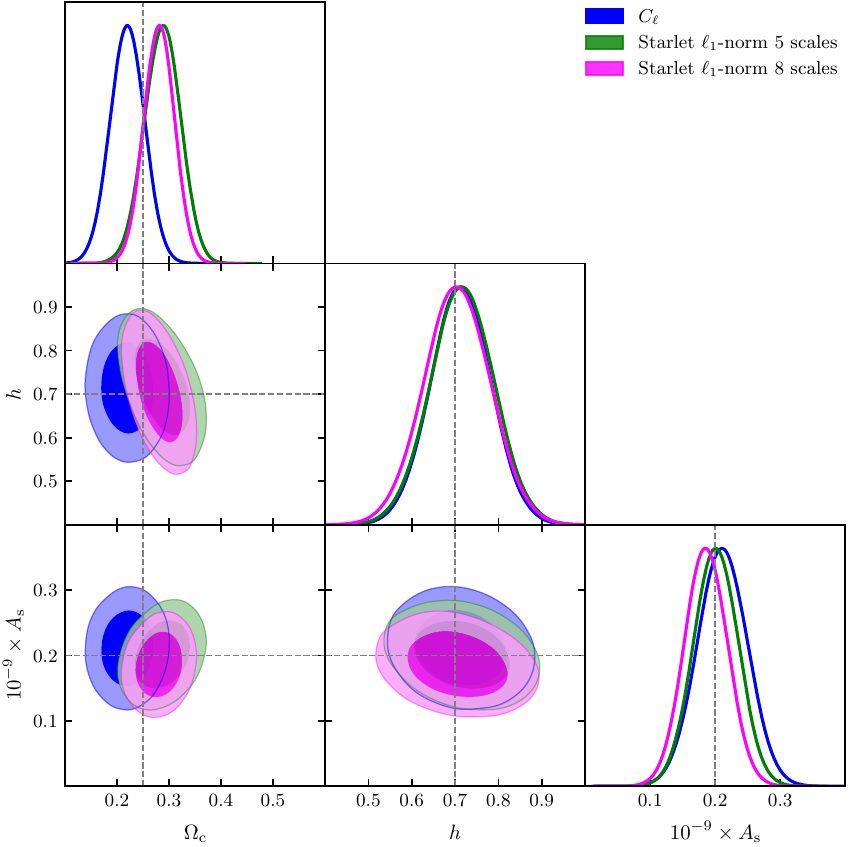}
    \end{minipage}%
    \begin{minipage}{0.25\textwidth}
        \includegraphics[width=\textwidth]{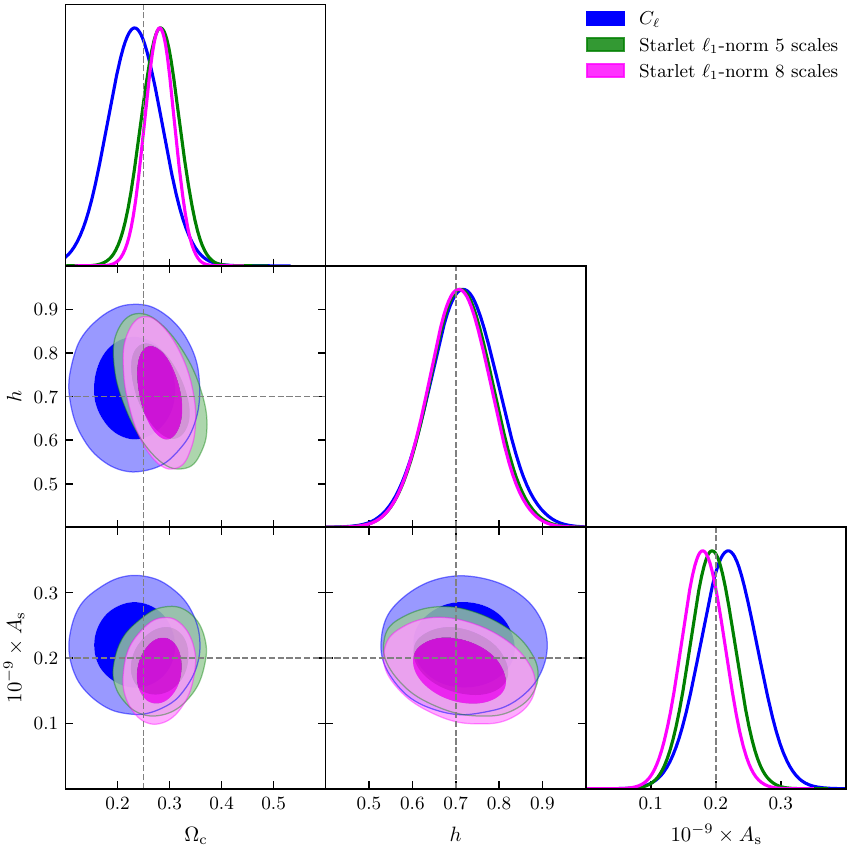}
    \end{minipage}
    \caption{Posterior distributions for $\Omega_{\rm c}$, $h$, $A_{\rm s}$, and another parameter obtained from the angular power spectrum, $C_\ell$, 
starlet $\ell_1$-norm for beamed maps $0.5^\circ$ with sky fraction $f_{\mathrm{sky}} = 0.19$: multiplicative stripes ($A=1.5$, $w=2.4^\circ$; \textit{first panel}), multiplicative stripes ($A=1.5$, $w=2.4^\circ$) + instrumental noise (\textit{second panel}), additive stripes (\textit{third panel}), and additive stripes + noise (\textit{fourth panel}). 
Contours correspond to $68\%$ and $95\%$ confidence levels.}

    \label{fig:SBI_zebras}
    
\end{figure*}

We assessed the impact of complex multiplicative and additive stripe patterns on posterior constraints for beamed, masked maps. Figure \ref{fig:SBI_zebras} (first panel) shows the effect of multiplicative stripes ($A=1.5$, $w=2.4^\circ$) in the absence of instrumental noise. The $C_\ell$ gains power across all scales (magenta vs. orange curve in Fig. \ref{fig:SS_2}, left panel), which explains the strong bias and artificial tightening observed in the posterior distributions. In contrast, additive stripes (third panel in Fig. \ref{fig:SBI_zebras}) primarily modify a specific scale (green vs. orange in the left panel of Fig. \ref{fig:SS_2}), so the bias in $C_\ell$ is milder, as most scales remain unaffected. For the starlet $\ell_1$-norm, both five and eight-scale decompositions remain mostly insensitive to these effects and the overall multi-scale information is preserved, resulting in low biased parameter estimates even in the presence of stripe patterns.

The starlet $\ell_1$-norm with five scales yields slightly tighter contours in $(\Omega_{\rm c}, A_{\rm s})$, while the eight-scale version shows broader distributions, suggesting some sensitivity to the choice of scale decomposition. The second and fourth panels of Fig. \ref{fig:SBI_zebras} show the effect of including instrumental noise. The $C_\ell$ posteriors broaden noticeably, reflecting the reduced constraining power when noise is added. In contrast, the starlet $\ell_1$-norm (both five and eight scales) remains largely unaffected, producing well-centered and stable constraints. This highlights the robustness of the multi-scale starlet approach against instrumental noise, in addition to its ability to mitigate mask-induced systematics. These results demonstrate that the multi-scale nature of the starlet $\ell_1$-norm effectively mitigates biases induced by both multiplicative and additive mask features, whereas the standard power spectrum is more sensitive to broad increases in power across scales. These results underline the advantage of starlet $\ell_1$-norm in scenarios where incomplete sky coverage or anisotropic features significantly affect the performance of traditional two-point 
statistics.

\subsection{Effect of Galactic synchrotron and extragalactic point sources residuals}

\begin{figure*}
    \centering
    \includegraphics[width=0.28\textwidth]{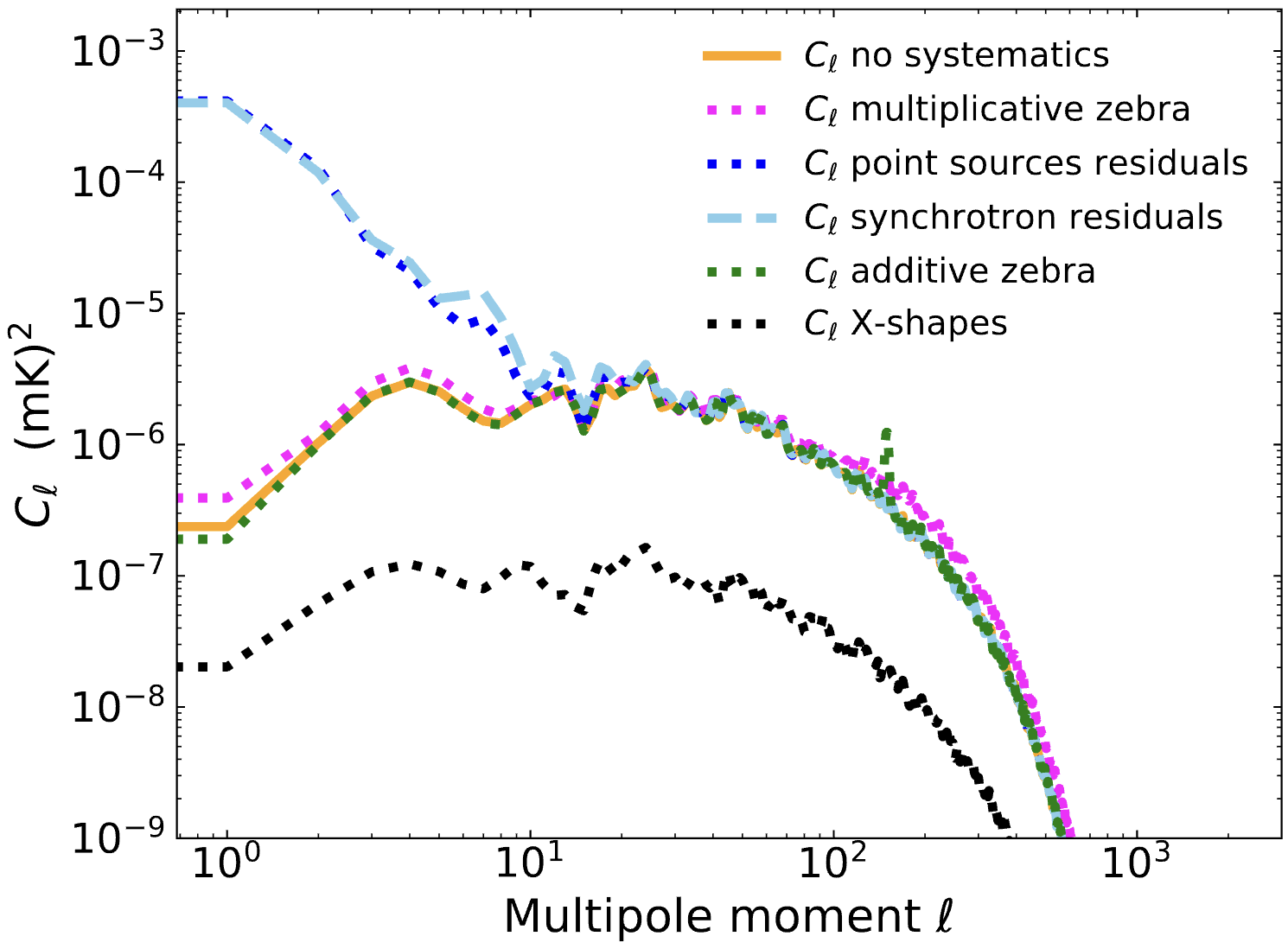}
    \includegraphics[width=0.29\textwidth]{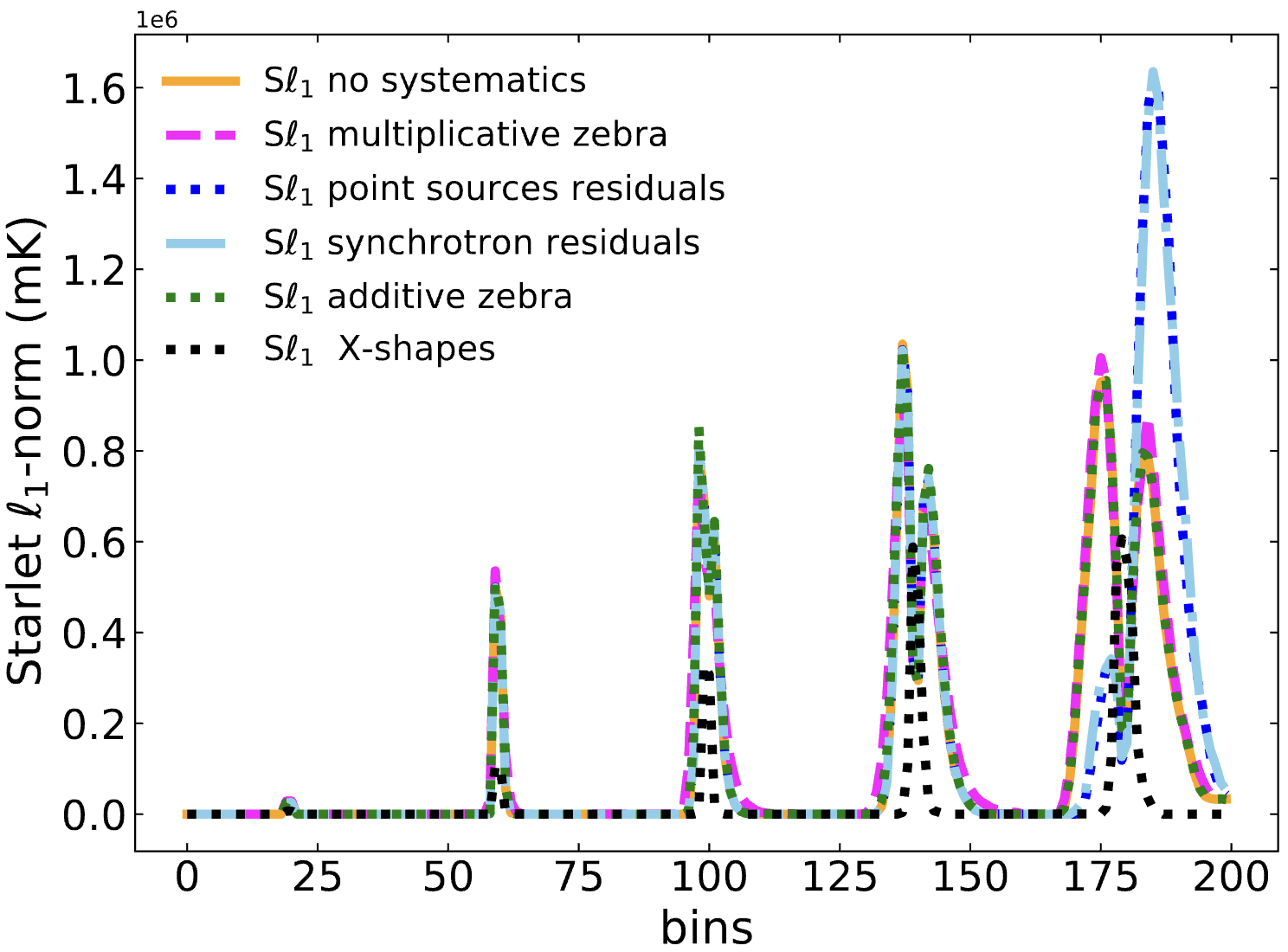}
    \includegraphics[width=0.36\textwidth]{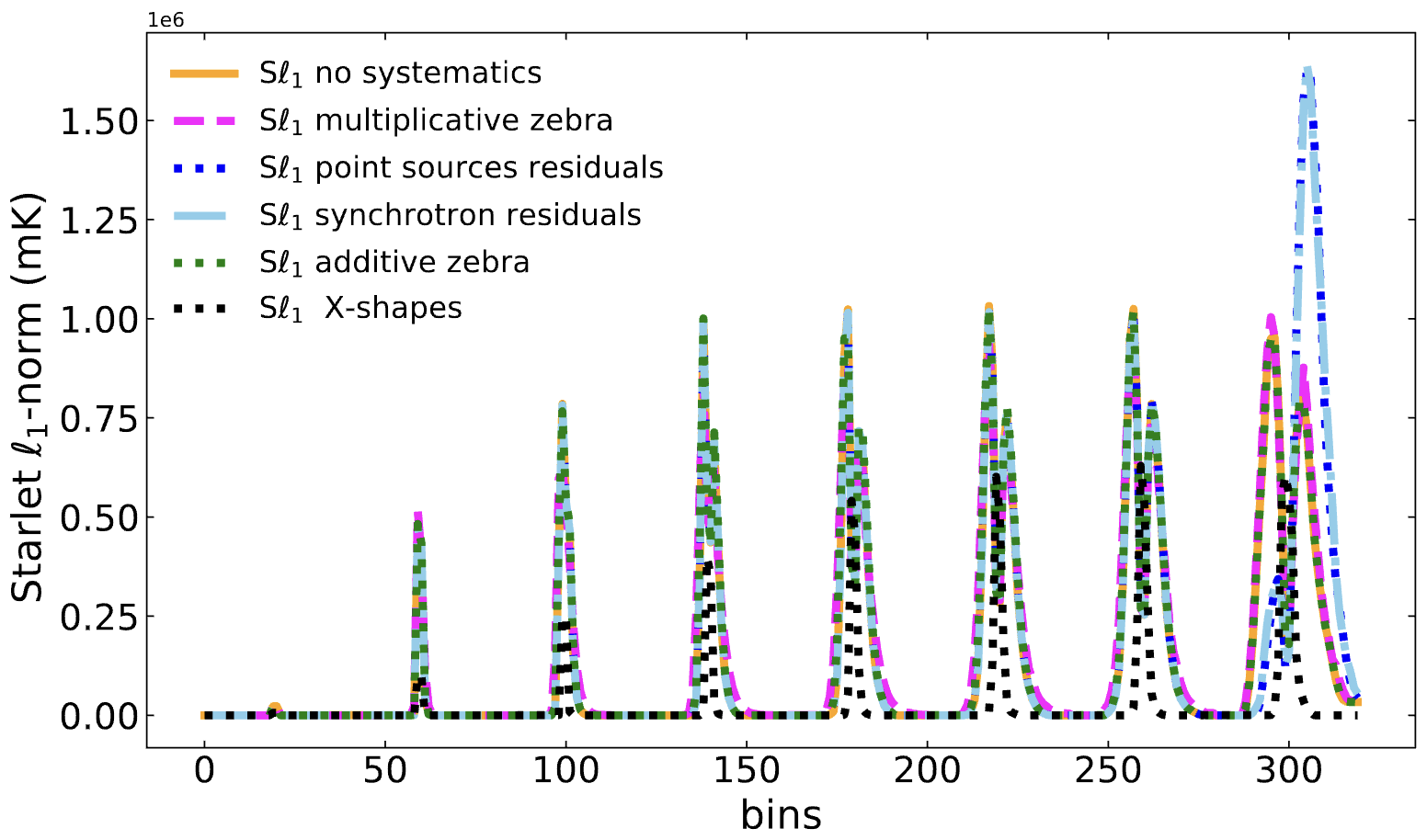}
    \caption{C$_\ell$ (\textit{first panel}), starlet $\ell_1$-norm with five scales (\textit{second panel}) and starlet $\ell_1$-norm with eight scales (\textit{third panel}), map without systematic effects (orange), map with multiplicative stripes (magenta), map with additive stripes (green), map with extragalactic point sources residuals (blue), map with Galactic synchrotron residuals (sky blue) and map with X-shape pattern (black). All maps are $f_{\rm sky}$=0.19 and beamed at $0.5^\circ$.}
    \label{fig:SS_2}
\end{figure*}

We investigate the impact of residual Galactic synchrotron emission and extragalactic point sources on parameter constraints for beamed, masked maps. The maps are shown in Figs. \ref{fig:synch_all} and \ref{fig:PS_all} (we note that while the Galactic synchrotron residual is visible by eye for higher contamination the extragalactic point sources are unnoticeable by eye) and the extracted summary statistics in Figs. \ref{fig:comp_synch} and \ref{fig:comp_PS}. Both Galactic residual synchrotron emission and extragalactic residual point sources are large-scale effects, as is shown in Fig. \ref{fig:SS_2}, affecting mainly low $\ell$ and the coarse scale of the starlet $\ell_{1}$-norm.

\begin{figure*}
    \centering
    \includegraphics[width=0.3\textwidth]{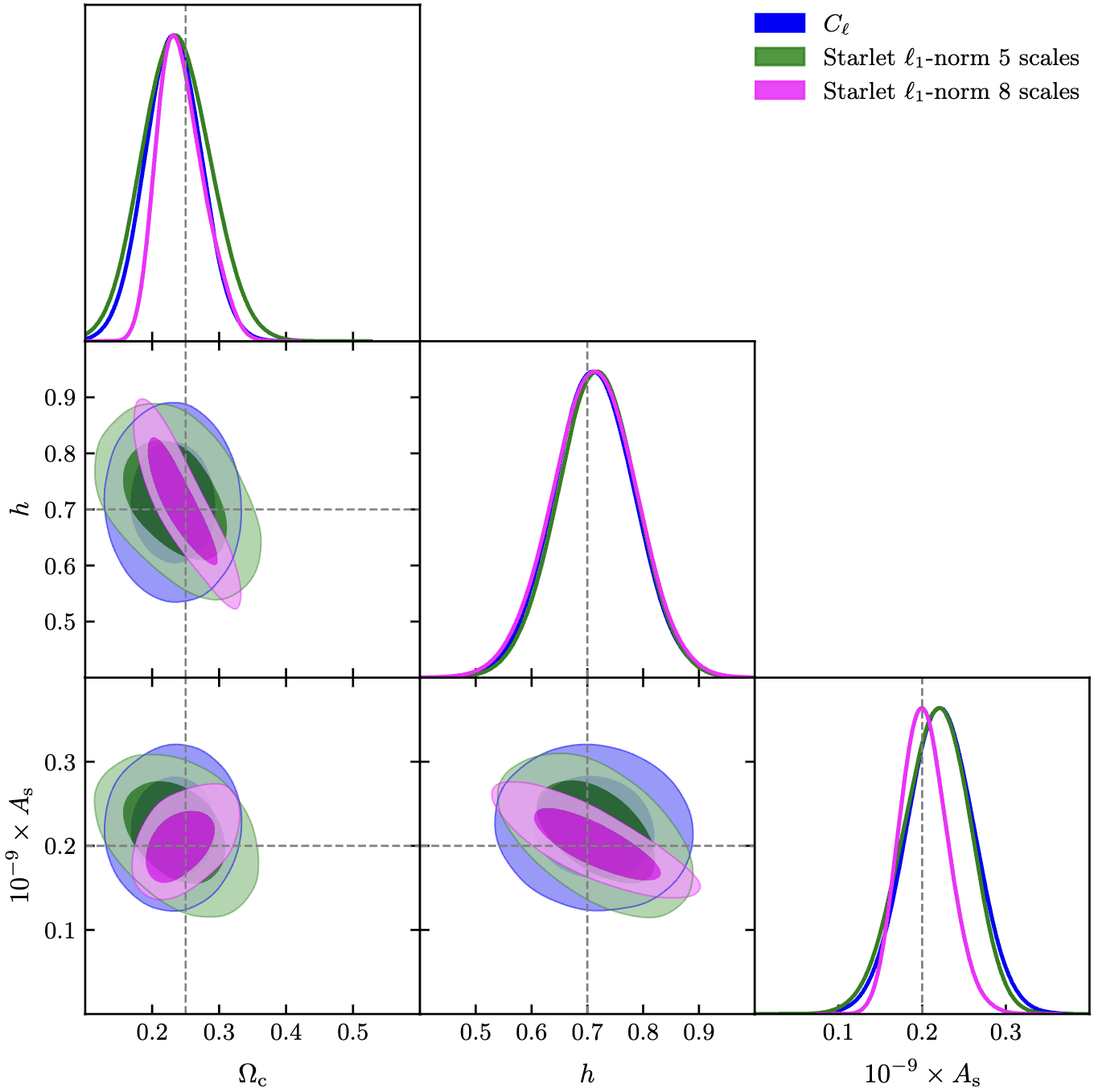}
    \includegraphics[width=0.3\textwidth]{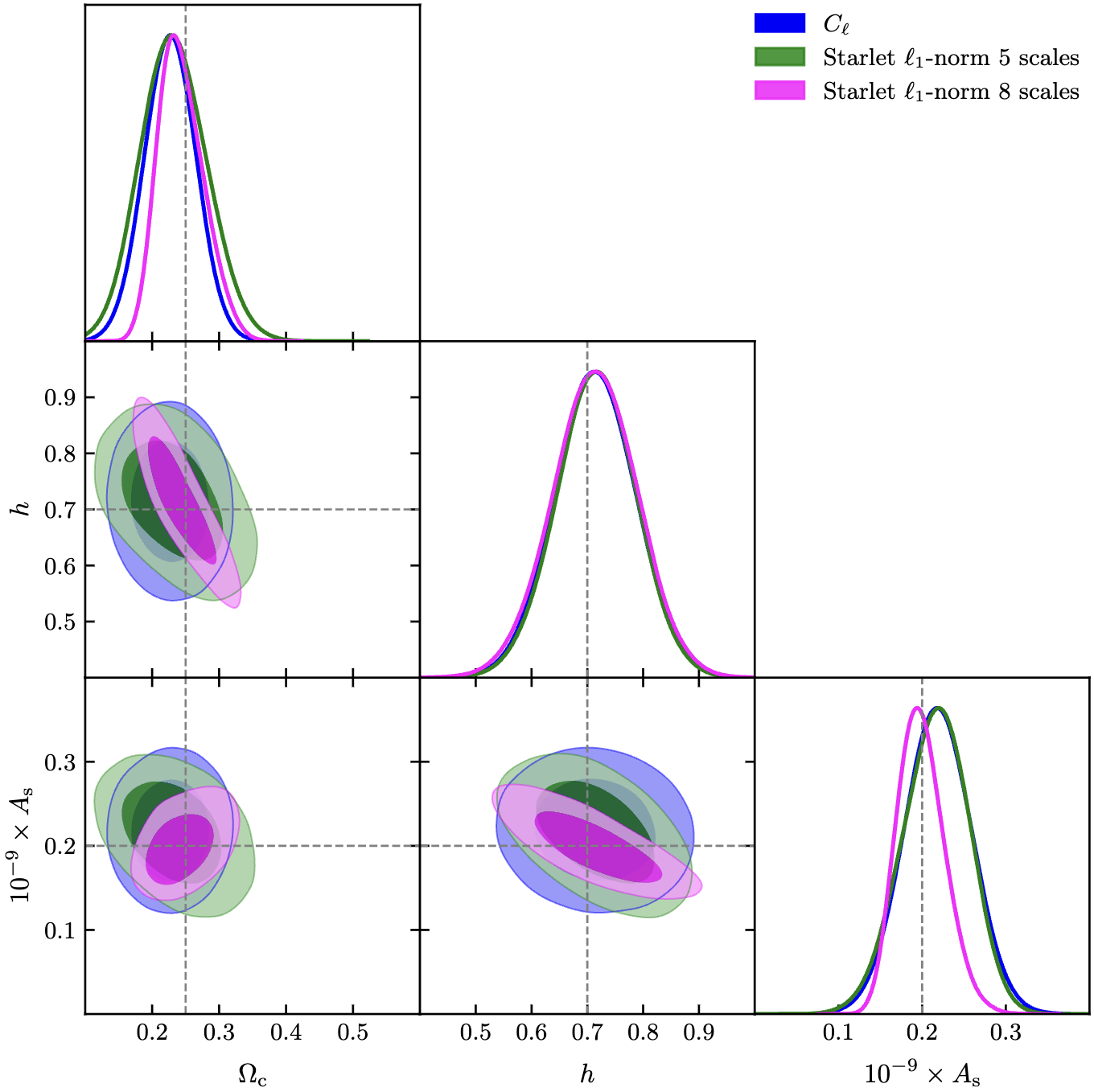}
    \includegraphics[width=0.3\textwidth]{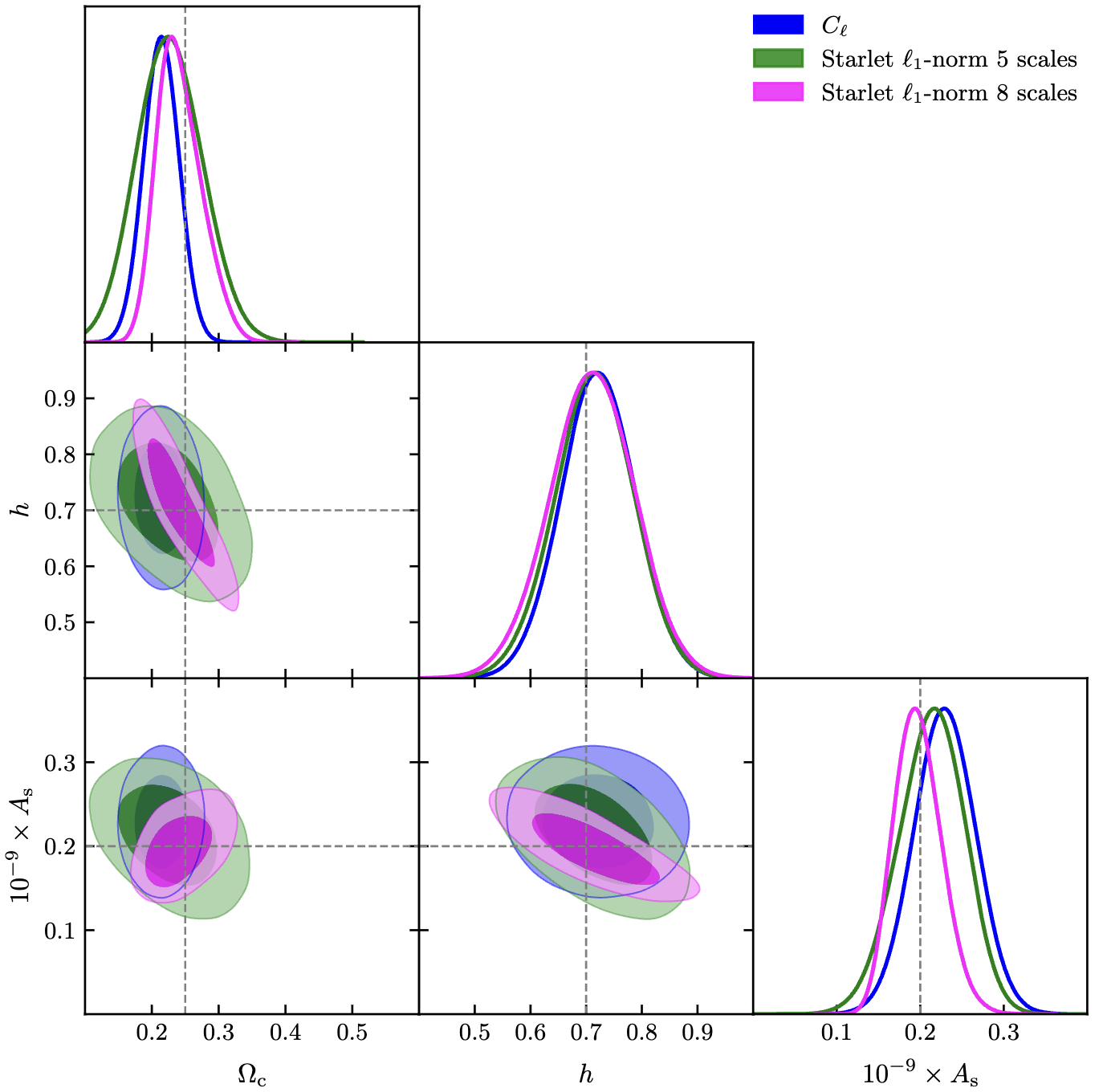}
    \caption{
    Posterior distributions for $\Omega_{\rm c}$, $h$, and $A_{\rm s}$ obtained from the angular power spectrum, $C_\ell$, 
    starlet $\ell_1$-norm (without coarse scale) for different Galactic synchrotron residual fractions: 0.0005 \% (\textit{first panel}), 0.001 \% (\textit{second panel}), and 0.002 \% (\textit{third panel}).  The contours correspond to $68\%$ and $95\%$ confidence levels.
    }
    \label{fig:synch_no_coarse}

\end{figure*}

Figure \ref{fig:synch_no_coarse} shows that the constraints derived from the starlet $\ell_{1}$-norm are overall consistent with those from the standard $C_{\ell}$ analysis, confirming the reliability of the method. When a larger number of starlet scales is included, the posteriors become noticeably tighter, particularly for 
$A_{\mathrm{s}}$ and $h$, illustrating the gain in constraining power obtained 
by capturing information across multiple spatial scales. Even when the Galactic synchrotron residuals are increased, the starlet-based inference remains robust, 
showing only a mild degradation in precision, whereas the C$_\ell$ exhibits a slight bias and tighter contours, showing its lack of reliability in the case of residual Galactic synchrotron contaminants. These results emphasize the 
capability of the starlet $\ell_{1}$-norm to exploit complementary, non-Gaussian information beyond that accessible to the angular power spectrum. When we include the coarse scale (in Figs. \ref{fig:synch} and \ref{fig:PS}), we find that the starlet $\ell_1$-norm exhibits a stronger bias than the angular power spectrum as the fraction of residual contaminants (Galactic synchrotron or extragalactic point sources) increases.

\begin{figure*}
    \centering
    \includegraphics[width=0.3\textwidth]{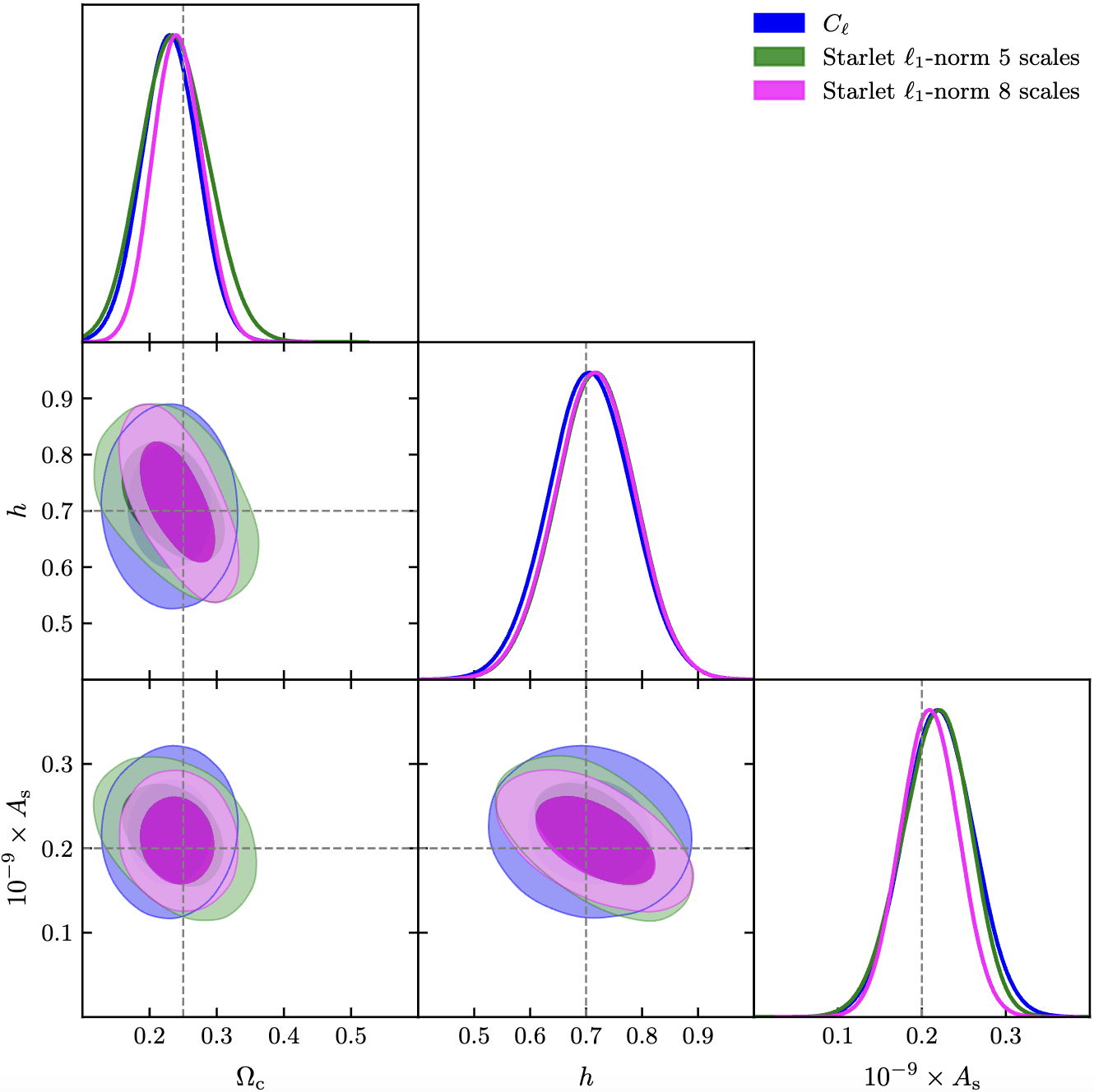}
    \includegraphics[width=0.3\textwidth]{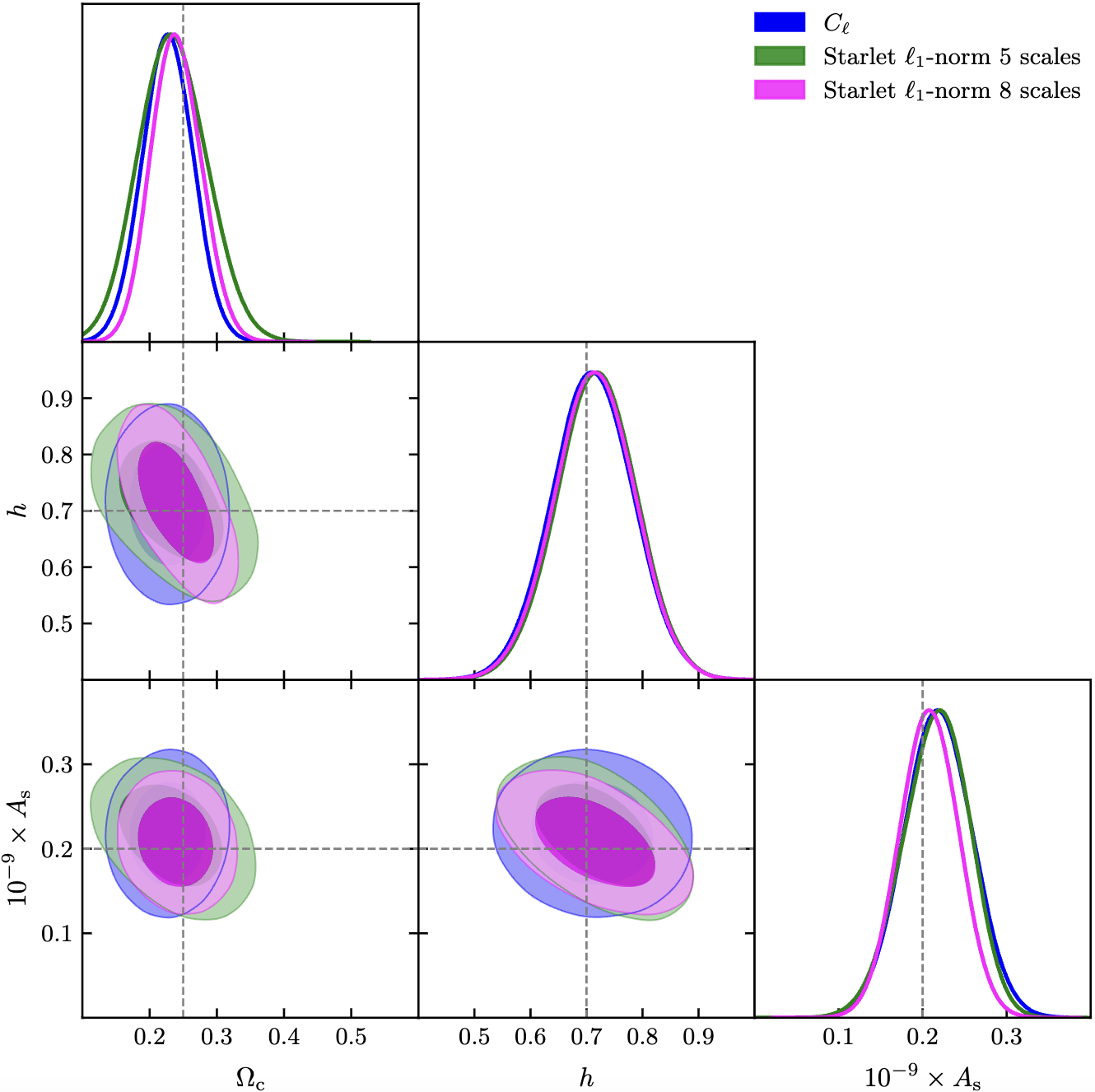}
    \includegraphics[width=0.3\textwidth]{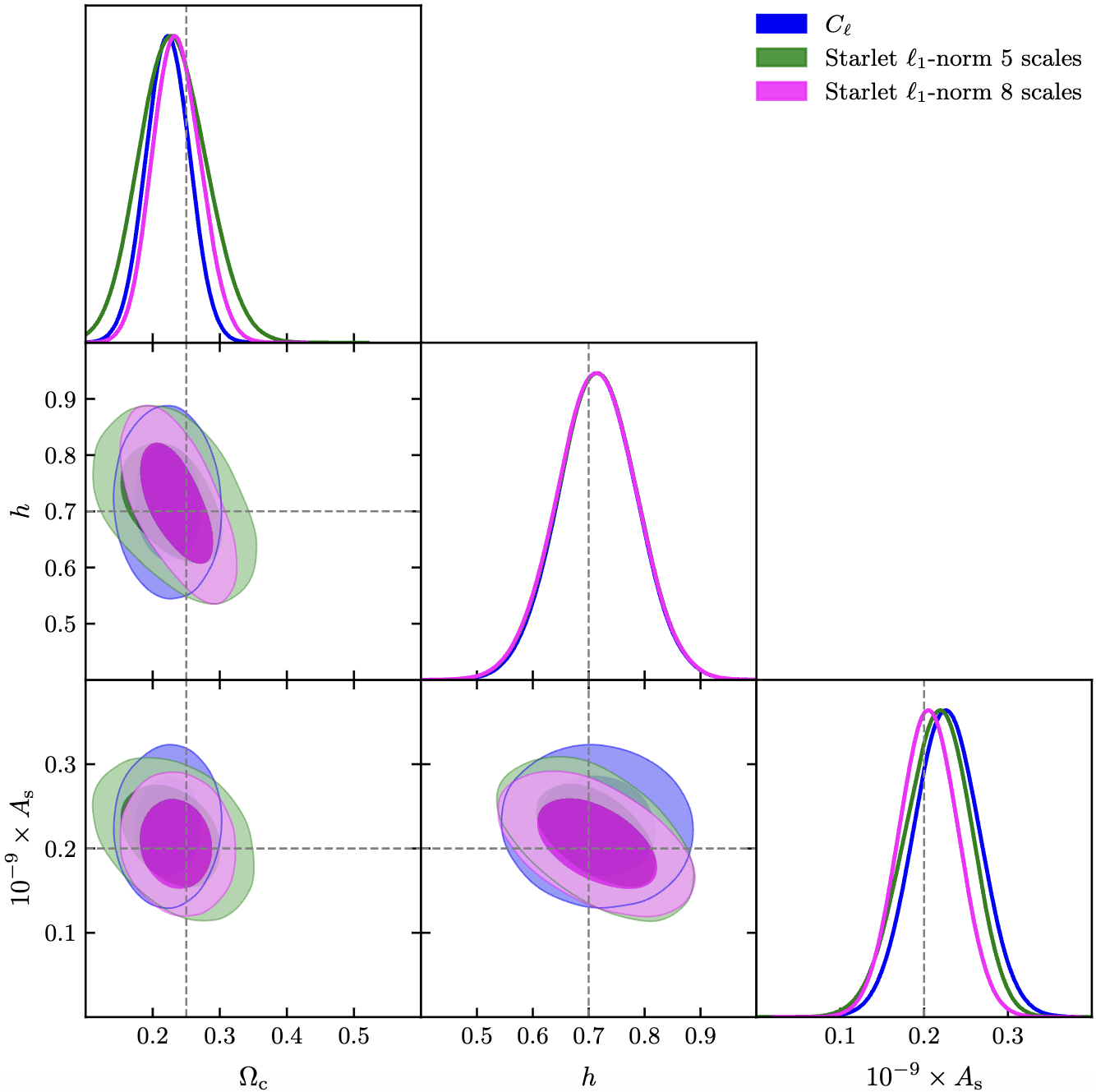}
    \caption{
    Posterior distributions for $\Omega_{\rm c}$, $h$, and $A_{\rm s}$ obtained from the angular power spectrum, $C_\ell$, 
    starlet $\ell_1$-norm (without coarse scale) for different extragalactic point sources residual fractions: 0.001 \% (\textit{first panel}), 0.0025 \% (\textit{second panel}), and 0.005 \% (\textit{third panel}).    The contours correspond to $68\%$ and $95\%$ confidence levels.
    }
    \label{fig:PS_no_coarse}
    
\end{figure*}

Such an enhanced sensitivity indicates that the starlet $\ell_1$-norm could be employed as a tool to diagnose the presence and amount of residual contamination after foreground cleaning. Since foreground cleaning does not perfectly remove Galactic synchrotron emission and extragalactic point sources, the level of residual contamination in observational data remains uncertain. Another interesting feature is that in Figs. \ref{fig:synch_no_coarse} and \ref{fig:PS_no_coarse}, the $C_\ell$ posteriors (blue contours) for extragalactic point source and Galactic synchrotron residuals are very similar, indicating a limited ability to distinguish between these contaminants. In contrast, the starlet $\ell_1$-norm with eight scales (magenta contours) produces clearly distinct posterior patterns for the two types of residuals, demonstrating its capacity to differentiate between Galactic synchrotron and extragalactic point source contamination.

\subsection{Effect of scanning strategy}
\begin{figure*} 
    \centering
    \begin{minipage}{0.25\textwidth}
        \includegraphics[width=\textwidth]{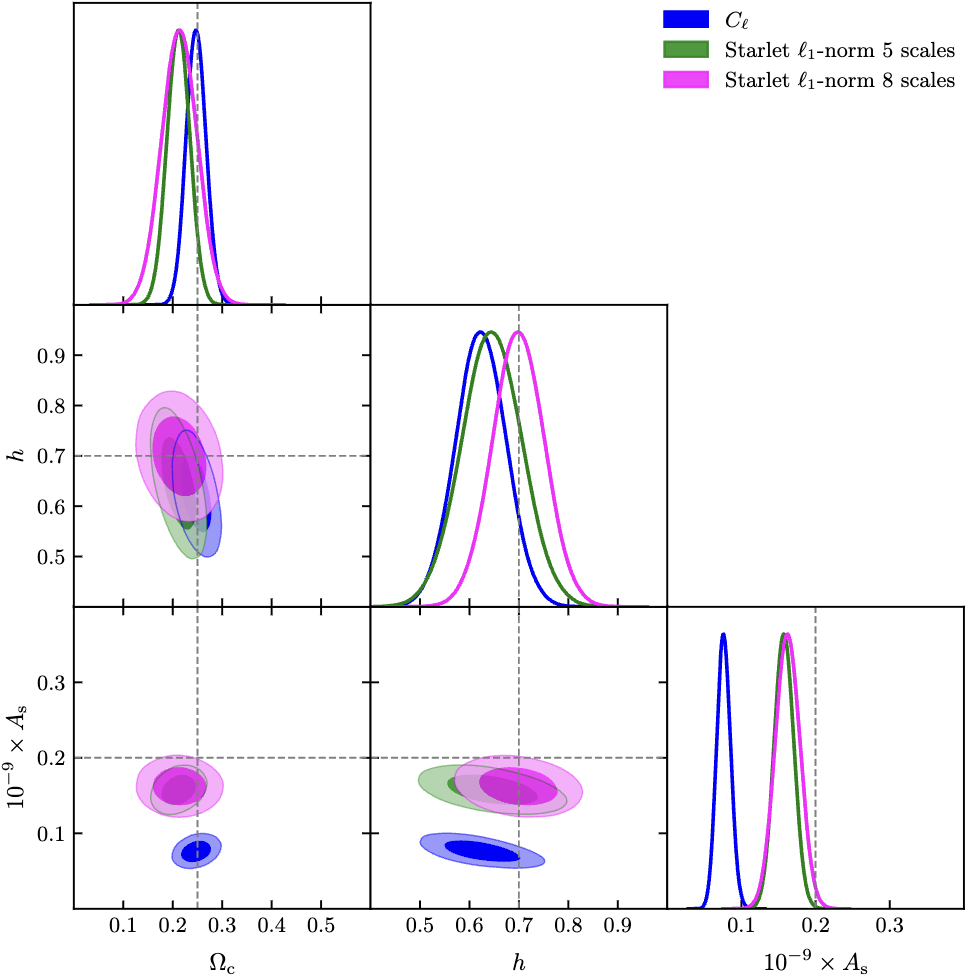}
    \end{minipage}%
    \begin{minipage}{0.25\textwidth}
        \includegraphics[width=\textwidth]{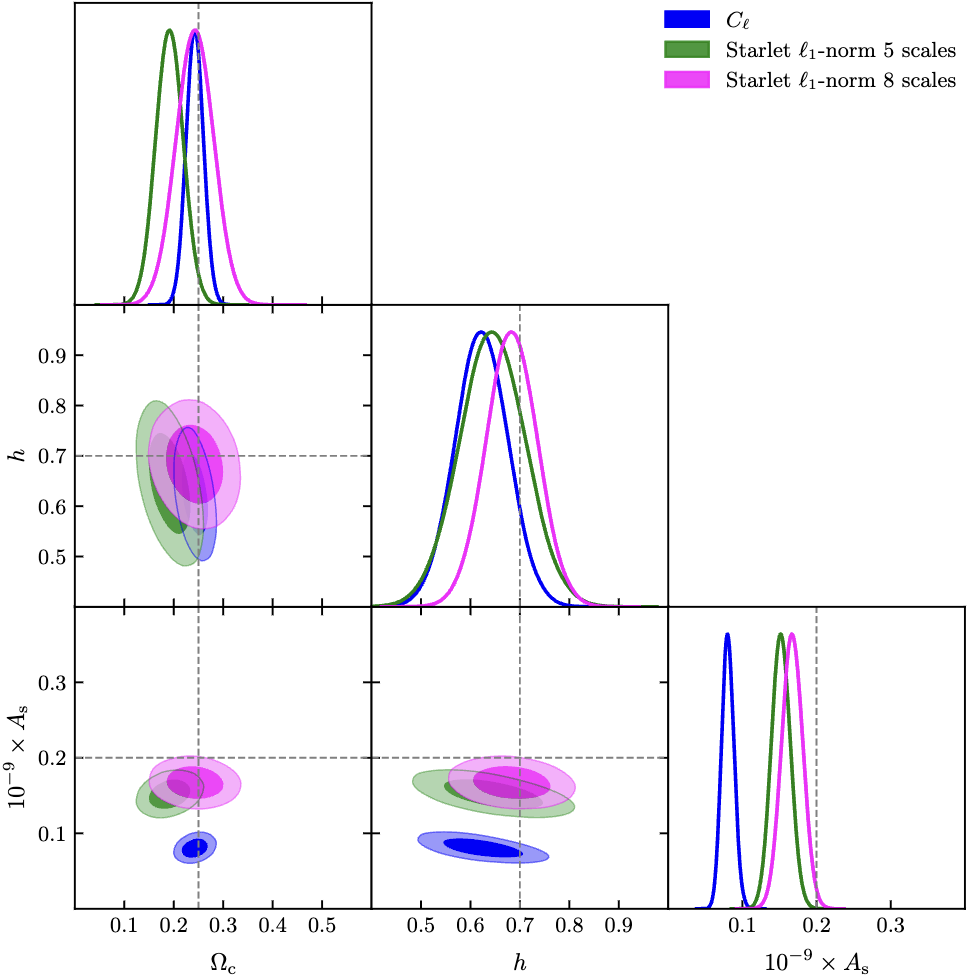}
    \end{minipage}%
    \begin{minipage}{0.25\textwidth}
        \includegraphics[width=\textwidth]{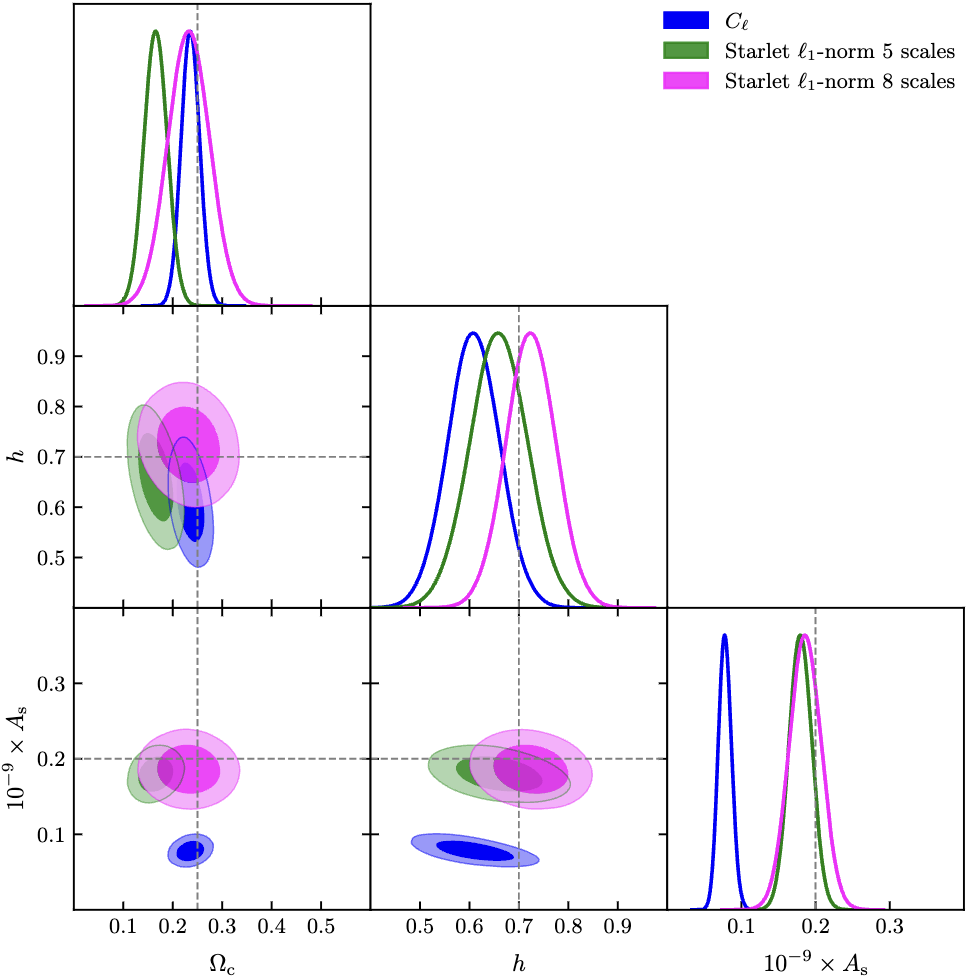}
    \end{minipage}%
    \begin{minipage}{0.25\textwidth}
        \includegraphics[width=\textwidth]{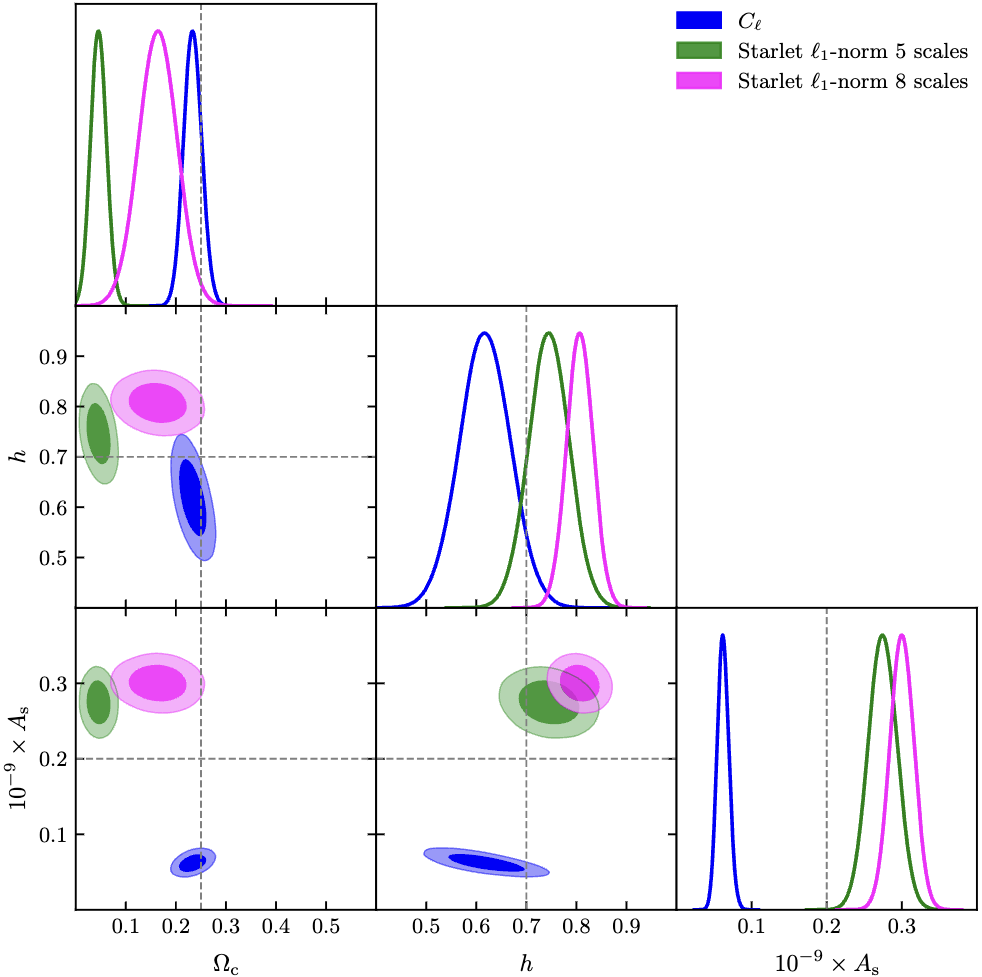}
    \end{minipage}
    \caption{Posterior distributions for $\Omega_{\rm c}$, $h$, $A_{\rm s}$, and another parameter obtained from the angular power spectrum, $C_\ell$, 
    starlet $\ell_1$-norm for different multiplicative X shape weights: 10X (\textit{first panel}), 25X (\textit{second panel}), 35X (\textit{third panel}), and 50X (\textit{fourth panel}).
    The contours correspond to $68\%$ and $95\%$ confidence levels.}
    \label{fig:scan_SBI}
\end{figure*}

To study inhomogeneous noise caused by scanning strategy, we generated synthetic coverage-weighted \hi\ maps by applying X-shaped scanning patterns on our masked lognormal maps as shown in Sect. \ref{subsec:Xshape}.
Figure \ref{fig:scan_SBI} shows that while the standard $C_{\ell}$ 
analysis exhibits a progressive bias and artificial tightening of the posteriors 
as the X-shape amplitude increases, the starlet $\ell_{1}$-norm yields more stable 
and consistent constraints across all cases. The configuration using eight starlet 
scales maintains well-behaved contours even for the largest X-shape weights, 
highlighting the robustness of the starlet representation to multiplicative 
distortions. This result demonstrates that the $\ell_{1}$-norm statistic 
effectively captures complementary, non-Gaussian information from the maps and 
remains resilient to weighting effects across scales. We also performed the analysis for the additive weighted scanning pattern (results shown in Fig. \ref{fig:scan_add_SBI}), since it is a large-scale effect (we show that it affects low $\ell$ in $C_\ell$ and the coarse scales of the starlets $\ell_{1}$-norm in Fig. \ref{fig:comp_scan_add}), it yields similar results to the residual contamination from Galactic synchrotron and extragalactic point sources.

Figures \ref{fig:comp_synch}, \ref{fig:comp_PS}, and \ref{fig:comp_scan_add} show that the coarse-scale $\ell_1$-norm exhibits pronounced non-Gaussian features, with reversed peaks and larger amplitude differences compared to the training set (Fig. \ref{fig:l1_norm}), representing patterns that were not encountered during training and explaining the observed bias seen in Figs. \ref{fig:scan_add_SBI}, \ref{fig:PS}, and \ref{fig:synch}. This sensitivity arises from different physical contributions: diffuse Galactic synchrotron residuals shift large-scale fluctuations, extragalactic point sources add localized high-amplitude coefficients that influence coarse-scale averages, and a weighted scanning pattern modifies coverage, each producing measurable deviations in the coarse-scale $\ell_1$-norm. Because the coarse scale spans a substantial fraction of the data vector, the starlet $\ell_1$-norm is particularly effective at detecting these large-scale effects, whereas the $C_\ell$, which is dominated by higher-$\ell$ modes, would be influenced very little by the lowest-$\ell$ modes even if they were included, and is therefore less sensitive to these deviations.

Despite using only for example 18 and 35 points for the five and eight-scale decompositions, respectively (see Fig. \ref{tab:table_bins} for the different scenarios we have analyzed in this section), compared to 2037 points for $C_\ell$, the starlet $\ell_1$-norm achieves tighter parameter constraints. This demonstrates that a compact multi-scale, non-Gaussian representation can encode more cosmological information than the full power spectrum.

\section{Conclusions}\label{sec:conclusion}

In this work, we have investigated the starlet $\ell_{1}$-norm as an alternative summary statistic to the angular power spectrum for neutral hydrogen intensity mapping. Using lognormal full-sky simulations at the redshift of interest, $z=0.425$, probed by MeerKLASS L-band observations \citep{MeerKLASS}, combined with SBI through the \texttt{JaxILI} framework, we assessed the robustness of both statistics under realistic observational conditions, including partial sky coverage, telescope beam convolution, thermal noise, and various systematic effects such as inhomogeneous noise caused by scanning strategy and foreground residuals.
Our results show that the starlet $\ell_{1}$-norm consistently outperforms the standard $C_{\ell}$ estimator in constraining cosmological parameters whenever small-scale non-Gaussian structures are preserved. The $\ell_{1}$-norm proves remarkably robust to instrumental noise, masking, and stripe-like artefacts, yielding unbiased and stable parameter estimates even in regimes where the angular power spectrum becomes biased or significantly degraded. Increasing the number of starlet scales further enhances the constraining power, achieving in the case of a $0.5^\circ$ beam with instrumental noise and intermediate mask $f_{sky}=0.19$ a 2.81× improvement in the FoM relative to the $C_\ell$ power spectrum and highlighting the importance of exploiting multi-scale information. In the presence of Galactic synchrotron and extragalactic point-source residuals, the starlet $\ell_{1}$-norm exhibits stronger sensitivity than $C_{\ell}$, suggesting that it may also serve as a diagnostic tool to identify and quantify residual foreground contamination after component separation.
These findings demonstrate that multi-scale, higher-order statistics such as the starlet $\ell_{1}$-norm provide a powerful complement to traditional two-point analyses for \hi\ intensity mapping. Its ability to capture non-Gaussian information and to mitigate the impact of complex observational systematics makes it a promising tool for cosmological inference with ongoing \hi\ surveys such as MeerKLASS and with future surveys planned with the SKAO.

\begin{acknowledgements}
We thank Florent Mertens for useful discussion.
This work was supported by the TITAN ERA Chair project (contract no. 101086741) within the Horizon Europe Framework Program of the European Commission, the  Agence Nationale de la Recherche (ANR-22-CE31-0014-01 TOSCA), and the French Programme National de Cosmologie et Galaxies (PNCG project CIMES).

\end{acknowledgements}

\bibliographystyle{aa}
\bibliography{main}

@ARTICLE{Bull15,
       author = {{Bull}, Philip and {Ferreira}, Pedro G. and {Patel}, Prina and {Santos}, M{\'a}rio G.},
        title = "{Late-time Cosmology with 21 cm Intensity Mapping Experiments}",
      journal = {\apj},
         year = 2015,
        month = apr,
       volume = {803},
       number = {1},
          eid = {21},
        pages = {21},
          doi = {10.1088/0004-637X/803/1/21},
archivePrefix = {arXiv},
       eprint = {1405.1452},
 primaryClass = {astro-ph.CO},
       adsurl = {https://ui.adsabs.harvard.edu/abs/2015ApJ...803...21B}
}

@inproceedings{Papamakarios2017,
 author = {Papamakarios, George and Pavlakou, Theo and Murray, Iain},
 booktitle = {Advances in Neural Information Processing Systems},
 editor = {I. Guyon and U. Von Luxburg and S. Bengio and H. Wallach and R. Fergus and S. Vishwanathan and R. Garnett},
 pages = {},
 publisher = {Curran Associates, Inc.},
 title = {Masked Autoregressive Flow for Density Estimation},
 url = {https://proceedings.neurips.cc/paper_files/paper/2017/file/6c1da886822c67822bcf3679d04369fa-Paper.pdf},
 volume = {30},
 year = {2017}
}

@ARTICLE{Ajani2021,
       author = {{Ajani}, Virginia and {Starck}, Jean-Luc and {Pettorino}, Valeria},
        title = "{Starlet {\ensuremath{\ell}}$_{1}$-norm for weak lensing cosmology}",
      journal = {\aap},
         year = 2021,
        month = jan,
       volume = {645},
          eid = {L11},
        pages = {L11},
          doi = {10.1051/0004-6361/202039988},
archivePrefix = {arXiv},
       eprint = {2101.01542},
 primaryClass = {astro-ph.CO},
       adsurl = {https://ui.adsabs.harvard.edu/abs/2021A&A...645L..11A}
}

@ARTICLE{Ajani2023,
       author = {{Ajani}, Virginia and {Harnois-D{\'e}raps}, Joachim and {Pettorino}, Valeria and {Starck}, Jean-Luc},
        title = "{Starlet higher order statistics for galaxy clustering and weak lensing}",
      journal = {\aap},
         year = 2023,
        month = apr,
       volume = {672},
          eid = {L10},
        pages = {L10},
          doi = {10.1051/0004-6361/202245510},
archivePrefix = {arXiv},
       eprint = {2211.10519},
 primaryClass = {astro-ph.CO},
       adsurl = {https://ui.adsabs.harvard.edu/abs/2023A&A...672L..10A}
}

@BOOK{Peebles1980,
       author = {{Peebles}, P.~J.~E.},
        title = "{The large-scale structure of the universe}",
         year = 1980,
       adsurl = {https://ui.adsabs.harvard.edu/abs/1980lssu.book.....P}
}

@ARTICLE{Bernardeau2002,
       author = {{Bernardeau}, F. and {Colombi}, S. and {Gazta{\~n}aga}, E. and {Scoccimarro}, R.},
        title = "{Large-scale structure of the Universe and cosmological perturbation theory}",
      journal = {\physrep},
         year = 2002,
        month = sep,
       volume = {367},
       number = {1-3},
        pages = {1-248},
          doi = {10.1016/S0370-1573(02)00135-7},
archivePrefix = {arXiv},
       eprint = {astro-ph/0112551},
 primaryClass = {astro-ph},
       adsurl = {https://ui.adsabs.harvard.edu/abs/2002PhR...367....1B}
}

@ARTICLE{Casas2022,
       author = {{Casas}, Santiago and {Carucci}, Isabella P. and {Pettorino}, Valeria and {Camera}, Stefano and {Martinelli}, Matteo},
        title = "{Constraining gravity with synergies between radio and optical cosmological surveys}",
      journal = {Physics of the Dark Universe},
         year = 2023,
        month = feb,
       volume = {39},
          eid = {101151},
        pages = {101151},
          doi = {10.1016/j.dark.2022.101151},
archivePrefix = {arXiv},
       eprint = {2210.05705},
 primaryClass = {astro-ph.CO},
       adsurl = {https://ui.adsabs.harvard.edu/abs/2023PDU....3901151C}
}

@ARTICLE{Cranmer2020,
       author = {{Cranmer}, Kyle and {Brehmer}, Johann and {Louppe}, Gilles},
        title = "{The frontier of simulation-based inference}",
      journal = {Proceedings of the National Academy of Science},
         year = 2020,
        month = dec,
       volume = {117},
       number = {48},
        pages = {30055-30062},
          doi = {10.1073/pnas.1912789117},
archivePrefix = {arXiv},
       eprint = {1911.01429},
 primaryClass = {stat.ML},
       adsurl = {https://ui.adsabs.harvard.edu/abs/2020PNAS..11730055C}
}

@ARTICLE{VillaescusaNavarro2018,
       author = {{Villaescusa-Navarro}, Francisco and {Genel}, Shy and {Castorina}, Emanuele and {Obuljen}, Andrej and {Spergel}, David N. and {Hernquist}, Lars and {Nelson}, Dylan and {Carucci}, Isabella P. and {Pillepich}, Annalisa and {Marinacci}, Federico and {Diemer}, Benedikt and {Vogelsberger}, Mark and {Weinberger}, Rainer and {Pakmor}, R{\"u}diger},
        title = "{Ingredients for 21 cm Intensity Mapping}",
      journal = {\apj},
         year = 2018,
        month = oct,
       volume = {866},
       number = {2},
          eid = {135},
        pages = {135},
          doi = {10.3847/1538-4357/aadba0},
archivePrefix = {arXiv},
       eprint = {1804.09180},
 primaryClass = {astro-ph.CO},
       adsurl = {https://ui.adsabs.harvard.edu/abs/2018ApJ...866..135V}
}

@ARTICLE{Pourtsidou2017,
       author = {{Pourtsidou}, Alkistis and {Bacon}, David and {Crittenden}, Robert},
        title = "{H I and cosmological constraints from intensity mapping, optical and CMB surveys}",
      journal = {\mnras},
         year = 2017,
        month = oct,
       volume = {470},
       number = {4},
        pages = {4251-4260},
          doi = {10.1093/mnras/stx1479},
archivePrefix = {arXiv},
       eprint = {1610.04189},
 primaryClass = {astro-ph.CO},
       adsurl = {https://ui.adsabs.harvard.edu/abs/2017MNRAS.470.4251P}
}

@ARTICLE{Alsing2019,
       author = {{Alsing}, Justin and {Charnock}, Tom and {Feeney}, Stephen and {Wandelt}, Benjamin},
        title = "{Fast likelihood-free cosmology with neural density estimators and active learning}",
      journal = {\mnras},
         year = 2019,
        month = sep,
       volume = {488},
       number = {3},
        pages = {4440-4458},
          doi = {10.1093/mnras/stz1960},
archivePrefix = {arXiv},
       eprint = {1903.00007},
 primaryClass = {astro-ph.CO},
       adsurl = {https://ui.adsabs.harvard.edu/abs/2019MNRAS.488.4440A}
}

@ARTICLE{Pritchard2012,
       author = {{Pritchard}, Jonathan R. and {Loeb}, Abraham},
        title = "{21 cm cosmology in the 21st century}",
      journal = {Reports on Progress in Physics},
         year = 2012,
        month = aug,
       volume = {75},
       number = {8},
          eid = {086901},
        pages = {086901},
          doi = {10.1088/0034-4885/75/8/086901},
archivePrefix = {arXiv},
       eprint = {1109.6012},
 primaryClass = {astro-ph.CO},
       adsurl = {https://ui.adsabs.harvard.edu/abs/2012RPPh...75h6901P}
}

@ARTICLE{Ansari2012,
       author = {{Ansari}, R. and {Campagne}, J.~E. and {Colom}, P. and {Le Goff}, J.~M. and {Magneville}, C. and {Martin}, J.~M. and {Moniez}, M. and {Rich}, J. and {Y{\`e}che}, C.},
        title = "{21 cm observation of large-scale structures at z \raisebox{-0.5ex}\textasciitilde 1. Instrument sensitivity and foreground subtraction}",
      journal = {\aap},
         year = 2012,
        month = apr,
       volume = {540},
          eid = {A129},
        pages = {A129},
          doi = {10.1051/0004-6361/201117837},
archivePrefix = {arXiv},
       eprint = {1108.1474},
 primaryClass = {astro-ph.CO},
       adsurl = {https://ui.adsabs.harvard.edu/abs/2012A&A...540A.129A}
}

@INPROCEEDINGS{Santos2015,
       author = {{Santos}, M. and {Bull}, P. and {Alonso}, D. and {Camera}, S. and {Ferreira}, P. and {Bernardi}, G. and {Maartens}, R. and {Viel}, M. and {Villaescusa-Navarro}, F. and {Abdalla}, F.~B. and {Jarvis}, M. and {Metcalf}, R.~B. and {Pourtsidou}, A. and {Wolz}, L.},
        title = "{Cosmology from a SKA HI intensity mapping survey}",
    booktitle = {Advancing Astrophysics with the Square Kilometre Array (AASKA14)},
         year = 2015,
        month = apr,
          eid = {19},
        pages = {19},
          doi = {10.22323/1.215.0019},
archivePrefix = {arXiv},
       eprint = {1501.03989},
 primaryClass = {astro-ph.CO},
       adsurl = {https://ui.adsabs.harvard.edu/abs/2015aska.confE..19S}
}

@ARTICLE{Battye2013,
       author = {{Battye}, R.~A. and {Browne}, I.~W.~A. and {Dickinson}, C. and {Heron}, G. and {Maffei}, B. and {Pourtsidou}, A.},
        title = "{H I intensity mapping: a single dish approach}",
      journal = {\mnras},
         year = 2013,
        month = sep,
       volume = {434},
       number = {2},
        pages = {1239-1256},
          doi = {10.1093/mnras/stt1082},
archivePrefix = {arXiv},
       eprint = {1209.0343},
 primaryClass = {astro-ph.CO},
       adsurl = {https://ui.adsabs.harvard.edu/abs/2013MNRAS.434.1239B}
}

@ARTICLE{Sheean2021,
       author = {{Jolicoeur}, Sheean and {Maartens}, Roy and {De Weerd}, Eline M. and {Umeh}, Obinna and {Clarkson}, Chris and {Camera}, Stefano},
        title = "{Detecting the relativistic bispectrum in 21cm intensity maps}",
      journal = {\jcap},
         year = 2021,
        month = jun,
       volume = {2021},
       number = {6},
          eid = {039},
        pages = {039},
          doi = {10.1088/1475-7516/2021/06/039},
archivePrefix = {arXiv},
       eprint = {2009.06197},
 primaryClass = {astro-ph.CO},
       adsurl = {https://ui.adsabs.harvard.edu/abs/2021JCAP...06..039J}
}

@ARTICLE{Carucci2024,
       author = {{Carucci}, I.~P. and {Bernal}, J.~L. and {Cunnington}, S. and {Santos}, M.~G. and {Wang}, J. and {Fonseca}, J. and {Grainge}, K. and {Irfan}, M.~O. and {Li}, Y. and {Pourtsidou}, A. and {Spinelli}, M. and {Wolz}, L.},
        title = "{Hydrogen intensity mapping with MeerKAT: Preserving cosmological signal by optimising contaminant separation}",
      journal = {\aap},
         year = 2025,
        month = nov,
       volume = {703},
          eid = {A222},
        pages = {A222},
          doi = {10.1051/0004-6361/202453461},
archivePrefix = {arXiv},
       eprint = {2412.06750},
 primaryClass = {astro-ph.CO},
       adsurl = {https://ui.adsabs.harvard.edu/abs/2025A&A...703A.222C}
}

@ARTICLE{MeerKLASS,
       author = {{MeerKLASS Collaboration}, Matilde and {Bernal}, Jos{\'e} L. and {Bull}, Philip and {Camera}, Stefano and {Carucci}, Isabella P. and {Chen}, Zhaoting and {Cunnington}, Steven and {Engelbrecht}, Brandon N. and {Fonseca}, Jos{\'e} and {Grainge}, Keith and {Irfan}, Melis O. and {Li}, Yichao and {Mazumder}, Aishrila and {Paul}, Sourabh and {Pourtsidou}, Alkistis and {Santos}, Mario G. and {Spinelli}, Marta and {Wang}, Jingying and {Witzemann}, Amadeus and {Wolz}, Laura},
        title = "{MeerKLASS L-band deep-field intensity maps: entering the H I dominated regime}",
      journal = {\mnras},
         year = 2025,
        month = mar,
       volume = {537},
       number = {4},
        pages = {3632-3661},
          doi = {10.1093/mnras/staf195},
archivePrefix = {arXiv},
       eprint = {2407.21626},
 primaryClass = {astro-ph.CO},
       adsurl = {https://ui.adsabs.harvard.edu/abs/2025MNRAS.537.3632M}
}

@ARTICLE{Starck_2006,
       author = {{Starck}, J.-L. and {Moudden}, Y. and {Abrial}, P. and {Nguyen}, M.},
        title = "{Wavelets, ridgelets and curvelets on the sphere}",
      journal = {\aap},
         year = 2006,
        month = feb,
       volume = {446},
       number = {3},
        pages = {1191-1204},
          doi = {10.1051/0004-6361:20053246},
archivePrefix = {arXiv},
       eprint = {astro-ph/0509883},
 primaryClass = {astro-ph},
       adsurl = {https://ui.adsabs.harvard.edu/abs/2006A&A...446.1191S}
}

@ARTICLE{foreground,
       author = {{Bowman}, Judd D. and {Morales}, Miguel F. and {Hewitt}, Jacqueline N.},
        title = "{Foreground Contamination in Interferometric Measurements of the Redshifted 21 cm Power Spectrum}",
      journal = {\apj},
         year = 2009,
        month = apr,
       volume = {695},
       number = {1},
        pages = {183-199},
          doi = {10.1088/0004-637X/695/1/183},
archivePrefix = {arXiv},
       eprint = {0807.3956},
 primaryClass = {astro-ph},
       adsurl = {https://ui.adsabs.harvard.edu/abs/2009ApJ...695..183B}
}

@ARTICLE{2021MNRAS,
       author = {{Li}, Yichao and {Santos}, Mario G. and {Grainge}, Keith and {Harper}, Stuart and {Wang}, Jingying},
        title = "{H I intensity mapping with MeerKAT: 1/f noise analysis}",
      journal = {\mnras},
         year = 2021,
        month = mar,
       volume = {501},
       number = {3},
        pages = {4344-4358},
          doi = {10.1093/mnras/staa3856},
archivePrefix = {arXiv},
       eprint = {2007.01767},
 primaryClass = {astro-ph.CO},
       adsurl = {https://ui.adsabs.harvard.edu/abs/2021MNRAS.501.4344L}
}

@ARTICLE{Wang21,
       author = {{Wang}, Jingying and {Santos}, Mario G. and {Bull}, Philip and {Grainge}, Keith and {Cunnington}, Steven and {Fonseca}, Jos{\'e} and {Irfan}, Melis O. and {Li}, Yichao and {Pourtsidou}, Alkistis and {Soares}, Paula S. and {Spinelli}, Marta and {Bernardi}, Gianni and {Engelbrecht}, Brandon},
        title = "{H I intensity mapping with MeerKAT: calibration pipeline for multidish autocorrelation observations}",
      journal = {\mnras},
         year = 2021,
        month = aug,
       volume = {505},
       number = {3},
        pages = {3698-3721},
          doi = {10.1093/mnras/stab1365},
archivePrefix = {arXiv},
       eprint = {2011.13789},
 primaryClass = {astro-ph.CO},
       adsurl = {https://ui.adsabs.harvard.edu/abs/2021MNRAS.505.3698W}
}

@ARTICLE{Cunnington22,
       author = {{Cunnington}, Steven},
        title = "{Detecting the power spectrum turnover with H I intensity mapping}",
      journal = {\mnras},
         year = 2022,
        month = may,
       volume = {512},
       number = {2},
        pages = {2408-2425},
          doi = {10.1093/mnras/stac576},
archivePrefix = {arXiv},
       eprint = {2202.13828},
 primaryClass = {astro-ph.CO},
       adsurl = {https://ui.adsabs.harvard.edu/abs/2022MNRAS.512.2408C}
}

@ARTICLE{redbook,
       author = {{Square Kilometre Array Cosmology Science Working Group} and {Bacon}, David J. and {Battye}, Richard A. and {Bull}, Philip and {Camera}, Stefano and {Ferreira}, Pedro G. and {Harrison}, Ian and {Parkinson}, David and {Pourtsidou}, Alkistis and {Santos}, M{\'a}rio G. and {Wolz}, Laura and {Abdalla}, Filipe and {Akrami}, Yashar and {Alonso}, David and {Andrianomena}, Sambatra and {Ballardini}, Mario and {Bernal}, Jos{\'e} Luis and {Bertacca}, Daniele and {Bengaly}, Carlos A.~P. and {Bonaldi}, Anna and {Bonvin}, Camille and {Brown}, Michael L. and {Chapman}, Emma and {Chen}, Song and {Chen}, Xuelei and {Cunnington}, Steven and {Davis}, Tamara M. and {Dickinson}, Clive and {Fonseca}, Jos{\'e} and {Grainge}, Keith and {Harper}, Stuart and {Jarvis}, Matt J. and {Maartens}, Roy and {Maddox}, Natasha and {Padmanabhan}, Hamsa and {Pritchard}, Jonathan R. and {Raccanelli}, Alvise and {Rivi}, Marzia and {Roychowdhury}, Sambit and {Sahl{\'e}n}, Martin and {Schwarz}, Dominik J. and {Siewert}, Thilo M. and {Viel}, Matteo and {Villaescusa-Navarro}, Francisco and {Xu}, Yidong and {Yamauchi}, Daisuke and {Zuntz}, Joe},
        title = "{Cosmology with Phase 1 of the Square Kilometre Array Red Book 2018: Technical specifications and performance forecasts}",
      journal = {\pasa},
         year = 2020,
        month = mar,
       volume = {37},
          eid = {e007},
        pages = {e007},
          doi = {10.1017/pasa.2019.51},
archivePrefix = {arXiv},
       eprint = {1811.02743},
 primaryClass = {astro-ph.CO},
       adsurl = {https://ui.adsabs.harvard.edu/abs/2020PASA...37....7S}
}

@ARTICLE{Alonso2014,
       author = {{Alonso}, David and {Ferreira}, Pedro G. and {Santos}, Mario G.},
        title = "{Fast simulations for intensity mapping experiments}",
      journal = {\mnras},
         year = 2014,
        month = nov,
       volume = {444},
       number = {4},
        pages = {3183-3197},
          doi = {10.1093/mnras/stu1666},
archivePrefix = {arXiv},
       eprint = {1405.1751},
 primaryClass = {astro-ph.CO},
       adsurl = {https://ui.adsabs.harvard.edu/abs/2014MNRAS.444.3183A}
}

@INPROCEEDINGS{wolz2014,
       author = {{Wolz}, L. and {Abdalla}, F.~B. and {Alonso}, D. and {Blake}, C. and {Bull}, P. and {Chang}, T.~C. and {Ferreira}, P. and {Kuo}, C.~Y. and {Santos}, M. and {Shaw}, J.~R.},
        title = "{Foreground Subtraction in Intensity Mapping with the SKA}",
    booktitle = {Advancing Astrophysics with the Square Kilometre Array (AASKA14)},
         year = 2015,
        month = apr,
          eid = {35},
        pages = {35},
          doi = {10.22323/1.215.0035},
archivePrefix = {arXiv},
       eprint = {1501.03823},
 primaryClass = {astro-ph.CO},
       adsurl = {https://ui.adsabs.harvard.edu/abs/2015aska.confE..35W}
}

@ARTICLE{Mertens,
       author = {{Mertens}, Florent G. and {Bobin}, J{\'e}r{\^o}me and {Carucci}, Isabella P.},
        title = "{Retrieving the 21-cm signal from the Epoch of Reionization with learnt Gaussian process kernels}",
      journal = {\mnras},
         year = 2024,
        month = jan,
       volume = {527},
       number = {2},
        pages = {3517-3531},
          doi = {10.1093/mnras/stad3430},
archivePrefix = {arXiv},
       eprint = {2307.13545},
 primaryClass = {astro-ph.CO},
       adsurl = {https://ui.adsabs.harvard.edu/abs/2024MNRAS.527.3517M}
}

@ARTICLE{Sia,
       author = {{Gkogkou}, Athanasia and {Bonjean}, Victor and {Starck}, Jean-Luc and {Spinelli}, Marta and {Tsakalides}, Panagiotis},
        title = "{Foreground removal in HI 21 cm intensity mapping under frequency-dependent beam distortions}",
      journal = {\aap},
         year = 2026,
        month = jan,
       volume = {706},
          eid = {A3},
        pages = {A3},
          doi = {10.1051/0004-6361/202557229},
archivePrefix = {arXiv},
       eprint = {2511.13328},
 primaryClass = {astro-ph.CO},
       adsurl = {https://ui.adsabs.harvard.edu/abs/2026A&A...706A...3G}
}

@ARTICLE{2025_EoR_BN_BS,
       author = {{Cerardi}, Nicolas and {Giri}, Sambit K. and {Bianco}, Michele and {Piras}, Davide and {de Salis}, Emmanuel and {De Santis}, Massimo and {Selcuk-Simsek}, Merve and {Denzel}, Philipp and {Hess}, Kelley M. and {Toribio}, M. Carmen and {Kirsten}, Franz and {Ghorbel}, Hatem},
        title = "{Implicit inference of the reionization history with higher-order statistics of the 21-cm signal}",
      journal = {arXiv e-prints},
         year = 2025,
        month = nov,
          eid = {arXiv:2511.11568},
        pages = {arXiv:2511.11568},
          doi = {10.48550/arXiv.2511.11568},
archivePrefix = {arXiv},
       eprint = {2511.11568},
 primaryClass = {astro-ph.CO},
       adsurl = {https://ui.adsabs.harvard.edu/abs/2025arXiv251111568C}
}

@ARTICLE{2025_WST_CD_EoR,
       author = {{Shimabukuro}, Hayato and {Liu}, Shihang and {Li}, Bohua},
        title = "{Probing Fuzzy Dark Matter in the 21 cm Signal via Wavelet Scattering Transform}",
      journal = {arXiv e-prints},
         year = 2025,
        month = nov,
          eid = {arXiv:2512.00402},
        pages = {arXiv:2512.00402},
          doi = {10.48550/arXiv.2512.00402},
archivePrefix = {arXiv},
       eprint = {2512.00402},
 primaryClass = {astro-ph.CO},
       adsurl = {https://ui.adsabs.harvard.edu/abs/2025arXiv251200402S}
}

@ARTICLE{2020_EoR_3PCF,
       author = {{Jennings}, W.~D. and {Watkinson}, C.~A. and {Abdalla}, F.~B.},
        title = "{Analysing the Epoch of Reionization with three-point correlation functions and machine learning techniques}",
      journal = {\mnras},
         year = 2020,
        month = nov,
       volume = {498},
       number = {3},
        pages = {4518-4532},
          doi = {10.1093/mnras/staa2598},
archivePrefix = {arXiv},
       eprint = {2011.14157},
 primaryClass = {astro-ph.CO},
       adsurl = {https://ui.adsabs.harvard.edu/abs/2020MNRAS.498.4518J}
}

@ARTICLE{Carucci25,
       author = {{Carucci}, I.~P. and {Bernal}, J.~L. and {Cunnington}, S. and {Santos}, M.~G. and {Wang}, J. and {Fonseca}, J. and {Grainge}, K. and {Irfan}, M.~O. and {Li}, Y. and {Pourtsidou}, A. and {Spinelli}, M. and {Wolz}, L.},
        title = "{Hydrogen intensity mapping with MeerKAT: Preserving cosmological signal by optimising contaminant separation}",
      journal = {\aap},
         year = 2025,
        month = nov,
       volume = {703},
          eid = {A222},
        pages = {A222},
          doi = {10.1051/0004-6361/202453461},
archivePrefix = {arXiv},
       eprint = {2412.06750},
 primaryClass = {astro-ph.CO},
       adsurl = {https://ui.adsabs.harvard.edu/abs/2025A&A...703A.222C}
}

@ARTICLE{2021JOSS....6.3783Z,
       author = {{Zonca}, Andrea and {Thorne}, Ben and {Krachmalnicoff}, Nicoletta and {Borrill}, Julian},
        title = "{The Python Sky Model 3 software}",
      journal = {The Journal of Open Source Software},
         year = 2021,
        month = nov,
       volume = {6},
       number = {67},
          eid = {3783},
        pages = {3783},
          doi = {10.21105/joss.03783},
archivePrefix = {arXiv},
       eprint = {2108.01444},
 primaryClass = {astro-ph.IM},
       adsurl = {https://ui.adsabs.harvard.edu/abs/2021JOSS....6.3783Z}
}

\begin{appendix}
\section{Figure of merit}\label{sec:FoM}
We quantify the constraining power of each statistic using the FoM. Given the posterior samples obtained via SBI, we compute the empirical covariance matrix
\begin{equation}
    \mathbf{C} = \frac{1}{N-1} \sum_{i=1}^{N} (\theta_i - \bar{\theta}) (\theta_i - \bar{\theta})^\mathrm{T}\,,
\end{equation}
where \(\theta_i\) denotes the \(i\)-th sample and \(\bar{\theta}\) the sample mean. The FoM is then defined as
\begin{equation}
    \mathrm{FoM} = \frac{1}{\sqrt{\det \mathbf{C}}}\,,
\end{equation}
corresponding to the inverse square root of the posterior volume in parameter space. This metric naturally incorporates parameter correlations and provides a compact measure of overall constraining power. All FoMs are subsequently normalized to the baseline statistic \(C_\ell\) to facilitate comparison between both methods.

\section{Additional figures}\label{sec:Figures}

\begin{figure*}
    \centering
    \includegraphics[width=\textwidth]{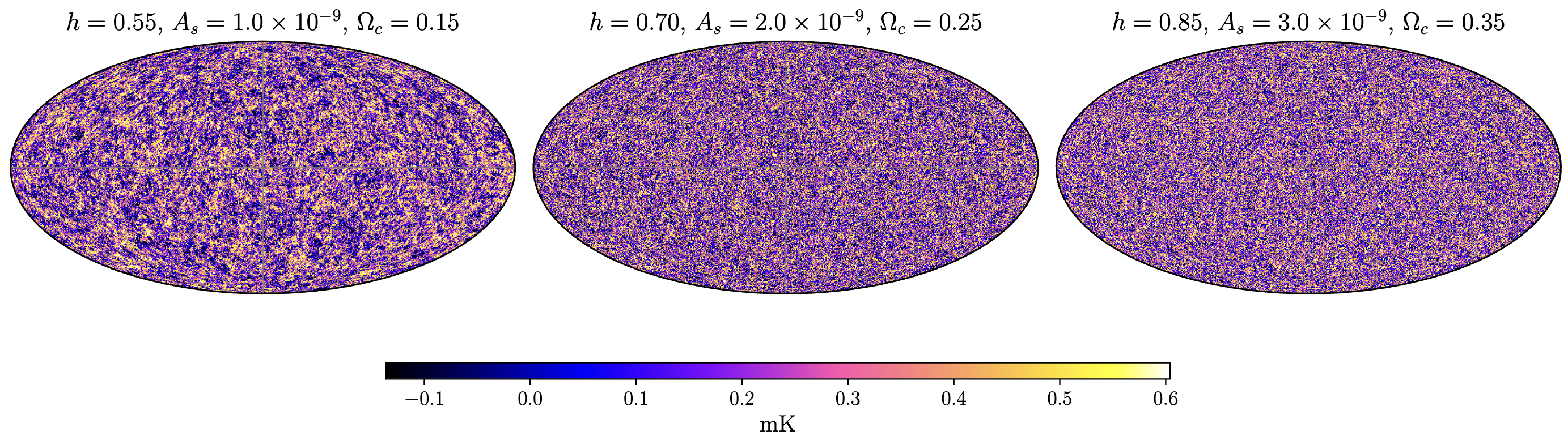}
    \caption{Mollweide projection generated from the code detailed in Sect. \ref{sec:sims} for different cosmologies.}
    \label{fig:mollweide}
\end{figure*}

\begin{figure*}
    \centering
    \includegraphics[width=\textwidth]{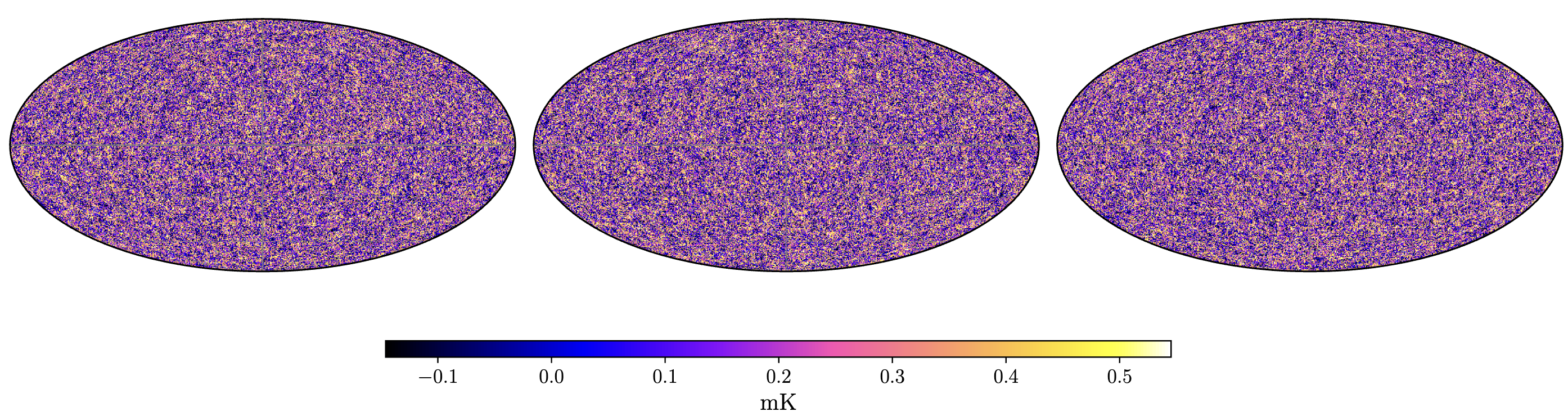}
    \caption{Mollweide projections of the simulated \hi\ intensity mapping (from code detailed in Sect. \ref{sec:sims}) signal at $z = 0.425$ for fiducial cosmology and different realizations.}

    \label{fig:map_realisations}
\end{figure*}

\begin{figure*}
    \centering
    \includegraphics[width=\textwidth]{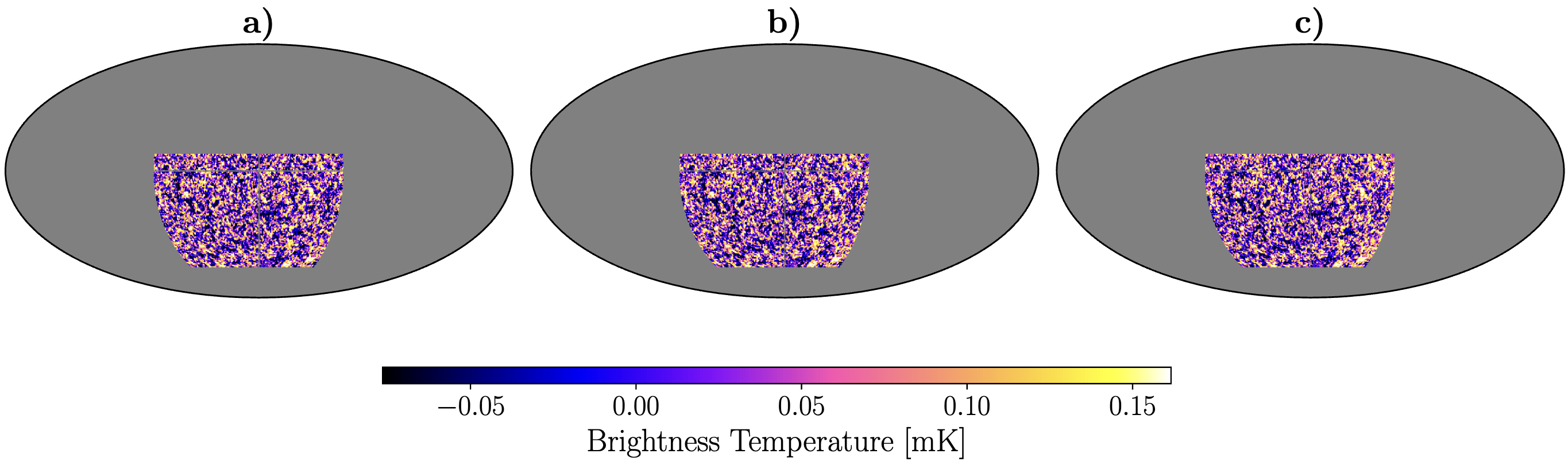}
    \caption{Mollweide projections of the simulated \hi\ intensity mapping signal at $z = 0.425$ for different fraction of Galactic synchrotron contaminants with telescope beam of 0.5 deg and mask of $f_{\rm sky}=0.19$.
\textit{Panel a}: 0.0005\%.
\textit{Panel b}: 0.001\%.
\textit{Panel c}: 0.002\%.}

    \label{fig:synch_all}
\end{figure*}

\begin{figure*}
    \centering
    \includegraphics[width=\textwidth]{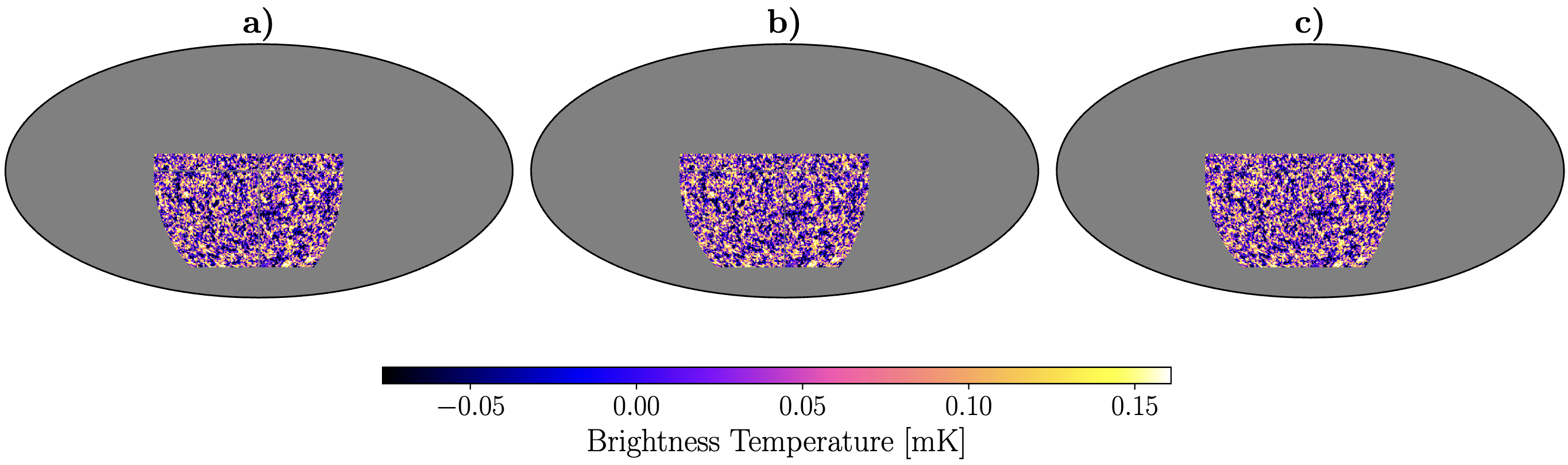}
    \caption{Mollweide projections of the simulated \hi\ intensity mapping signal at $z = 0.425$ for different fraction of extragalactic point sources contaminants with telescope beam of 0.5 deg and mask of $f_{\rm sky}=0.19$.
\textit{Panel a}: 0.001\%.
\textit{Panel b}: 0.0025\%.
\textit{Panel c}: 0.005\%.}

    \label{fig:PS_all}
\end{figure*}

\begin{figure*}
    \centering

    \begin{minipage}{0.48\textwidth}
        \centering
        \includegraphics[width=\linewidth]{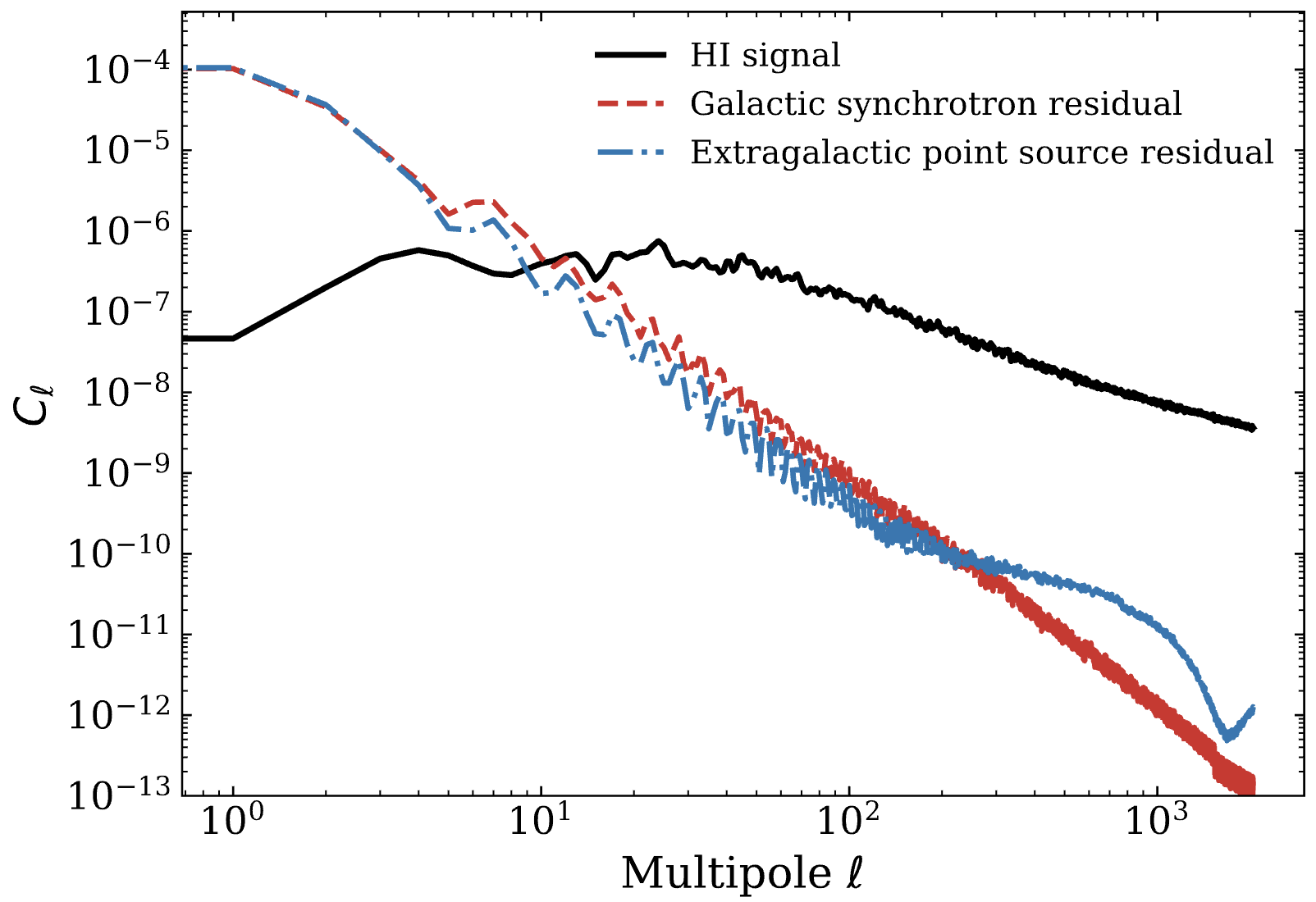}
        \caption{Angular power spectra of the masked residual templates: Galactic synchrotron (0.001\%) and extragalactic point sources (0.0025\%) together with a clean map with intermediary mask of $f_{sky}$ = 0.19 before any effects are applied.}
        \label{fig:PS_S}
    \end{minipage}
    \hfill
    \begin{minipage}{0.48\textwidth}
        \centering
        \caption{Bins used in the SBI.}
        \renewcommand{\arraystretch}{1.4}
        \begin{tabular}{lc}
\hline
Simulation setup & \# bins \\
\hline
Clean HI signal (5-scale) & 38 \\
Mask $f_{sky}$ = 0.6 (5-scale) & 36 \\
Mask $f_{sky}$ = 0.19 (5-scale) & 35 \\
Mask $f_{sky}$ = 0.014 (5-scale) & 26 \\
Mask $f_{sky}$ = 0.19, beam 1.34 $^\circ$  (5-scale)& 25 \\
Mask $f_{sky}$ = 0.19, beam 0.5 $^\circ$  (5-scale) & 32 \\
\hline
Mask $f_{sky}$ = 0.19, beam 0.5 $^\circ$, noise (5-scale) & 35 \\
Mask $f_{sky}$ = 0.19, beam 0.5 $^\circ$, noise (8-scale) & 57 \\
\hline
Mask $f_{sky}$ = 0.19, beam 0.5 $^\circ$, no coarse scale (5-scale) & 18 \\
Mask $f_{sky}$ = 0.19, beam 0.5 $^\circ$, no coarse scale (8-scale) & 35 \\
\hline
\end{tabular}
\tablefoot{Number of bins (out of 200 for 5-scale and 320 for 8-scale) used for the simulation based inference for different setting cases.}
\renewcommand{\arraystretch}{1}
        
        \label{tab:table_bins}
    \end{minipage}

\end{figure*}

\begin{figure*}
    \centering
    \includegraphics[width=\textwidth]{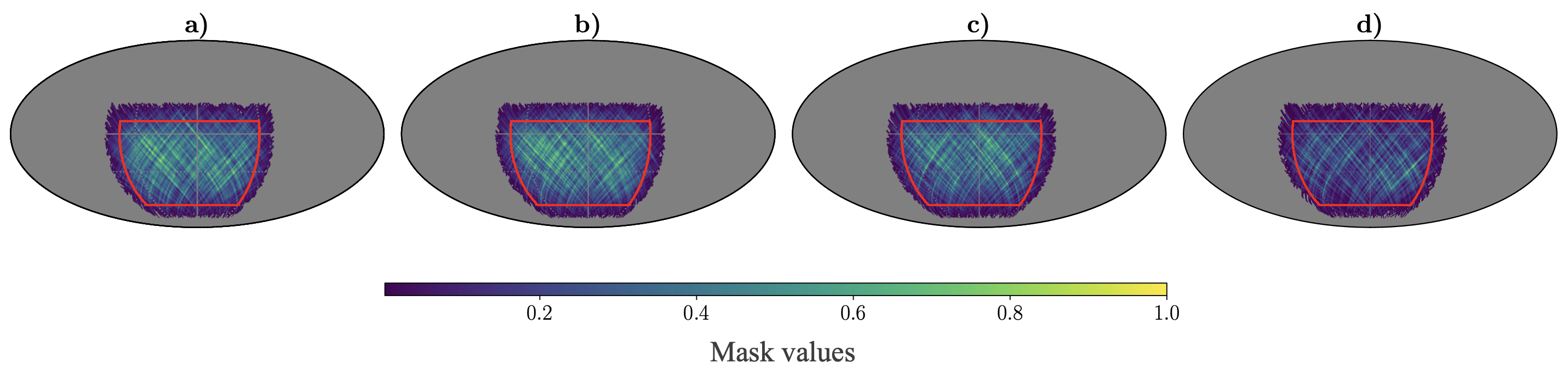}
    \caption{Mollweide projections of the weighted mask for different additive X-shape weight.
\textit{Panel a}: 10X. \textit{Panel b}: 25X. \textit{Panel c}: 35X. \textit{Panel d}: 50X.}
        
    \label{fig:scan_map_masks}
\end{figure*}

\begin{figure*}
    \centering
    \includegraphics[width=\textwidth]{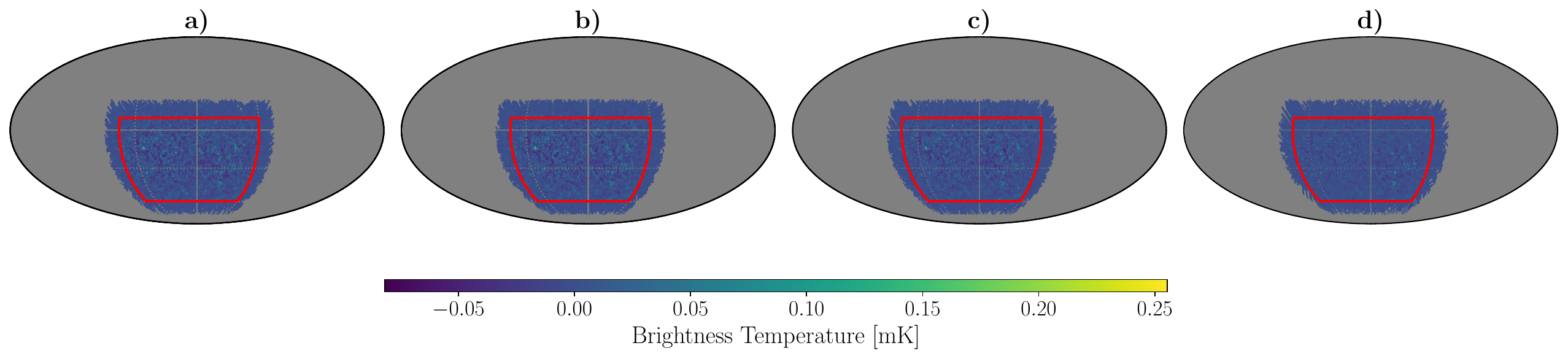}
    \caption{Mollweide projections of the simulated \hi\ intensity mapping signal at $z = 0.425$ for different multiplicative X-shape weight with telescope beam of 0.5 deg and mask of $f_{\rm sky}=0.19$.
\textit{Panel a}: 10X. \textit{Panel b}: 25X. \textit{Panel c}: 35X. \textit{Panel d}: 50X.}
    \label{fig:scan_map}
\end{figure*}

\begin{figure*}
    \centering
    \includegraphics[width=\textwidth]{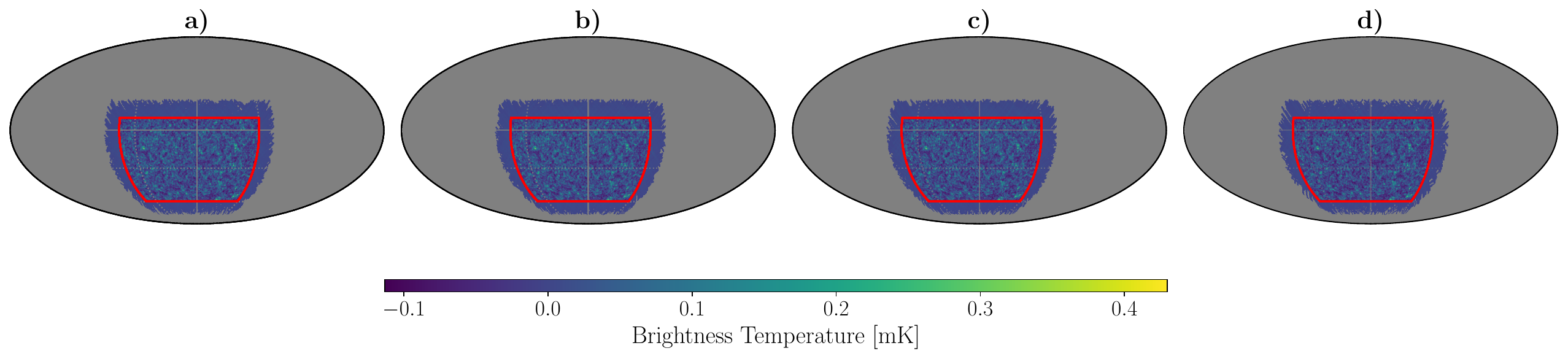}
    \caption{Mollweide projections of the simulated \hi\ intensity mapping signal at $z = 0.425$ for different additive X-shape weight with telescope beam of 0.5 deg and mask of $f_{\rm sky}=0.19$.
\textit{Panel a}: 10X. \textit{Panel b}: 25X. \textit{Panel c}: 35X. \textit{Panel d}: 50X.}
        
    \label{fig:scan_map_additive}
\end{figure*}

\begin{figure*}
    \centering
    \includegraphics[width=0.3\textwidth]{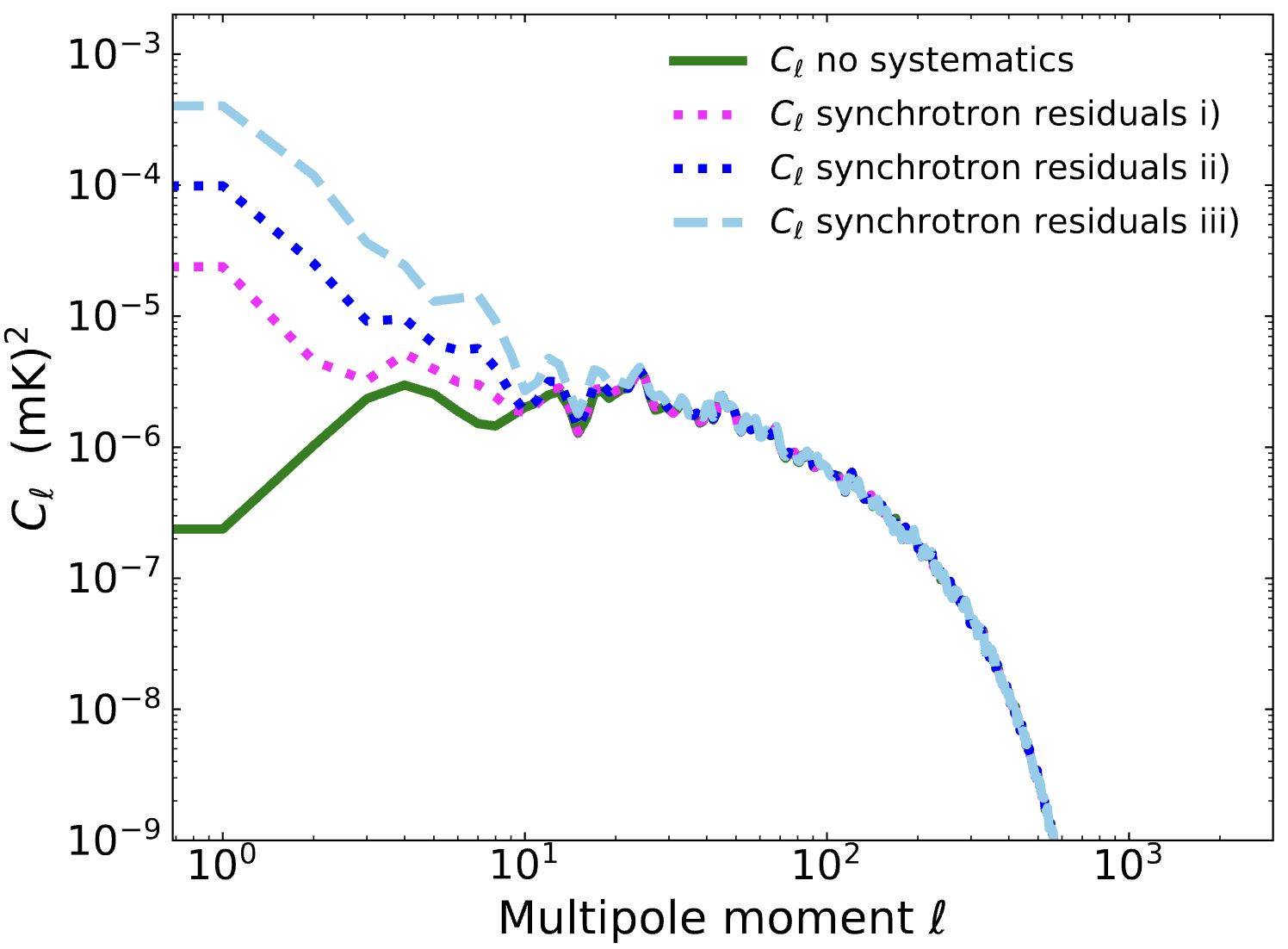}
    \includegraphics[width=0.31\textwidth]{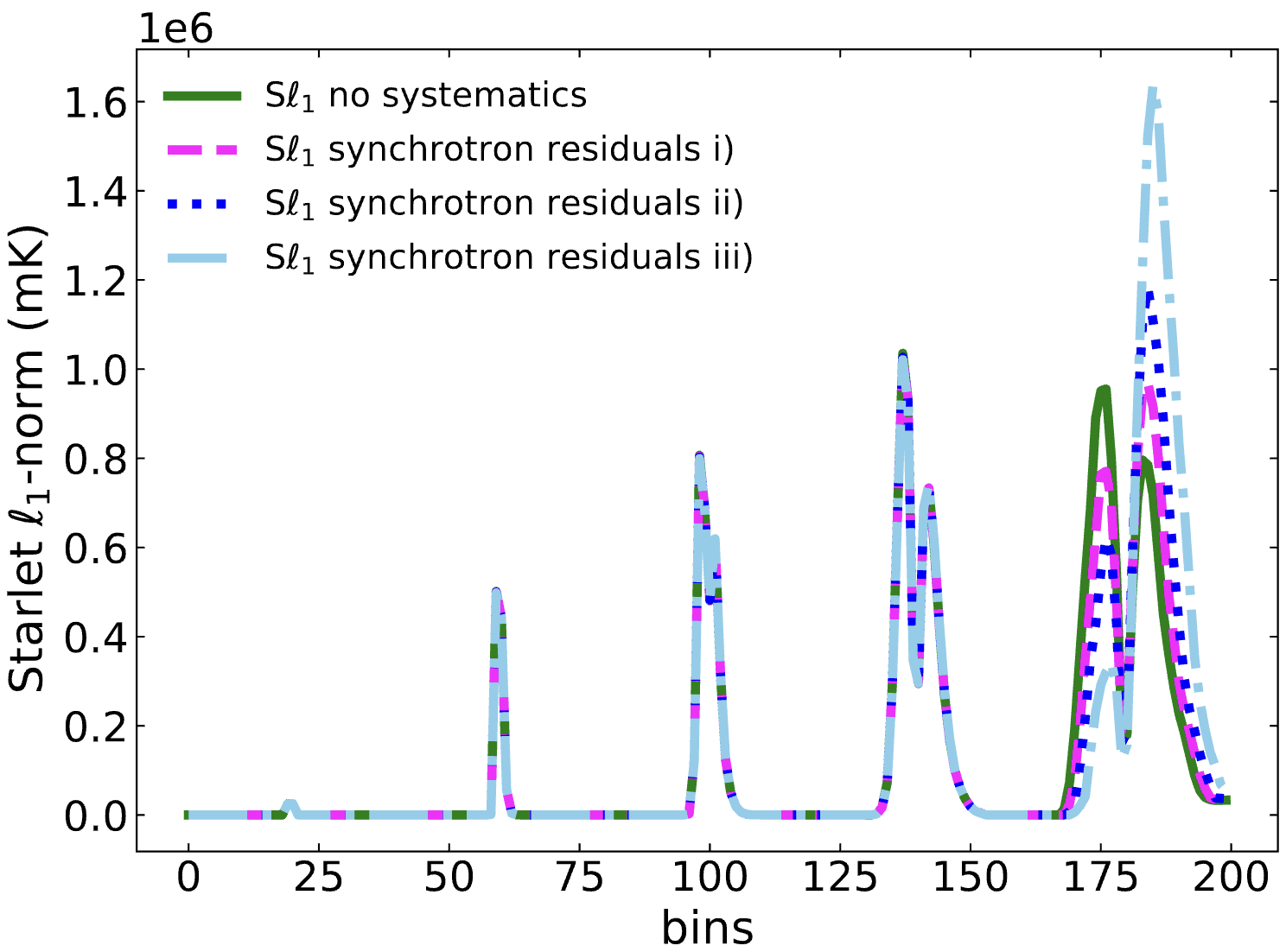}
    \includegraphics[width=0.38\textwidth]{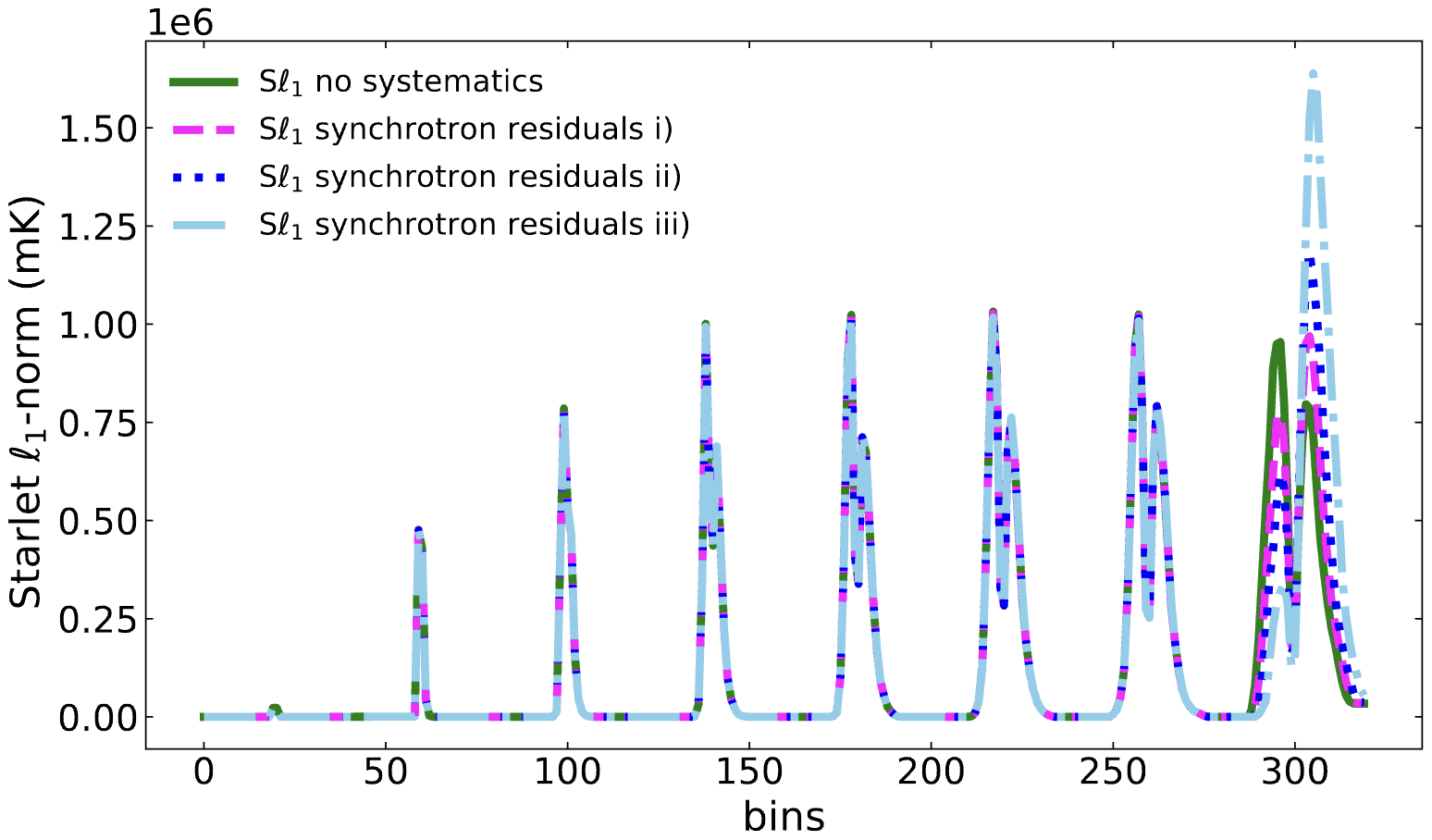}
    \caption{Comparison of the summary statistics for different Galactic synchrotron residual contaminations: C$_\ell$ (\textit{first panel}), starlet $\ell_1$-norm with 5 scales (\textit{second panel}), and starlet $\ell_1$-norm with 8 scales (\textit{third panel}) for the map without systematic effects (green) and maps with Galactic synchrotron contamination: (i) 0.0005\% (magenta), (ii) 0.001\% (blue), and (iii) 0.002\% (sky blue). All maps have $f_{\rm sky}$=0.19 and are beamed at $0.5^\circ$.}
    \label{fig:comp_synch}
\end{figure*}

\begin{figure*}
    \centering
    \includegraphics[width=0.3\textwidth]{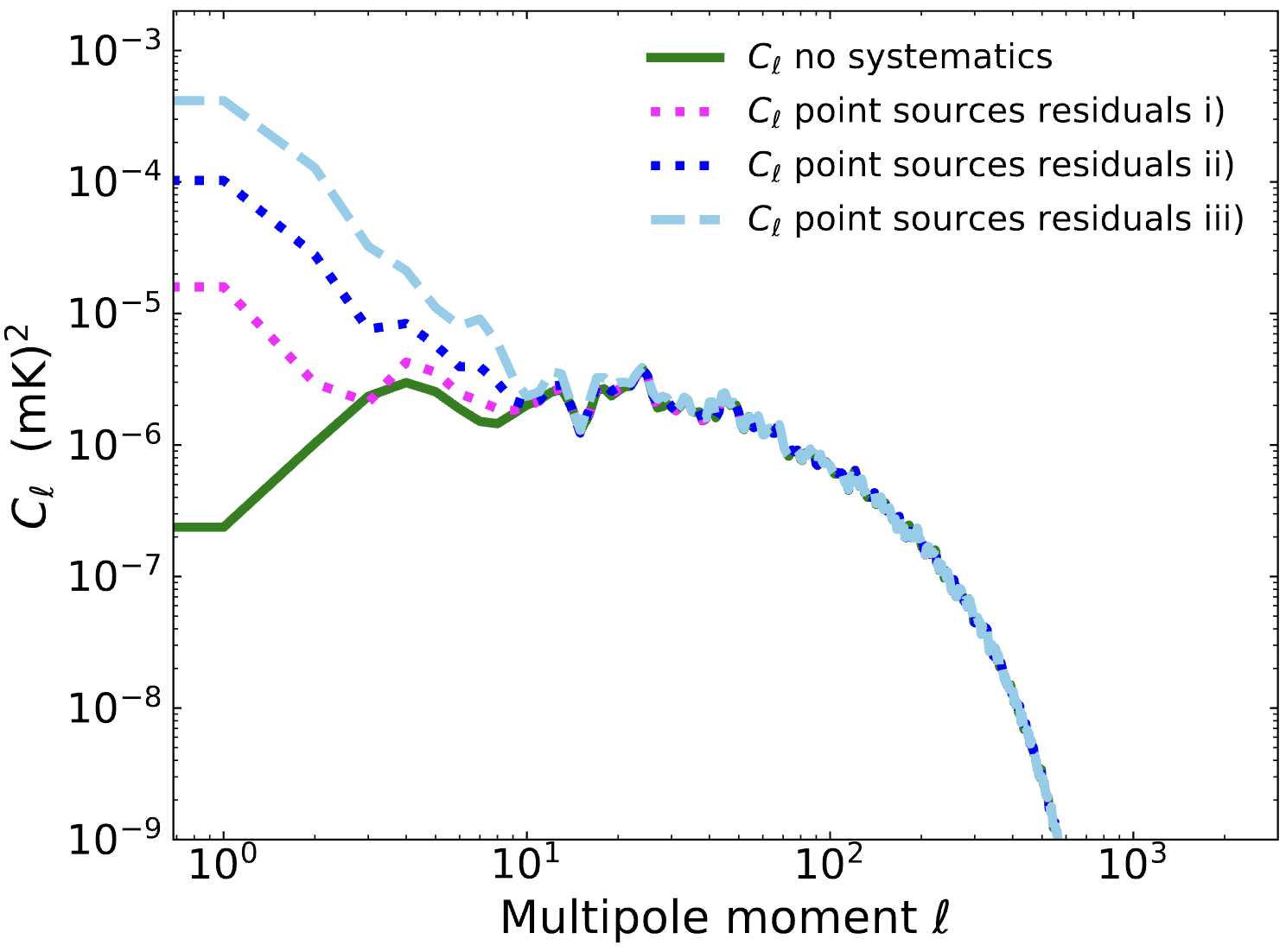}
    \includegraphics[width=0.31\textwidth]{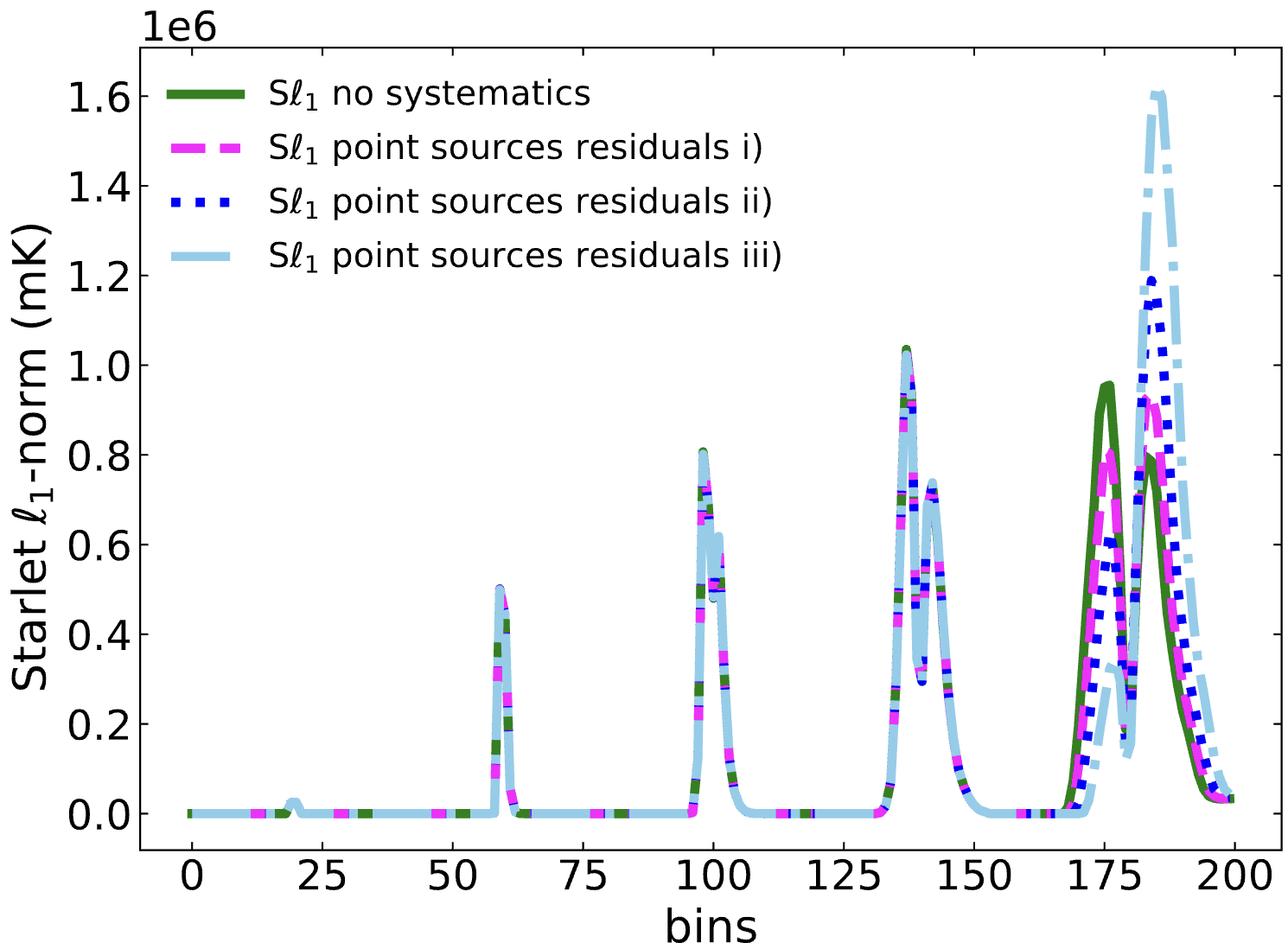}
    \includegraphics[width=0.38\textwidth]{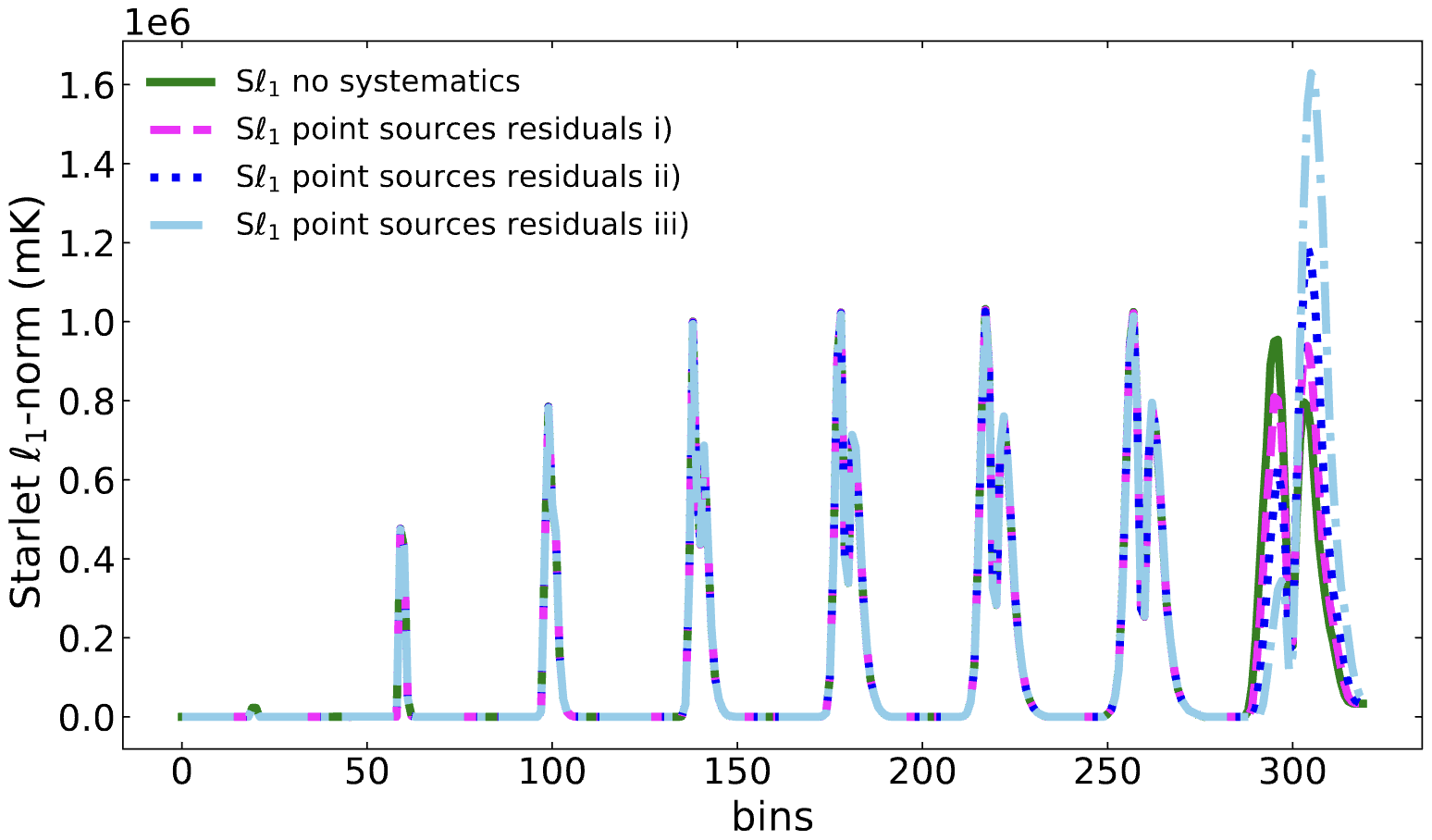}
    \caption{Comparison of the summary statistics for different residual extragalactic point sources contamination: C$_\ell$ (\textit{first panel}), starlet $\ell_1$-norm with 5 scales (\textit{second panel}), and starlet $\ell_1$-norm with 8 scales (\textit{third panel}) for the map without systematic effects (green) and maps with residual extragalactic point sources contamination: (i) 0.001\% (magenta), (ii) 0.0025\%  (blue), and (iii) 0.005\% (sky blue). All maps have $f_{\rm sky}$=0.19 and are beamed at $0.5^\circ$.}
    \label{fig:comp_PS}
\end{figure*}

\begin{figure*}
    \centering
    \includegraphics[width=0.3\textwidth]{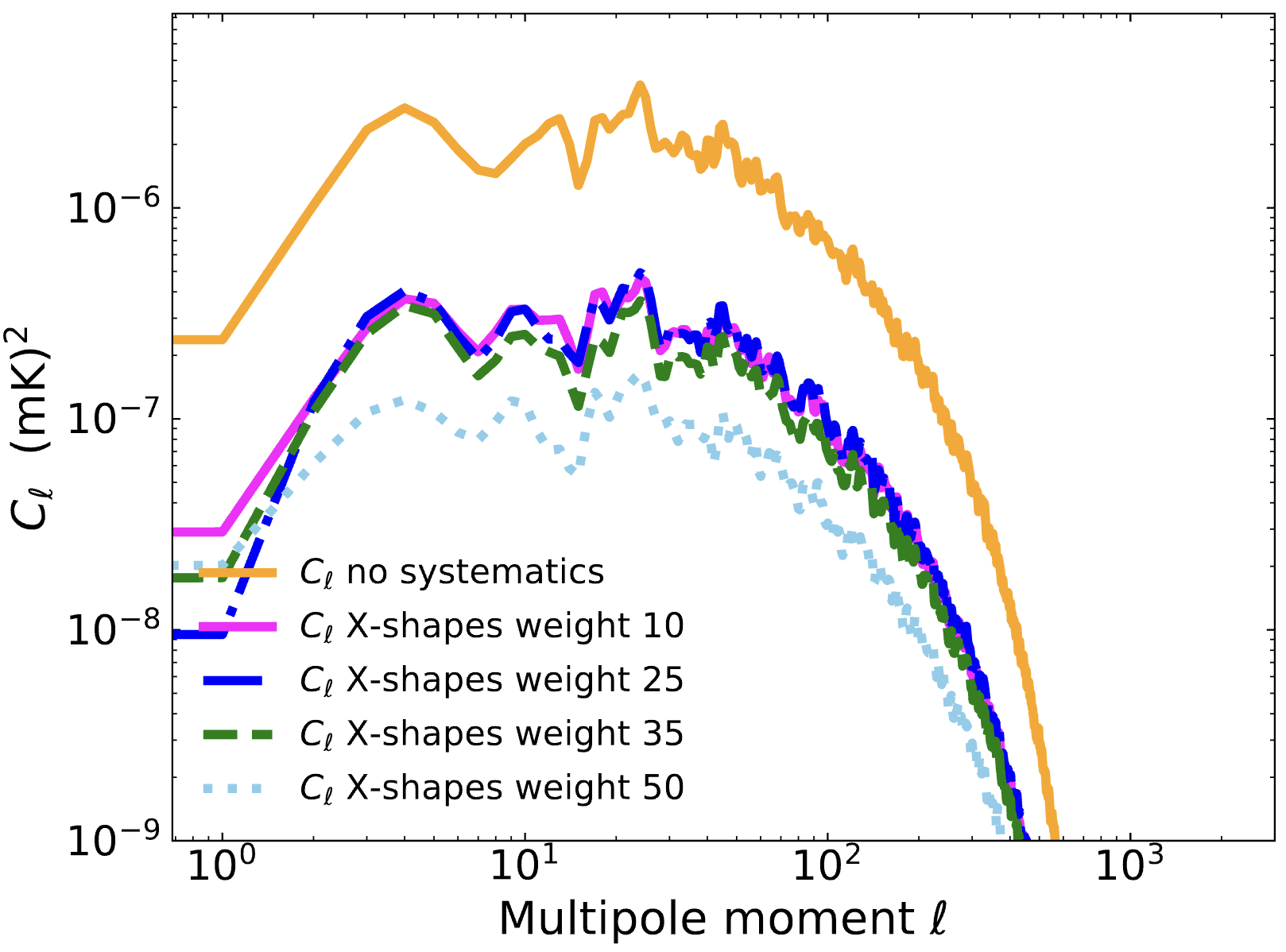}
    \includegraphics[width=0.31\textwidth]{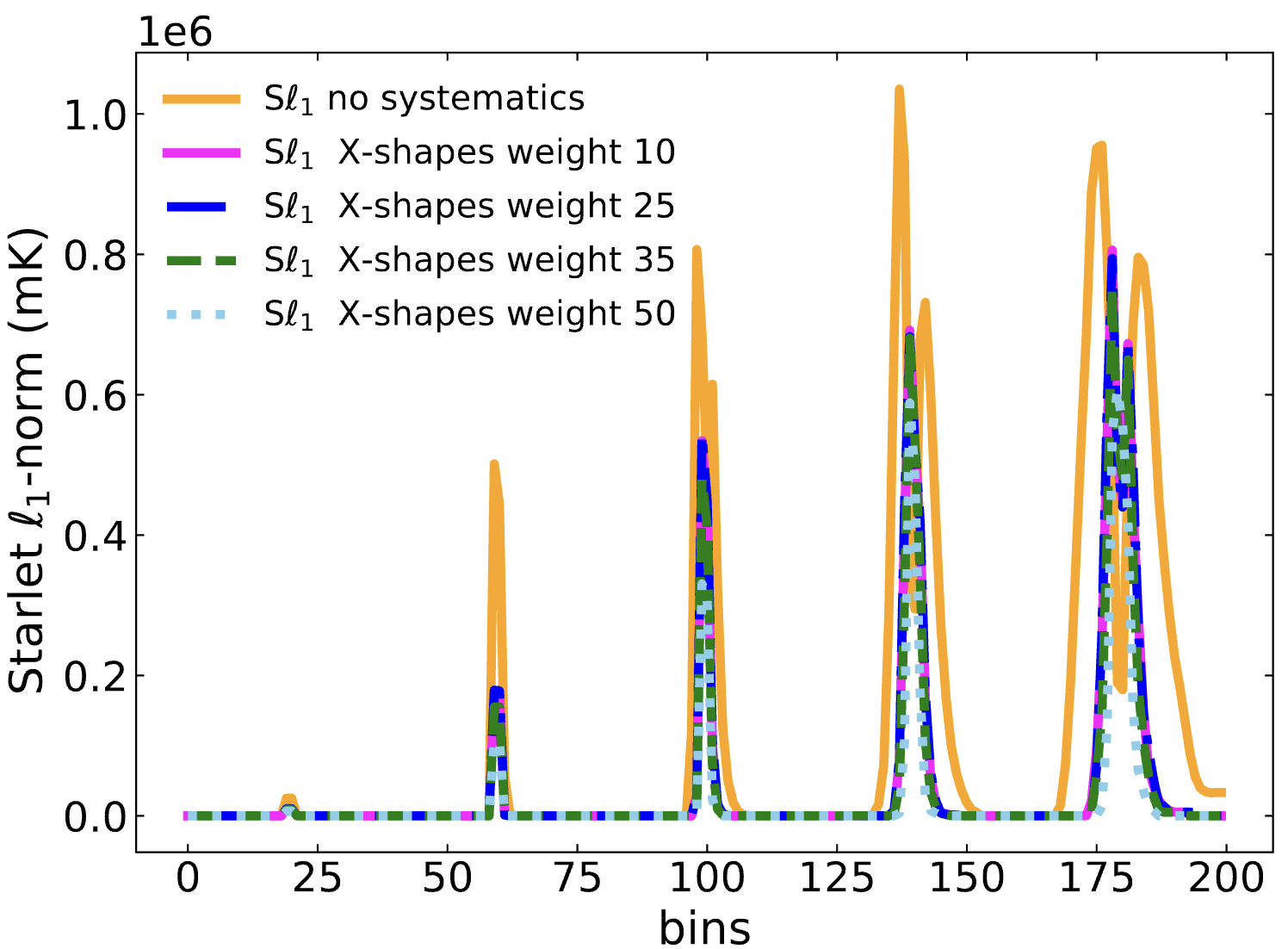}
    \includegraphics[width=0.38\textwidth]{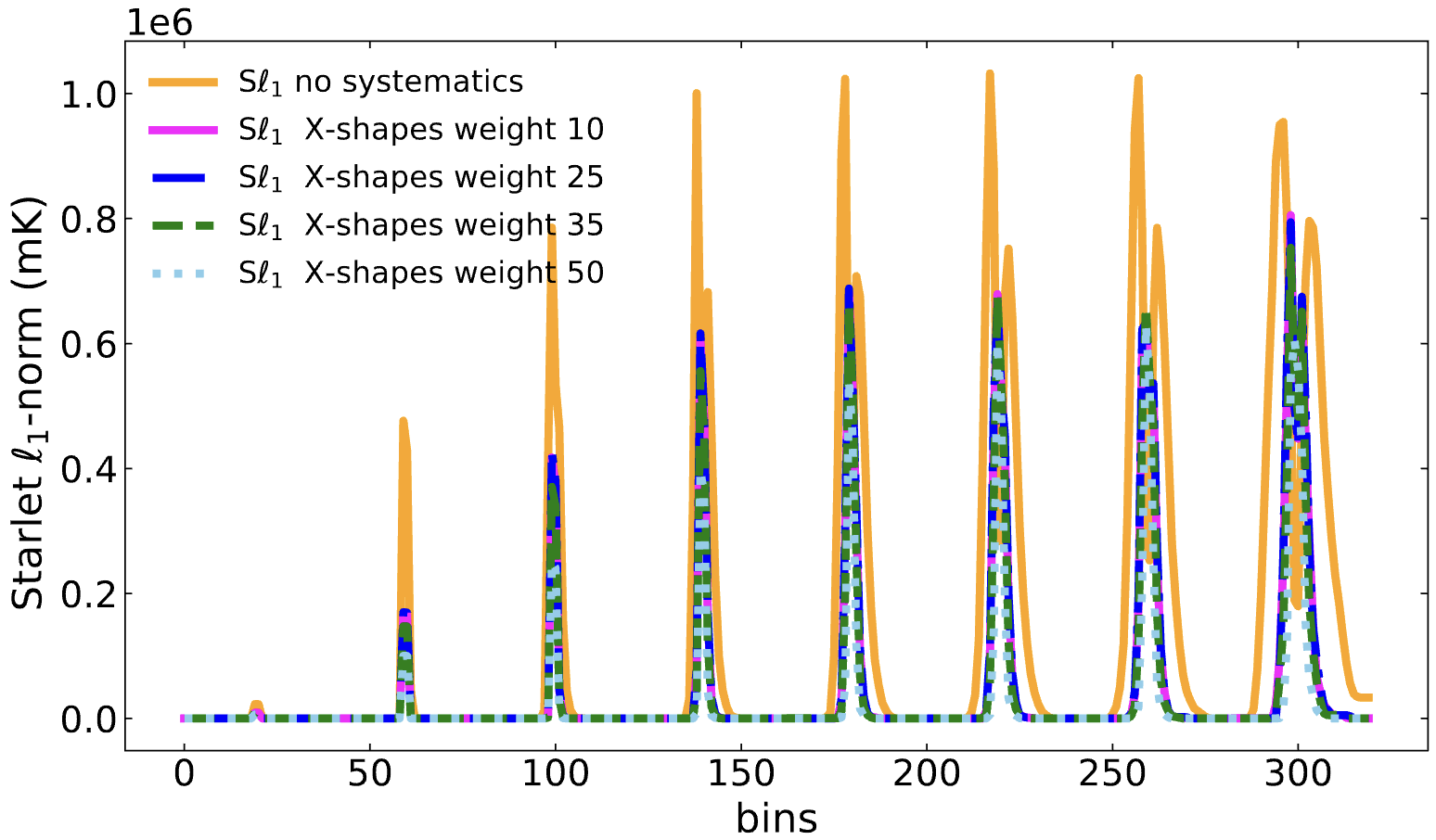}
    \caption{Comparison of the summary statistics for different multiplicative X-shaped weights. C$_\ell$ (\textit{first panel}), starlet $\ell_1$-norm with 5 scales (\textit{second panel}), and starlet $\ell_1$-norm with 8 scales (\textit{third panel}), for the map without systematic effects (orange) and maps with multiplicative X-shaped weights: 10 (magenta), 25 (blue), 35 (green), and 50 (sky blue). All maps have $f_{\rm sky}$=0.19 and are beamed at $0.5^\circ$.}
    \label{fig:comp_scan}
\end{figure*}

\begin{figure*}
    \centering
    \includegraphics[width=0.3\textwidth]{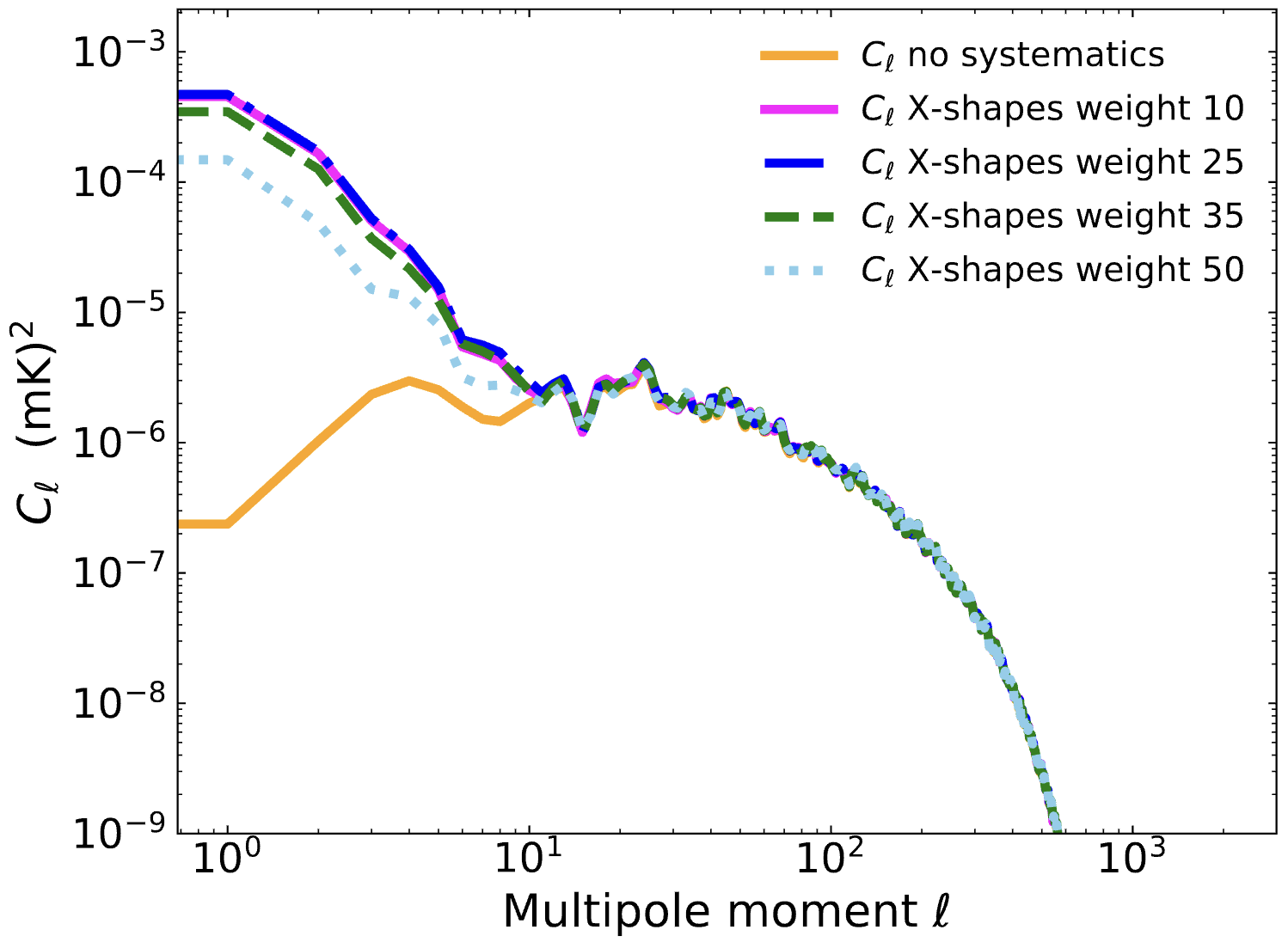}
    \includegraphics[width=0.31\textwidth]{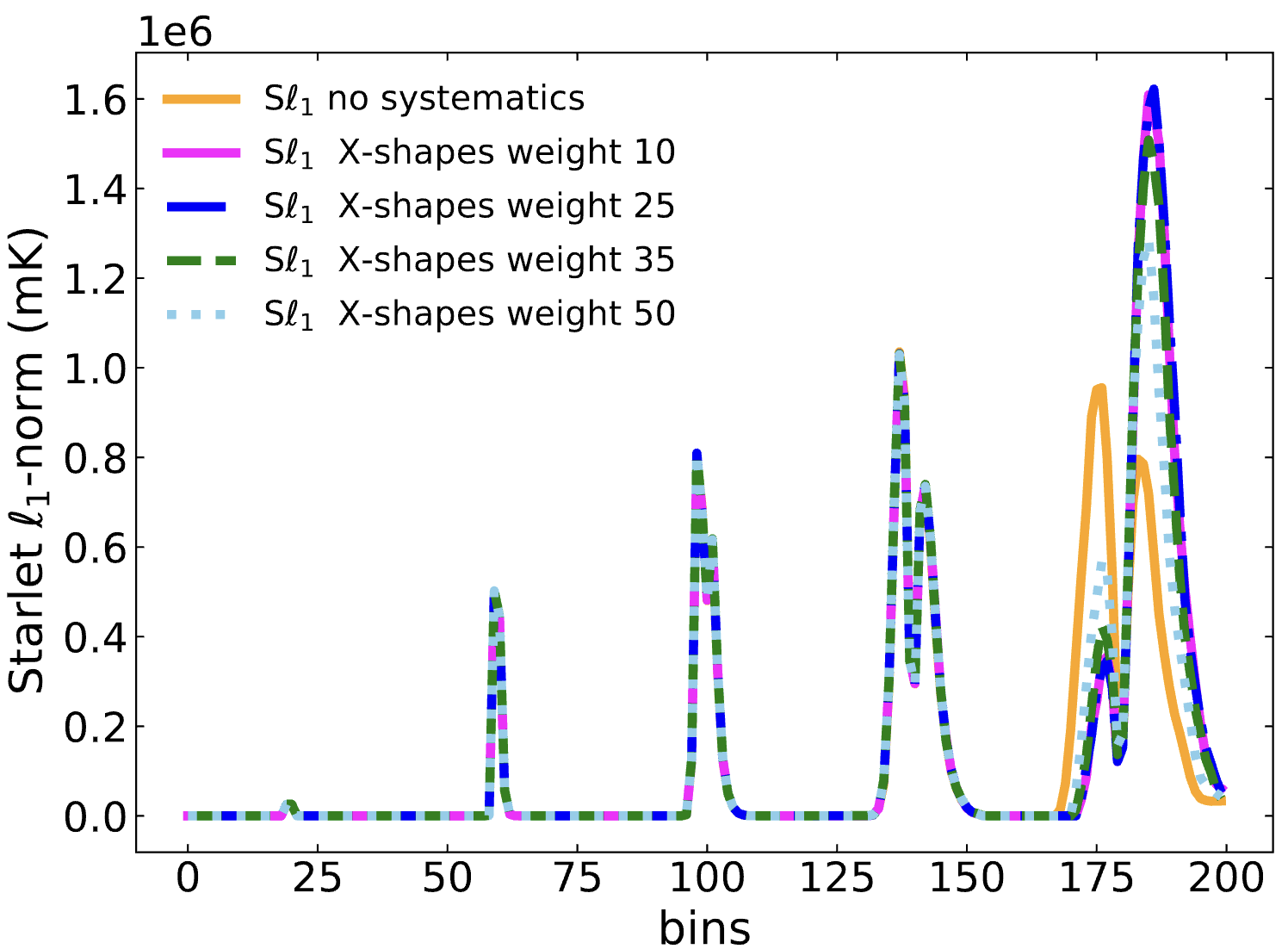}
    \includegraphics[width=0.38\textwidth]{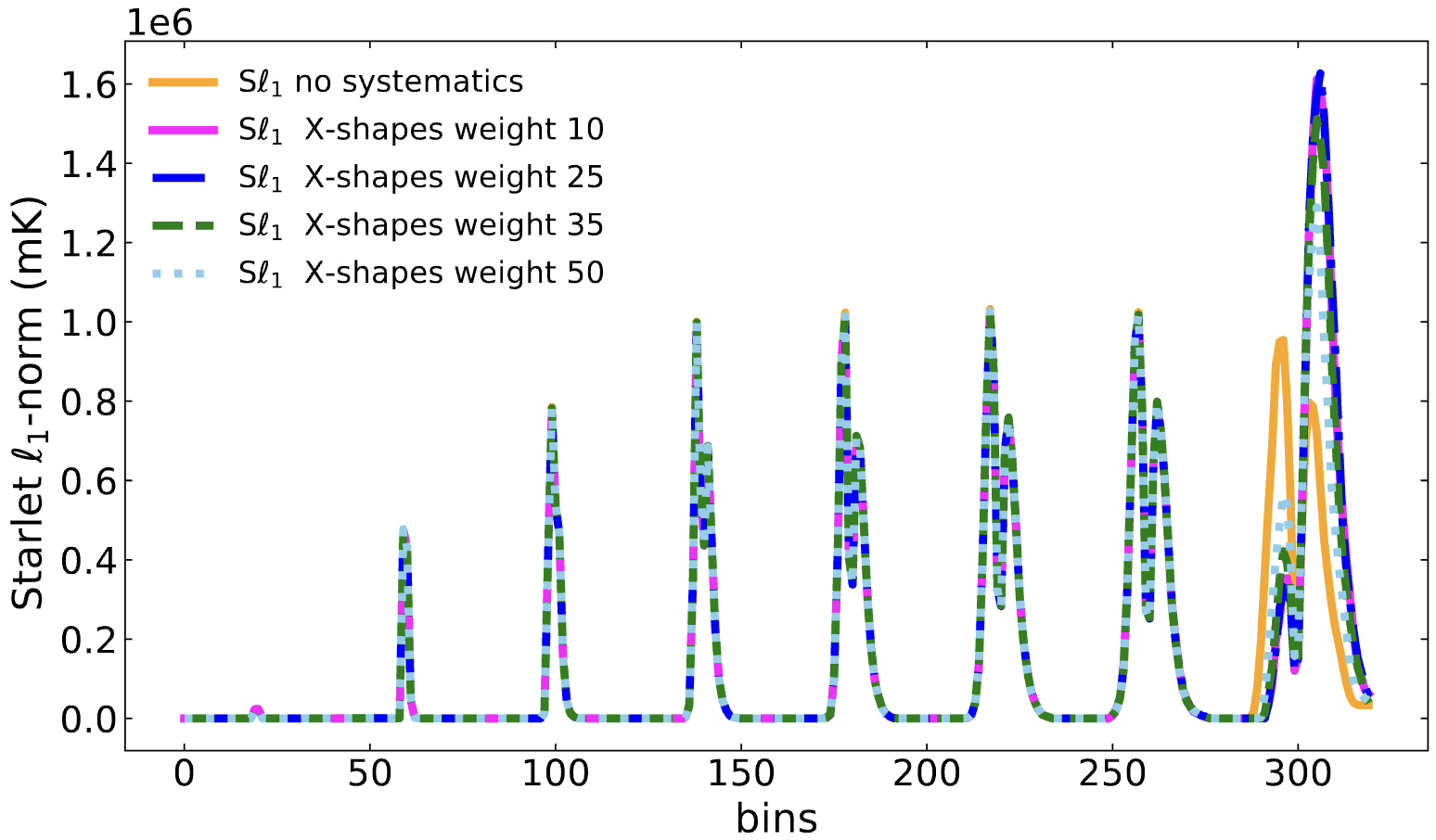}
    \caption{Comparison of the summary statistics for different additive X-shaped weights. C$_\ell$ (\textit{first panel}), starlet $\ell_1$-norm with 5 scales (\textit{second panel}), and starlet $\ell_1$-norm with 8 scales (\textit{third panel}), for the map without systematic effects (orange) and maps with additive X-shaped weights: 10 (magenta), 25 (blue), 35 (green), and 50 (sky blue). All maps have $f_{\rm sky}$=0.19 and are beamed at $0.5^\circ$.}
    \label{fig:comp_scan_add}
\end{figure*}

\begin{figure*}
    \centering
    \includegraphics[width=0.32\textwidth]{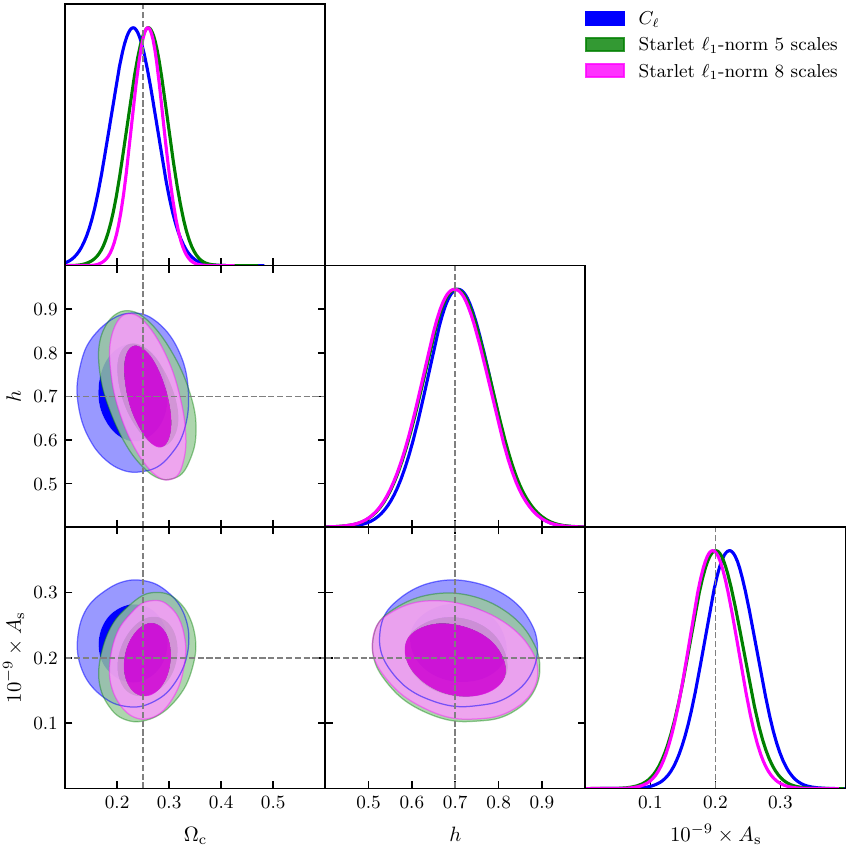}
    \includegraphics[width=0.32\textwidth]{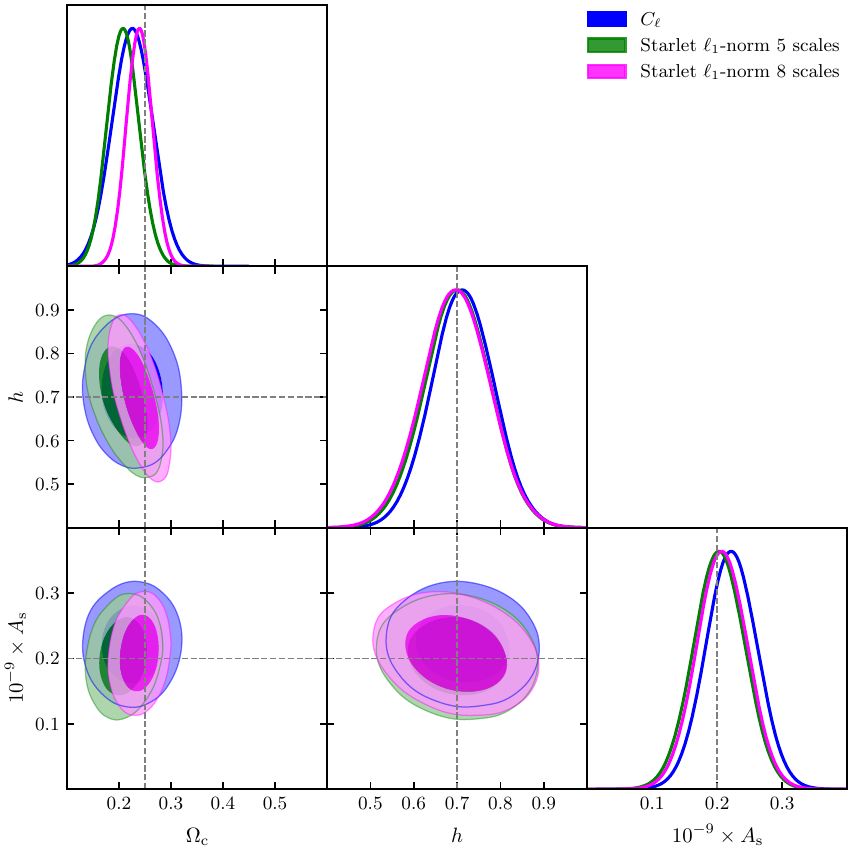}
    \includegraphics[width=0.32\textwidth]{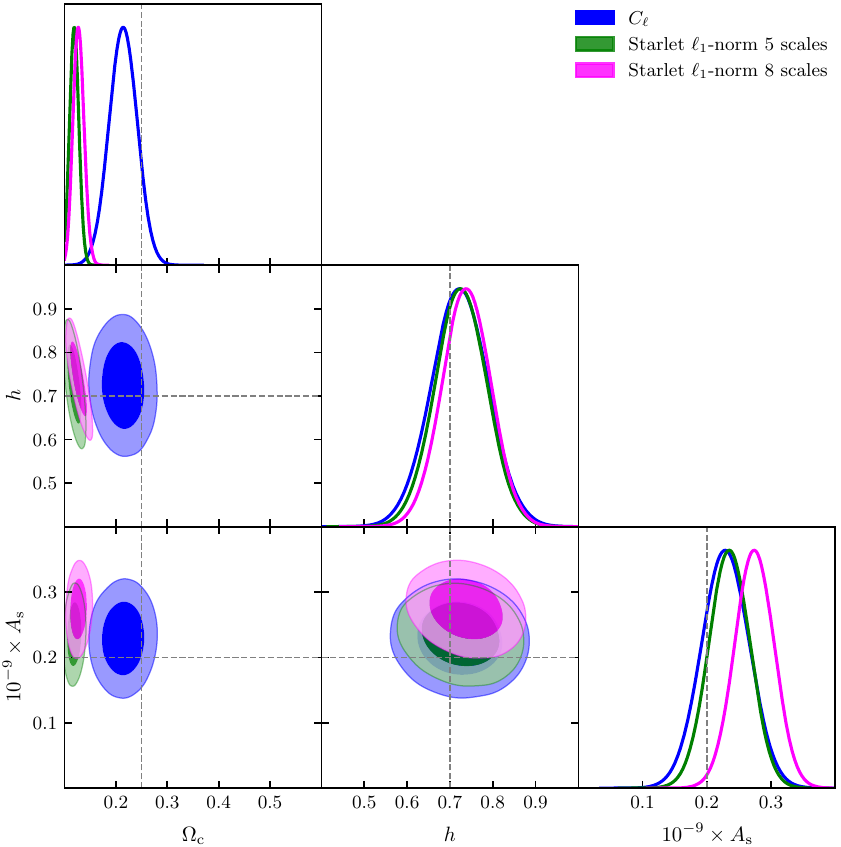}
    \caption{
    Posterior distributions for $\Omega_{\rm c}$, $h$, and $A_{\rm s}$ obtained from the angular power spectrum $C_\ell$, 
    starlet $\ell_1$-norm for different Galactic synchrotron residual fractions 0.0005 \% (\textit{first panel}), 0.001 \% (\textit{second panel}), and 0.002 \% (\textit{third panel}).    Contours correspond to $68\%$ and $95\%$ confidence levels.
    }
    \label{fig:synch}
\end{figure*}

\begin{figure*}
    \centering
    \includegraphics[width=0.32\textwidth]{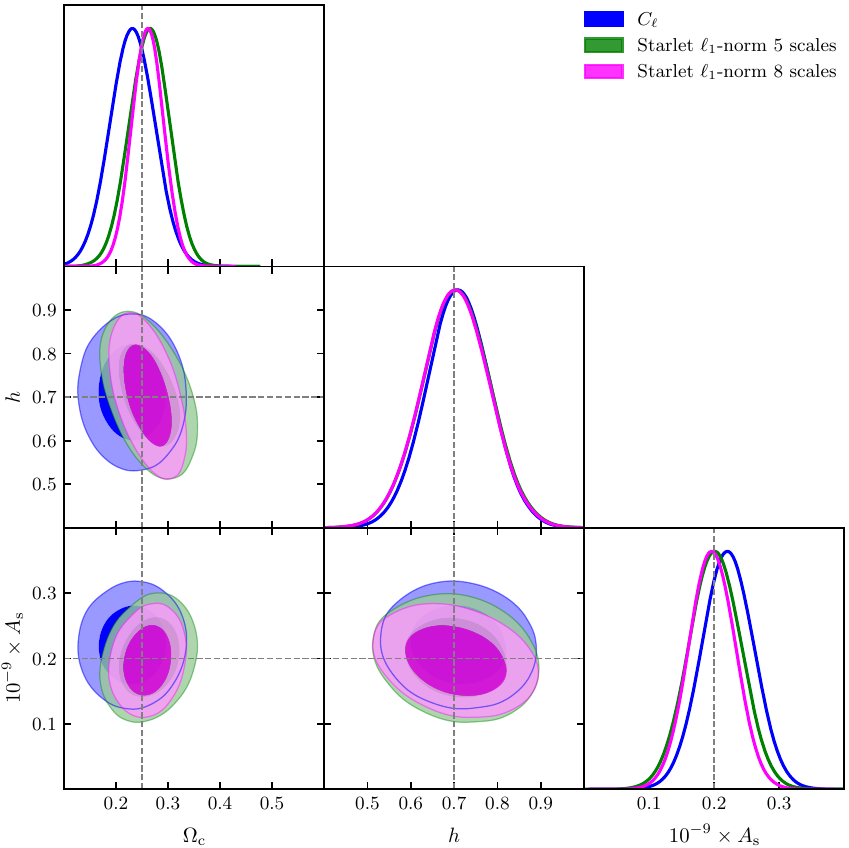}
    \includegraphics[width=0.32\textwidth]{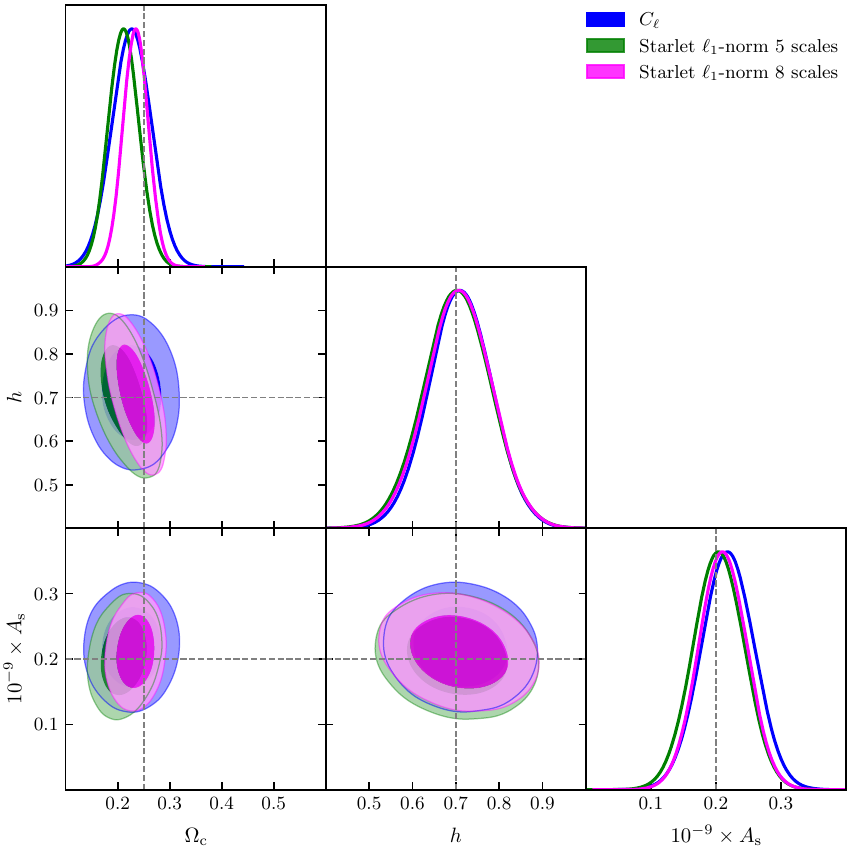}
    \includegraphics[width=0.32\textwidth]{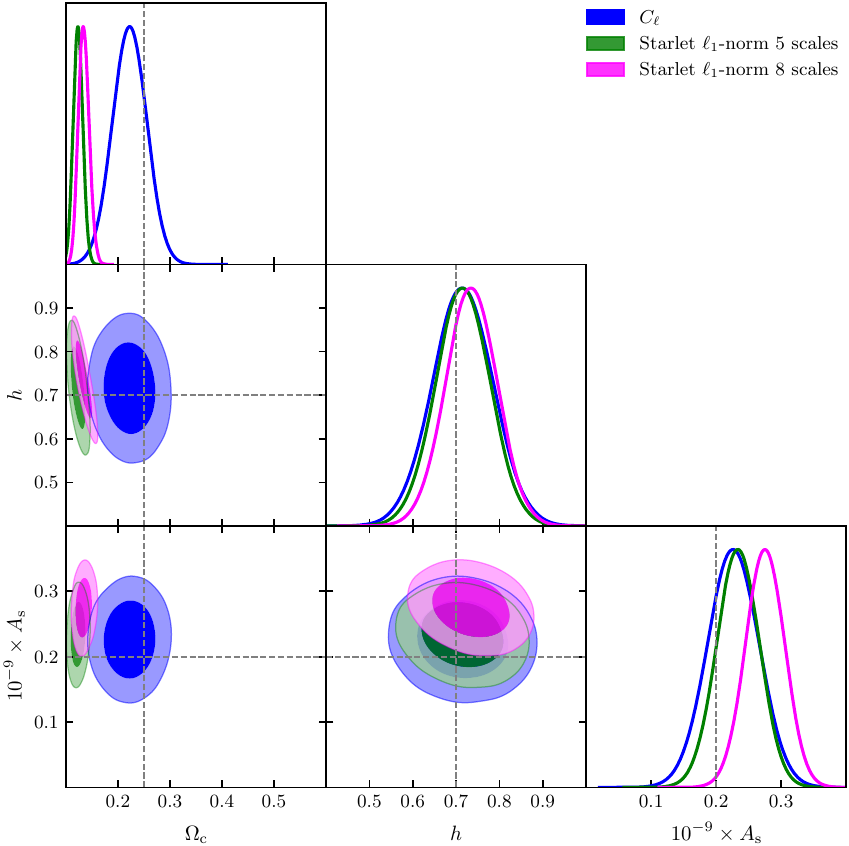}
    \caption{
    Posterior distributions for $\Omega_{\rm c}$, $h$, and $A_{\rm s}$ obtained from the angular power spectrum $C_\ell$, 
    starlet $\ell_1$-norm for different extragalactic point sources residual fractions 0.001 \% (\textit{first panel}), 0.0025 \% (\textit{second panel}), and 0.005 \% (\textit{third panel}).    Contours correspond to $68\%$ and $95\%$ confidence levels.
    }
    \label{fig:PS}
\end{figure*}

\begin{figure*}
    \centering
    \begin{minipage}{0.25\textwidth}
        \includegraphics[width=\textwidth]{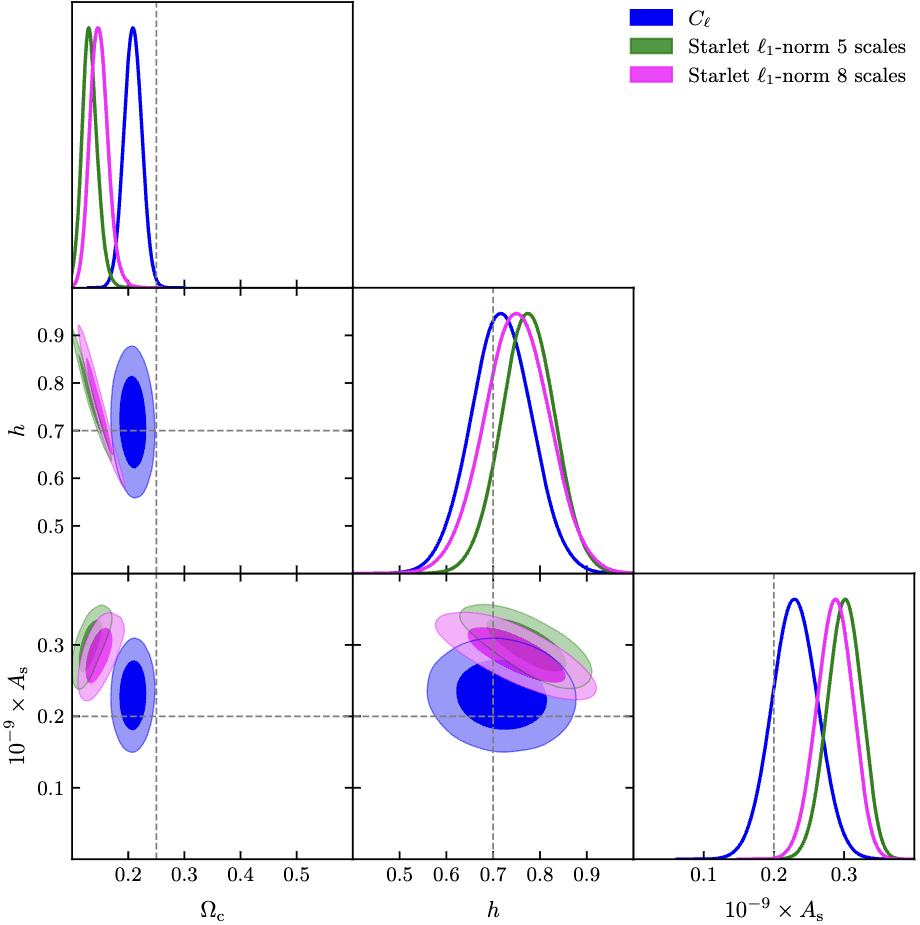}
    \end{minipage}%
    \begin{minipage}{0.25\textwidth}
        \includegraphics[width=\textwidth]{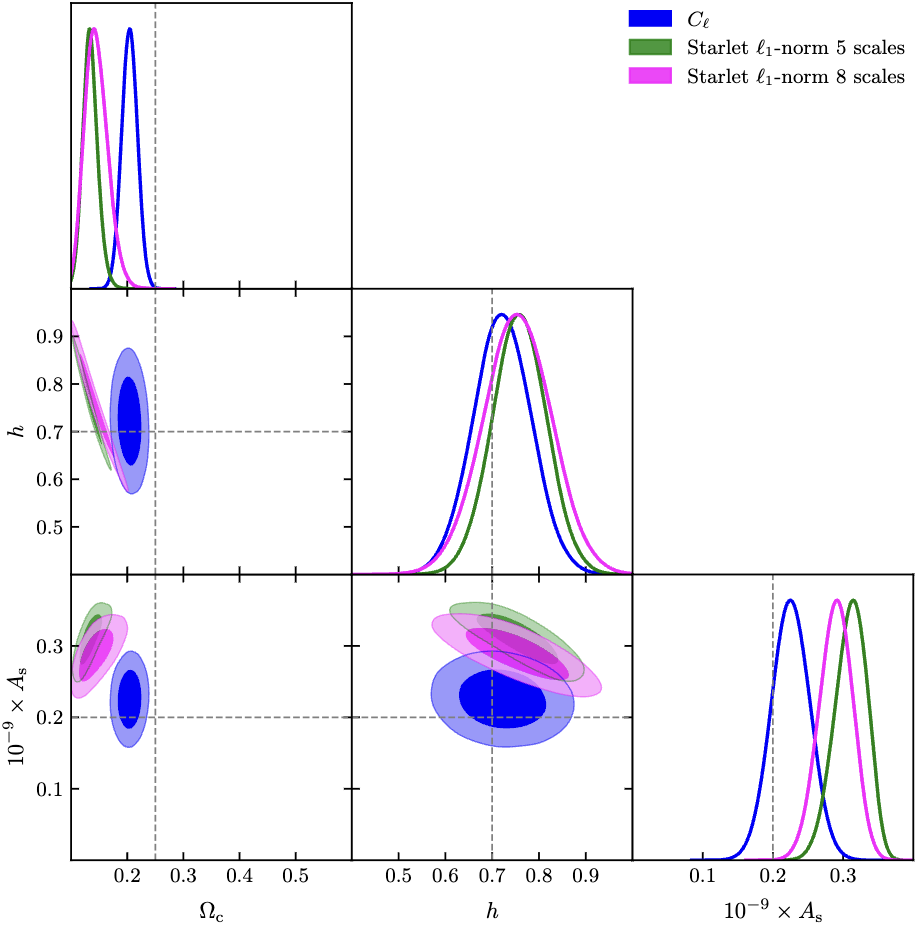}
    \end{minipage}%
    \begin{minipage}{0.25\textwidth}
        \includegraphics[width=\textwidth]{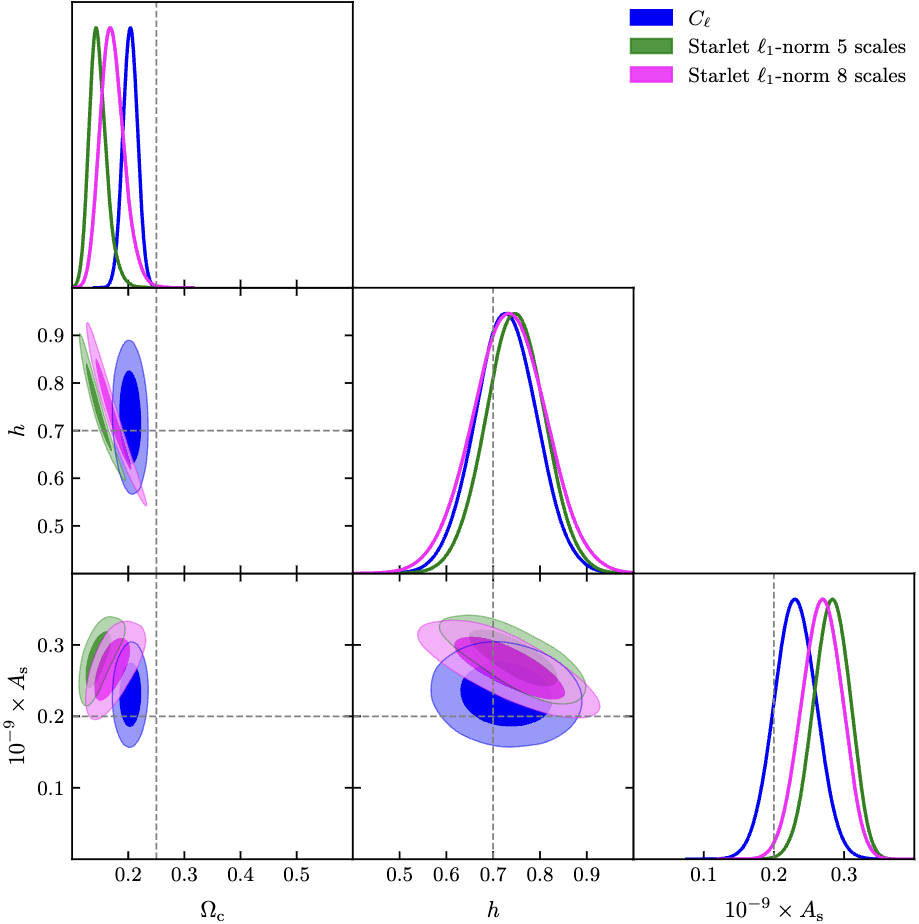}
    \end{minipage}%
    \begin{minipage}{0.25\textwidth}
        \includegraphics[width=\textwidth]{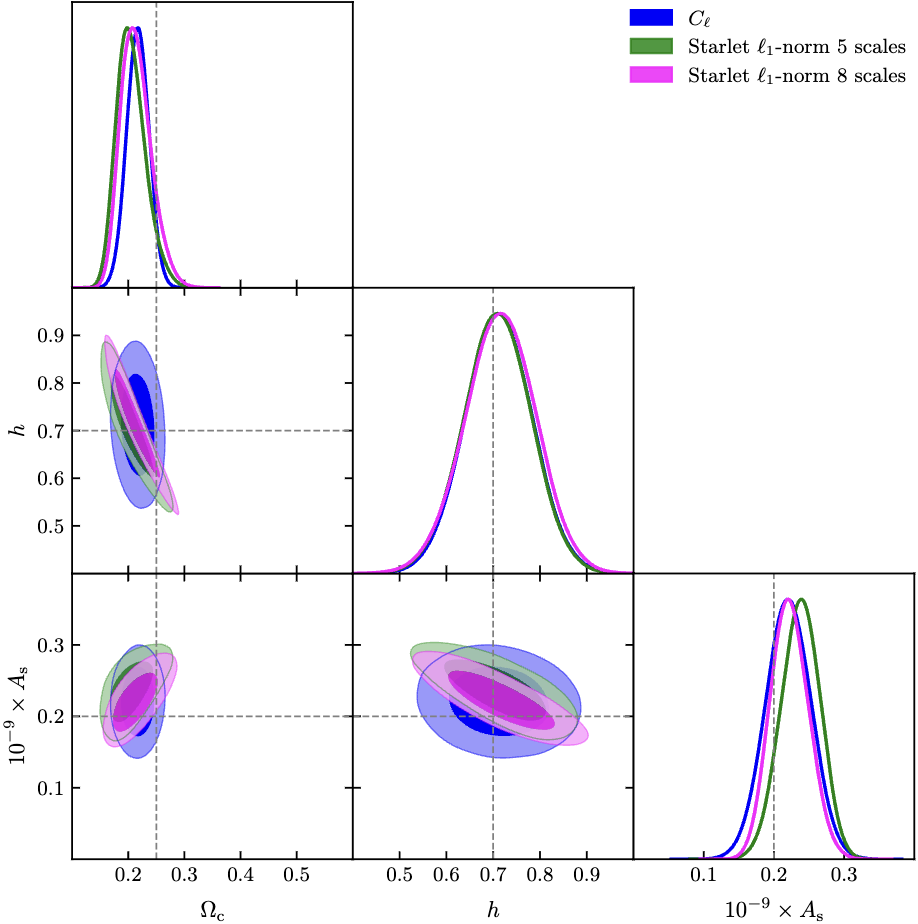}
    \end{minipage}
    \caption{Posterior distributions for $\Omega_{\rm c}$, $h$, $A_{\rm s}$, and another parameter obtained from the angular power spectrum $C_\ell$, 
starlet $\ell_1$-norm for different additive X shape weights: 10X (\textit{first panel}), 25X (\textit{second panel}), 35X (\textit{third panel}), and 50X (\textit{fourth panel}). 
Contours correspond to $68\%$ and $95\%$ confidence levels.}

    \label{fig:scan_add_SBI}
\end{figure*}

\section{Validation of the inference pipeline}\label{sec:validation}
To validate the reliability of our SBI framework, we performed a series of standard diagnostics, including parameter recovery checks, monitoring of the validation loss during training, coverage probability diagnostics, and posterior predictive checks. These diagnostics collectively assess whether the learned posterior distributions are statistically calibrated and informative.
\subsection{Parameter recovery}
\begin{figure*}
\centering
\includegraphics[width=\textwidth]{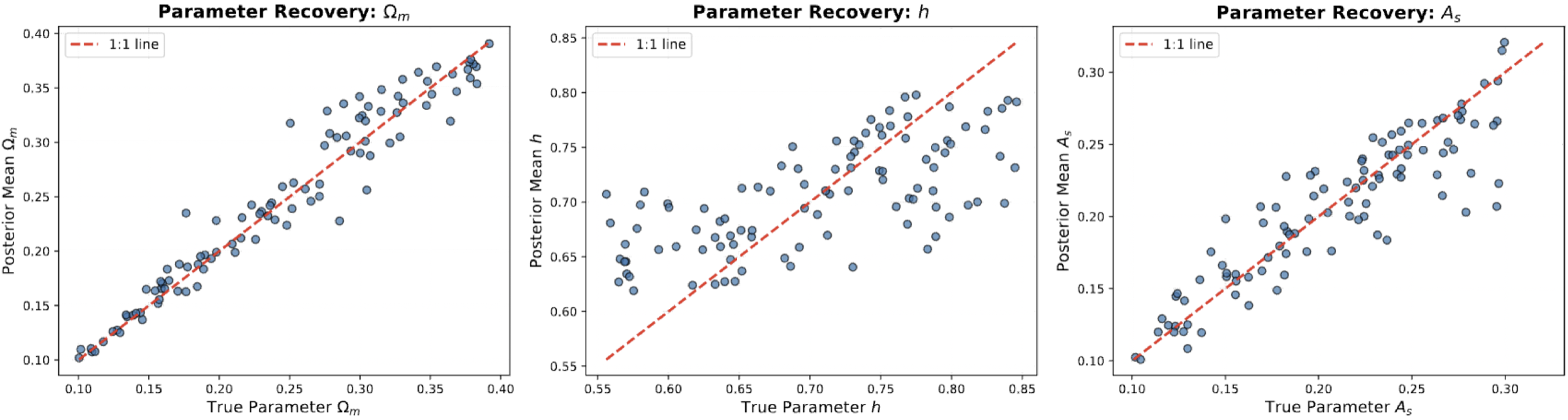}
\caption{Parameter recovery for $\Omega_{\rm c}$ (\textit{left}), $h$ (\textit{middle}), and $A_{\rm s}$ (\textit{right}) using the $C_\ell$ summary. The dashed red line indicates perfect recovery.}
\label{fig:recovery_Cl}
\end{figure*}

\begin{figure*}
\centering
\includegraphics[width=\textwidth]{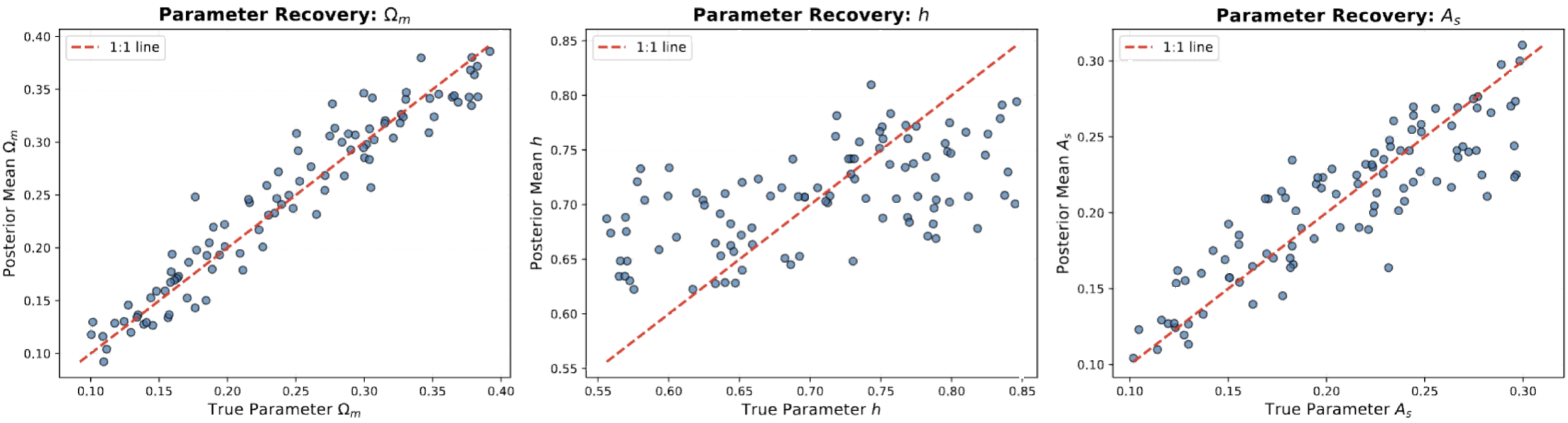}
\caption{Parameter recovery for $\Omega_{\rm c}$ (\textit{left}), $h$ (\textit{middle}), and $A_{\rm s}$ (\textit{right}) using the starlet $\ell_1$-norm summary. The dashed red line indicates perfect recovery.}
\label{fig:recovery_l1}
\end{figure*}

We drew a set of simulated observations from the prior and inferred posteriors using the trained model, comparing the true parameters to the posterior means and credible intervals. As shown in Figs. \ref{fig:recovery_Cl} and \ref{fig:recovery_l1}, the true values fall within the 68\% credible intervals in most cases, exhibiting strong correlation along the 1:1 line. Both summary statistics recover $\Omega_{\rm c}$ and $A_{\rm s}$ accurately, with predictions tightly clustered around the identity line. The Hubble parameter $h$ shows increased scatter, reflecting its weaker sensitivity to the summary statistics relative to $\Omega_{\rm c}$ and $A_{\rm s}$, consistent with expectations. Overall, these results confirm that the starlet $\ell_1$-norm and $C_\ell$ summary statistics both provide reliable and accurate parameter estimates, demonstrating the capability of our inference pipeline to recover the generative parameters across the prior range.
\subsection{Validation loss curve}
\begin{figure*}
\centering
\begin{subfigure}{0.48\textwidth}
\centering
\includegraphics[width=\linewidth]{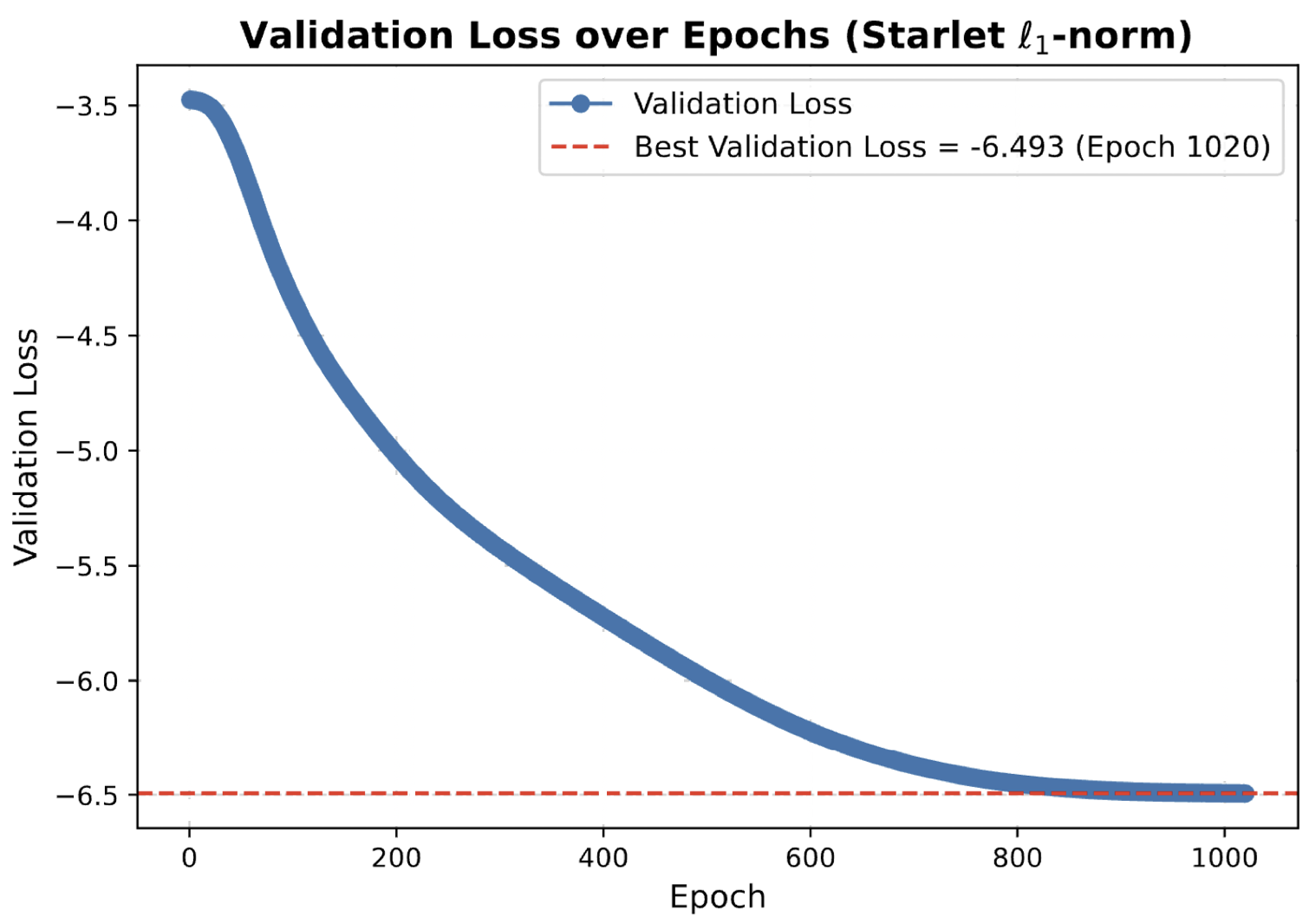}
\caption{}
\label{fig:l1_loss_curve}
\end{subfigure}
\hfill
\begin{subfigure}{0.48\textwidth}
\centering
\includegraphics[width=\linewidth]{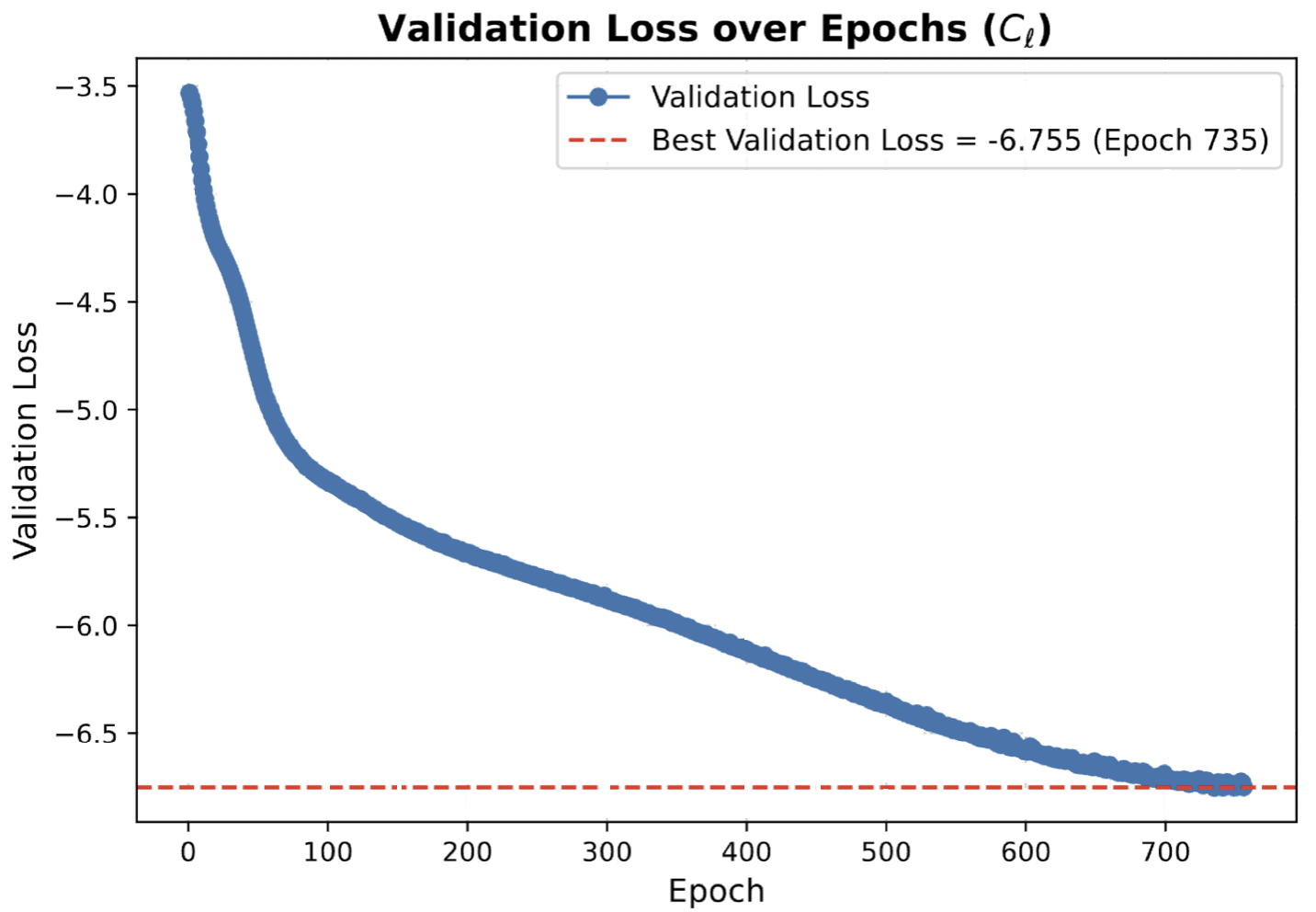}
\caption{}
\label{fig:Cl_loss_curve}
\end{subfigure}
\caption{Validation loss curves for the two summary statistics: starlet $\ell_1$-norm summary (\textit{panel a}) and $C_\ell$ summary (\textit{panel b}). Both models show stable convergence with no evidence of overfitting.}
\label{fig:val_loss_combined}
\end{figure*}

We tracked the validation loss throughout training to assess model convergence and prevent overfitting. The loss, shown in Fig. \ref{fig:val_loss_combined}, decreases steadily before saturating, with no signs of divergence, indicating stable optimization behavior.
\subsection{Coverage probability diagnostic}
\begin{figure*}
\centering
\includegraphics[width=0.48\textwidth]{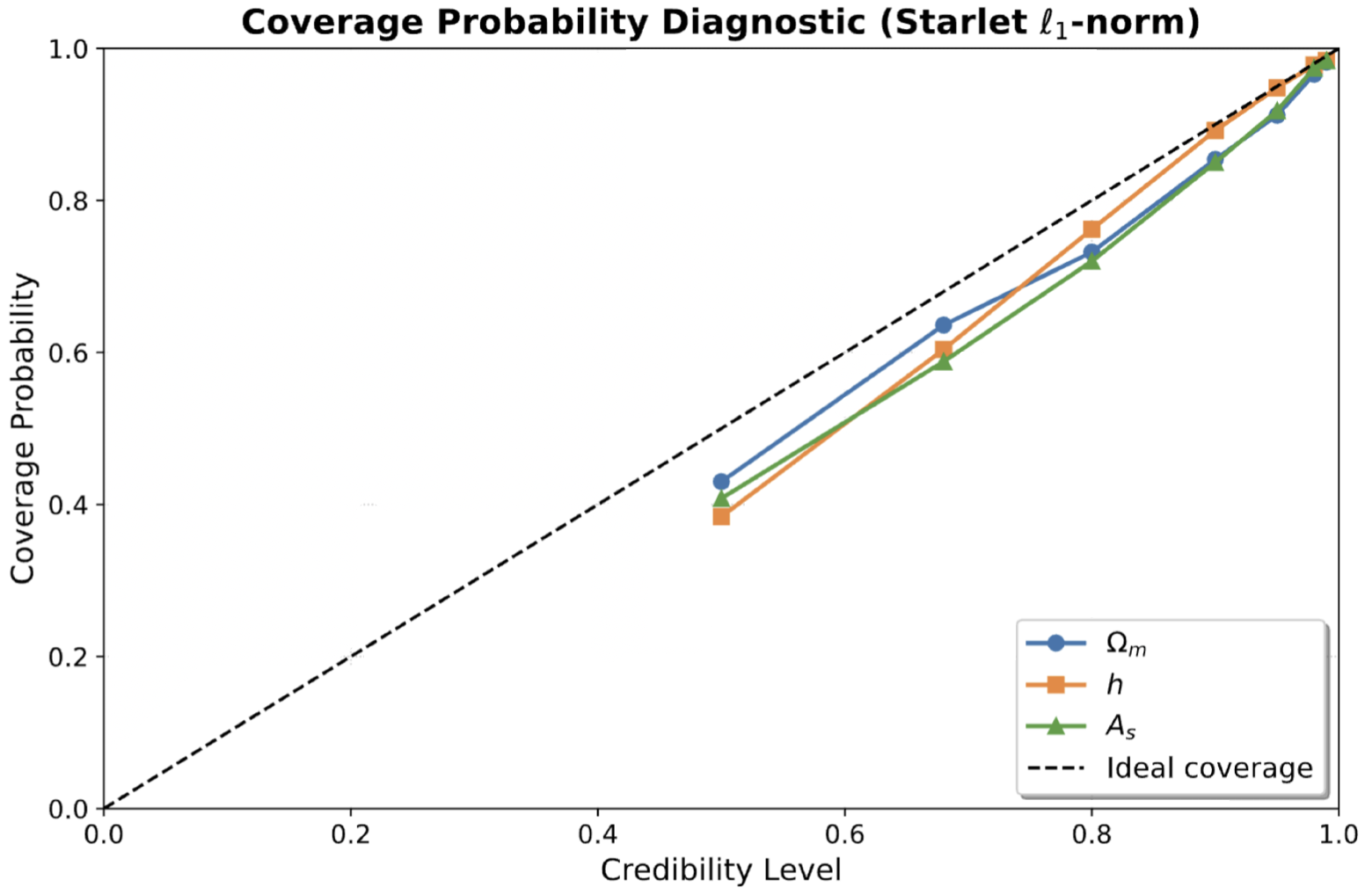}
\includegraphics[width=0.48\textwidth]{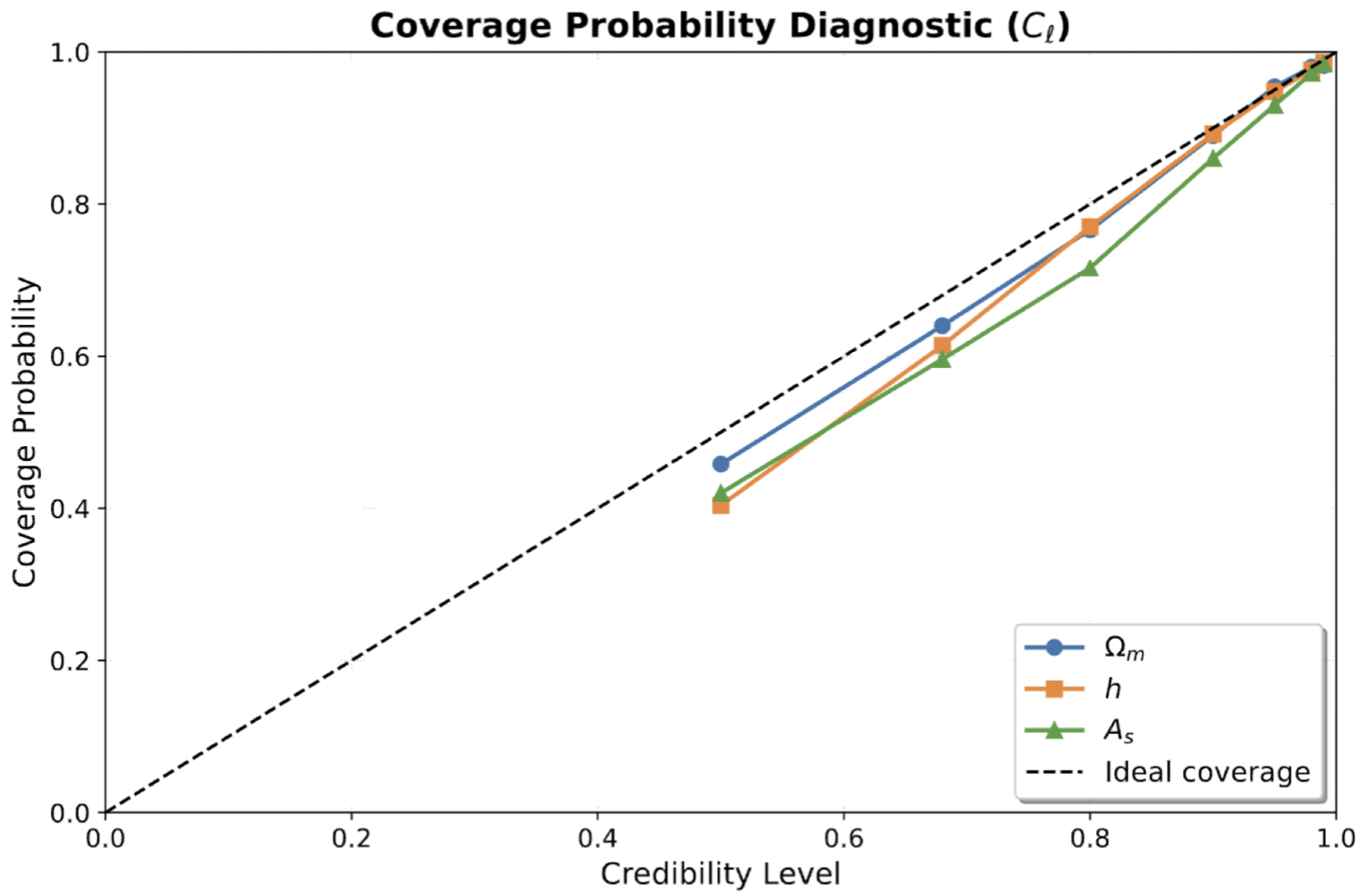}
\caption{Empirical coverage probability for Bayesian credible regions derived from two summary statistics: the starlet $\ell_1$-norm (\textit{left}) and the $C_\ell$ power spectrum (\textit{right}). The empirical coverage is shown as a function of nominal credibility for $\Omega_{\rm c}$ (blue circles), $h$ (orange squares), and $A_{\rm s}$ (green triangles). The dashed diagonal line indicates ideal coverage. Deviations from the diagonal correspond to under- or over-coverage.}
\label{fig:coverage_diagnostics}
\end{figure*}

To evaluate the statistical calibration of the posterior, we computed coverage probabilities for nominal credible levels. We generated a large number of mock datasets by sampling parameters from the prior and producing corresponding data through the forward model. For each dataset, we computed the posterior using our trained neural density estimator and derived marginal credible intervals at multiple levels (50\%, 68\%, 80\%, 90\%, 95\%, and 99\%). The coverage probability at each level was estimated as the fraction of cases in which the ground-truth parameter lies within the corresponding credible interval.
Figure \ref{fig:coverage_diagnostics} shows the empirical coverage probability as a function of nominal credibility level for the two summary statistics: the starlet $\ell_1$-norm (left panel) and the angular power spectrum $C_\ell$ (right panel). Each curve corresponds to one of the cosmological parameters: $\Omega_{\rm c}$, $h$, and $A_{\rm s}$. The ideal coverage line (dashed) represents perfect calibration. The results indicate that the method is generally well calibrated, particularly at higher credibility levels, with slight under-coverage at lower levels.
\subsection{Posterior predictive check}
\begin{figure*}
\centering
\begin{subfigure}{0.48\textwidth}
\centering
\includegraphics[width=\linewidth]{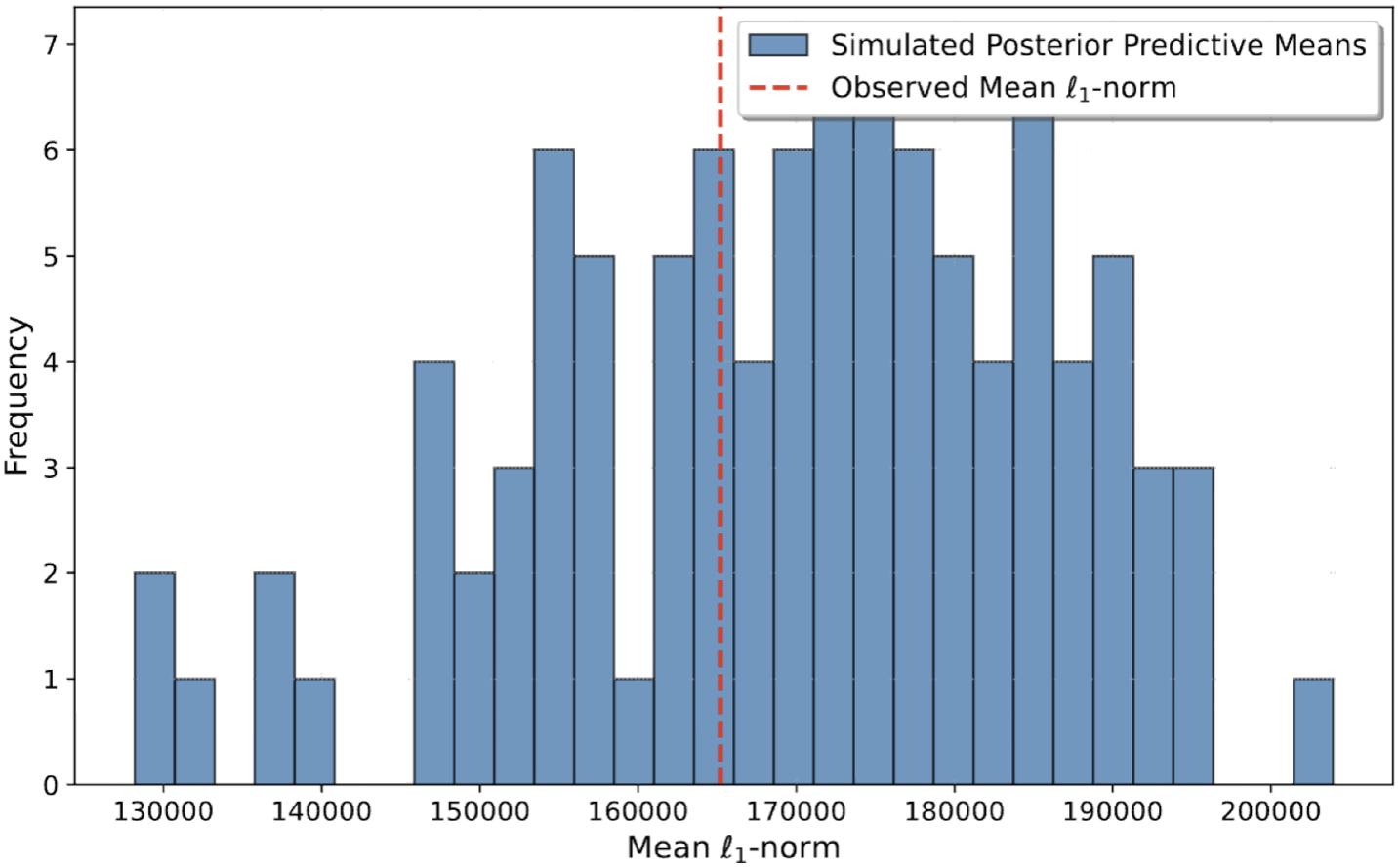}
\caption{}
\label{fig:ppc_l1}
\end{subfigure}
\hfill
\begin{subfigure}{0.48\textwidth}
\centering
\includegraphics[width=\linewidth]{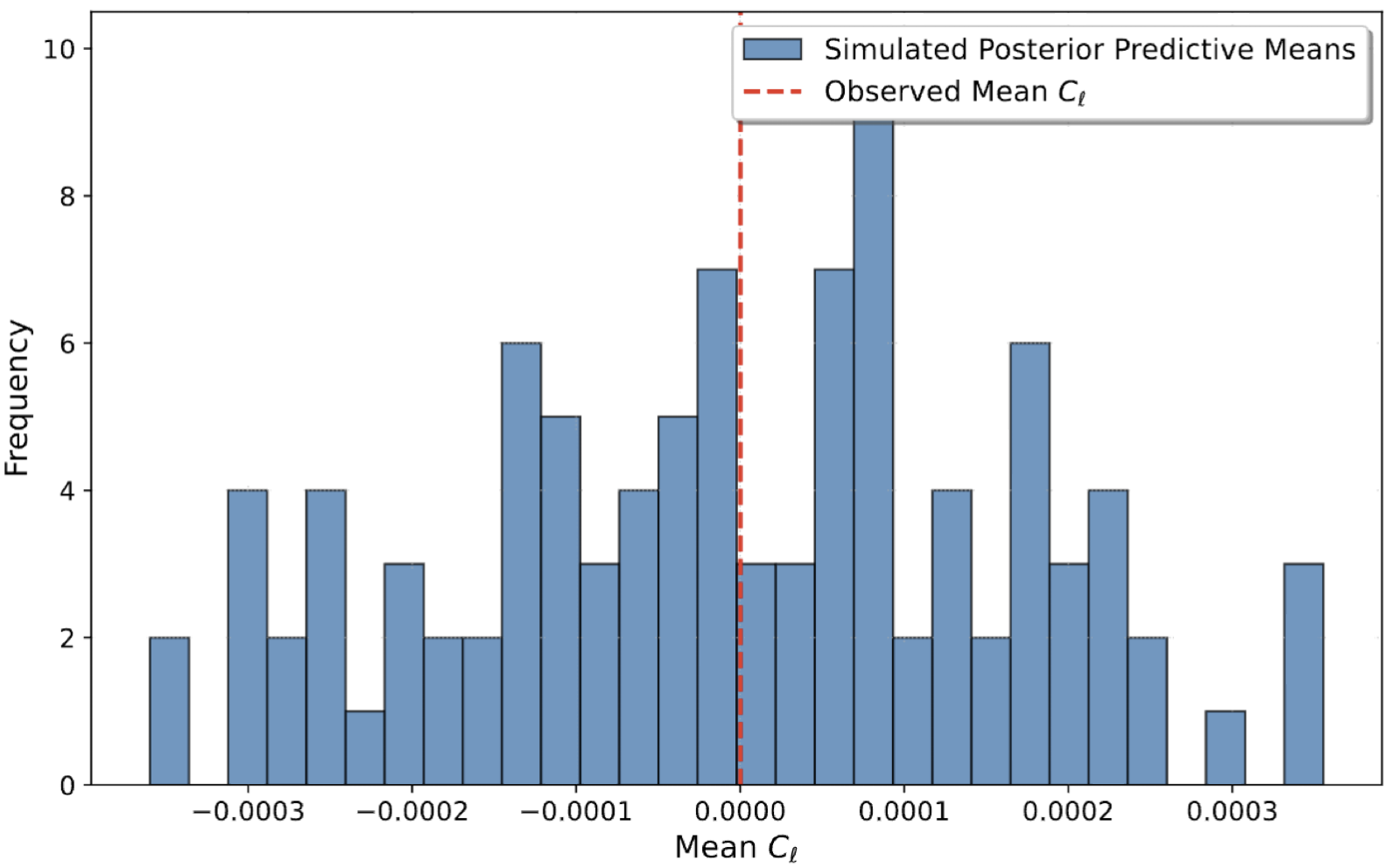}
\caption{}
\label{fig:ppc_cl}
\end{subfigure}
\caption{Posterior predictive checks for the two summary statistics: starlet $\ell_1$-norm (\textit{panel a}) and $C_\ell$ (\textit{panel b}). Histograms show the distribution of simulated summary statistics from posterior samples, with the observed value indicated as a dashed red line. The observed statistics lie well within the simulated distributions, suggesting an adequate description of the data.}
\label{fig:ppc_fig}
\end{figure*}

To assess the adequacy of our inferred posteriors and the consistency of the simulations with the observed data, we performed posterior predictive checks (PPCs) using both summary statistics. For each posterior sample, we generated mock data and computed the corresponding summary statistic. The distribution of these simulated statistics is shown in Fig. \ref{fig:ppc_fig}, with the observed value indicated as a dashed~red line.
In both cases, the observed summary lies well within the central bulk of the simulated distribution, indicating no significant tension between the observed data and the generative model conditioned on the inferred parameters. This suggests that the model is capable of producing data statistically consistent with the observations, providing further confidence in the fidelity of our inference.

\end{appendix}

\end{document}